\documentclass[fleqn,usenatbib]{mnras}

\usepackage[T1]{fontenc}

\DeclareRobustCommand{\VAN}[3]{#2}
\let\VANthebibliography\thebibliography
\def\thebibliography{\DeclareRobustCommand{\VAN}[3]{##3}\VANthebibliography}

\usepackage{graphicx}	
\usepackage{amsmath}	
\usepackage{mathtools} 
\usepackage{amssymb}
\usepackage{float}
\usepackage{txfonts}
\usepackage{xcolor}

\newcommand{\bz}{$\langle B_z \rangle$}
\newcommand{\nz}{$\langle N_z \rangle$}
\newcommand{\vsini}{$v \sin i$}
\newcommand{\kms}{km\,s$^{-1}$}

\newcommand{\msun}{M$_\odot$}

\newcommand{\teff}{$T_{\rm eff}$}
\newcommand{\ra}{$R_{\rm A}$}
\newcommand{\rk}{$R_{\rm K}$}

\title[Spectropolarimetry of $\rho$\,Oph\,C]{Revisiting the centrifugal breakout scaling relationship for H$\alpha$ emission in light of $\rho$\,Oph\,C, the coolest B-type star with an H$\alpha$-bright centrifugal magnetosphere}

\author[M. E. Shultz et al.]{
M. E. Shultz,$^{1}$\thanks{E-mail: matt.shultz@gmail.com}
E. Alecian,$^{2}$
I.\ Berry,$^{3}$
J.\ Krti{\v{c}}ka,$^{4}$
J.\ Labadie-Bartz,$^{5,6}$
P.\ Leto,$^{7}$
S.\ P.\ Owocki,$^{8}$
\newauthor{
A.\ ud-Doula,$^{9}$
and G.\ A.\ Wade$^{10}$
}
\\
$^{1}$Independent Researcher\\
$^{2}$Universit\'e Grenoble Alpes, IPAG, F-38000 Grenoble, France\\
$^{3}$Institute for Astronomy, University of Hawai'i at Mānoa, Honolulu, HI 96822 USA\\
$^{4}$Department of Theoretical Physics and Astrophysics, Faculty of Science, Masaryk University, CZ-611 37 Brno, Czech Republic\\
$^{5}$DTU Space, Technical University of Denmark, Elektrovej 327, Kgs., Lyngby, 2800, Denmark\\
$^{6}$LIRA, Paris Observatory, PSL University, CNRS, Sorbonne University, Universit\'e Paris, Cit\'e, CY Cergy University, 5 place Jules Janssen, 92195 Meudon, France\\
$^{7}$INAF - Osservatorio Astrofisico di Catania, Via S. Sofia 78, 95123 Catania, Italy\\
$^{8}$Department of Physics and Astronomy, University of Delaware, 217 Sharp Lab, Newark, Delaware, 19716, USA\\
$^{9}$Department of Physics, Penn State Scranton, Dunmore, PA 18512, USA\\
$^{10}$Department of Physics and Space Science, Royal Military College of Canada, Kingston, Ontario K7K 7B4, Canada\\
}

\date{Accepted XXX. Received YYY; in original form ZZZ}

\pubyear{\the\year{}}

\begin{document}
\label{firstpage}
\pagerange{\pageref{firstpage}--\pageref{lastpage}}
\maketitle

\begin{abstract}
The corotating magnetospheres of B-type stars are strongly affected by magnetic field strength and rotational velocity, which determine both the geometry and strength of visible and radio emission; however, H$\alpha$ emission largely disappears below a bolometric luminosity of $\log{(L / L_\odot)} \sim 2.8$. The nearby B5\,V star $\rho$\,Oph\,C (HD\,147932) is the least-luminous rapidly rotating magnetic B-type star to display magnetospheric H$\alpha$ emission. While $\rho$\,Oph\,C's radio magnetosphere has been extensively studied, its magnetic field and H$\alpha$ magnetosphere have been constrained only via snapshot observations. We present a new dataset of high-resolution ESPaDOnS spectropolarimetry covering the star's rotational period, which is revisited by combining {\em Hipparcos}, {\em K2} and {\em TESS} photometry. We find that the star's magnetic field strength and geometry are consistent with the model inferred from radio data. As expected from the magnetic model, there is very little variability in H$\alpha$. Comparison of the light curve to both equivalent width measurements from metallic lines and magnetospheric models suggests that the star's photometric variability is dominated by surface chemical abundance spots; however, there is complex rotationally modulated variability at about the mmag level which probably originates in the magnetosphere. $\rho$\,Oph\,C is the least luminous B-type star in which magnetospheric H$\alpha$ emission has yet been detected, and its emission is notably weaker than the predictions of the centrifugal breakout model. Its photometric eclipse depth is also much lower than predicted by the CBO model. However, its radio luminosity is consistent with CBO predictions. We speculate on this basis that $\rho$\,Oph\,C may be a transitional object between effective temperature regimes with hydrogen-dominated and decoupled metallic winds.
\end{abstract}




\section{Introduction}

Circumstellar magnetospheres form when the radiatively driven wind plasma of an early-type (OBA) star is trapped by a sufficiently strong magnetic field \citep[typically a few hundred to a few thousand gauss;][]{bm1997,ud2002,petit2013}. The outer extent of the magnetosphere is set by the point at which the ram pressure of the wind exceeds the magnetic pressure, which typically occurs from a few stellar radii out to tens of stellar radii \citep[][]{ud2002,petit2013}. These magnetospheres can be detected via a variety of diagnostics, each of which probes a distinct element of the magnetosphere: incoherent gyrosynchrotron radio emission reveals the high energy radio lobes \citep[e.g.][]{1987ApJ...322..902D,2021MNRAS.507.1979L}; coherent, pulsed radio electron cyclotron maser emission (ECME) is formed within auroral circuits \citep[e.g.][]{2000A&A...362..281T,2022ApJ...925..125D}; magnetically confined wind shocks are detectable from X-ray emission \citep[e.g.][]{bm1997,2014ApJS..215...10N}; ultraviolet emission and absorption probes the magnetically confined wind \citep[e.g.][]{henrichs2013}; and the relatively dense, cool plasma of the circumstellar environment shows up in H Balmer, Paschen, and Brackett line emission, as well as in photometric eclipses \citep[e.g.][]{lb1978,grun2012,petit2013,2015A&A...578A.112O}.

Due to magnetospheric braking \citep{ud2009}, magnetic hot stars are usually much more slowly rotating than their non-magnetic counterparts, with some stars having rotational periods of decades \citep[e.g.][]{2010MNRAS.407.1423M,2017MNRAS.471.2286S,2018MNRAS.475.5144S}. If rotation does not play an important role in the magnetosphere, the magnetically confined plasma persists within the magnetosphere over dynamical timescales: the magnetosphere is loaded by the wind, and drained by gravitational infall of the plasma back to the photosphere \citep{ud2002}. Such dynamical magnetospheres \citep[DMs;][]{petit2013} require high mass-loss rates to sustain sufficient density for H line emission, conditions that typically only pertain in magnetic O-type stars \citep[the magnetic B0 star $\xi^1$\,CMa is the only known B-type star with a DM detectable in H$\alpha$;][]{2017MNRAS.471.2286S}. 

Closer to the Zero-Age Main Sequence, magnetospheric braking has not had time to slow down the star's rotation; within this age range there are many rapidly rotating stars, with periods as short as 0.5~d and surface equatorial rotational velocities in excess of 300 \kms~\citep[][]{2019MNRAS.490..274S}. Due to magnetically enforced corotation of the confined wind plasma with the stellar surface, the magnetospheric plasma can experience rotational velocities of thousands of \kms, leading to significant centrifugal forces \citep[][]{town2005c,ud2008}. This leads to the formation of a centrifugal magnetosphere \citep[CM;][]{petit2013} in which rotational support prevents gravitational infall, allowing high densities and detectable H line emission. 

Since mass loading by the wind is not counteracted by gravitational drainage within the CM, the limiting density is set by the strength of the magnetic field. When the gas pressure exceeds the magnetic pressure, the plasma ceases to be guided by the magnetic field, and instead the magnetic field moves with the plasma; the centrifugal force then causes the the magnetic field to reconnect, and plasma is flung away from the star in centrifugal breakout (CBO) events \citep{ud2008}. CBO scaling has been shown to reproduce the broad characteristics of H line emission in magnetic early B-type stars: the threshold for emission onset; the scaling of emission strength; and the shape of H line emission profiles \citep{2020MNRAS.499.5379S,2020MNRAS.499.5366O}. CBO has also been shown to drive radio emission, which is seen only in rapidly rotating, strongly magnetized stars \citep{2021MNRAS.507.1979L,2022MNRAS.513.1429S,2022MNRAS.513.1449O}. A CBO model predicts both the onset of radio emission and the scaling of radio luminosity. CBO events involve magnetic reconnection, which accelerates a small population of electrons to weakly relativistic velocities, thereby populating the radio magnetosphere; thus, while H$\alpha$ emission probes and the cool, dense, pre-CBO plasma, radio emission probes the high-energy, low density, post-CBO plasma. While the original radio CBO scaling relationship was derived for gyrosynchrontron emission, there are indications that it is also the source of ECME \citep[e.g.][]{2021ApJ...921....9D,2022MNRAS.517.5756D}.

While CBO successfully reproduces the H line emission properties of stars that possess such emission, it cannot yet explain why emission disappears for stars with effective temperatures below about 17 kK \citep[or more or less equivalently, bolometric luminosities below $\log{L_{\rm bol}} \sim 2.8$;][]{2020MNRAS.499.5379S,2020MNRAS.499.5366O}. Cooler, less luminous stars have weaker winds, but in a strict CBO model, the mass-loss rate should not factor in at all: mass-loading should continue until such a point as the CBO-limited density is reached. Indeed, no such effective temperature cutoff is observed in radio emission: both gyrosynchrotron and ECME are seen even in relatively cool magnetic A-type stars \citep[][]{2021MNRAS.507.1979L,2022MNRAS.513.1429S,2022MNRAS.517.5756D}. Two hypotheses have been proposed to explain the lack of H emission below the temperature cutoff. First, magnetospheres might be `leaky' due to non-ideal MHD effects leading to cross-field transport: a slow leakage mechanism might only become relevant at low mass-loss rates \citep[e.g.][]{2018MNRAS.474.3090O}. Second, H ions might become decoupled from metallic ions in the winds of the cooler stars \citep[e.g.][]{1995A&A...301..823B}. In conventional winds, radiative driving operates via energy level transitions in ionized metals, which then drag along the ionized H. If H decouples, the magnetosphere would be populated entirely by metallic ions, a `metallic magnetosphere' which cannot produce H emission \citep{Owocki2026}.

Here we present an analysis of the nearby B5\,V star $\rho$\,Oph\,C (HD\,147932), which is the least luminous magnetic B-type star to display H line emission consistent with an origin in a CM. $\rho$\,Oph\,C's X-ray and radio emission properties have been extensively studied \citep{2016A&A...592A..88P,2020MNRAS.499L..72L}. However its magnetic field strength and geometry have remained only weakly constrained due to a paucity of spectropolarimetry, while its H line emission variability has gone unexplored due to the small number of high-resolution spectra \citep{alecian2014}. As a result of this, the magnetic field model inferred by \cite{2020MNRAS.499L..72L} from radio data has not yet been spectropolarimetrically confirmed.

Section \ref{sect:obs} describes the newly acquired high-resolution spectropolarimetric dataset, which samples the star's full rotational period, along with the new {\em TESS} light curve. Stellar parameters are determined in Sect.\ \ref{sect:stellarpars}. In Sect.\ \ref{sect:rot} we refine the rotational period using the photometric time series data. The magnetic field properties are determined in Sect.\ \ref{sect:magfld}. Magnetospheric H$\alpha$ emission properties are examined in Sect.\ \ref{sect:magnetospheres}, along with an investigation of the {\em TESS} light curve through the lens of magnetospheric eclipsing. The discussion in Sect.\ \ref{sect:discuss} situates $\rho$\,Oph\,C in the context of the full sample of early B-type stars with magnetospheric H line emission. Results are summarized in Sect.\ \ref{sect:summary}.

\section{Observations}\label{sect:obs}

\subsection{Spectropolarimetry}\label{subsect:esp}

\subsubsection{HARPSpol}

The High Accuracy Radial velocity Planet Searcher polarimeter (HARPSpol) is a high-resolution ($R \sim 110,000$) \'echelle spectropolarimeter mounted at the 3.6 m European Southern Observatory (ESO) La Silla telescope \citep{2011Msngr.143....7P}. It covers the wavelength region between 380 nm and 690 nm across 71 mostly overlapping spectral orders, with a gap between 526 and 534 nm. Spectropolarimetric sequences consist of 4 differently polarized sub-exposures yielding 4 Stokes $I$ spectra, 1 Stokes $V$, $Q$, or $U$ spectrum, and one $N$ spectrum. Reduction and analysis of HARPSpol data were described in detail by \cite{2016MNRAS.456....2W}. 

Two HARPSpol observations of $\rho$\,Oph\,C were reported by \cite{alecian2014} as part of the MiMeS Survey. Each measurement consisted of 2 spectropolarimetric sequences, one obtained on 19/02 and the other on 20/02/2013, each with a 900 s sub-exposure time, for a total exposure time for each measurement of 7200 s. Since this two-hour span corresponds to approximately 10\% of the period found by \cite{2018AJ....155..196R}, we have elected to analyze the HARPSpol observations individually, yielding instead a total of 4 spectropolarimetric measurements. The mean peak signal-to-noise ($S/N$) of these observations per spectral pixel in Stokes $I$ is 238, ranging between 192 and 278. 

\subsubsection{ESPaDonS}

As its name implies, the Echelle SpectroPolarimetric Device for the Observation of Stars (ESPaDOnS) is a high-resolution ($R \sim 65,000$ at 500~nm) \'echelle spectropolarimeter, covering the spectral region between about 370 and 1050 nm across 40 overlapping orders. It has been mounted at the Cassegrain focus of the 3.6~m Canada-France-Hawaii Telescope (CFHT) since 2004, and has been the primary work-horse for stellar magnetometry for almost two decades. It obtains spectropolarimetric sequences consisting of 4 differently polarized sub-exposures, yielding 4 unpolarized (Stokes $I$) spectra, and one polarized spectrum in either circular (Stokes $V$) or linear (Stokes $Q$ and $U$) polarization. By combining the sub-exposures in such a fashion that intrinsic source polarization cancels out, a diagnostic null $N$ spectrum is also obtained; this enables the noise level to be precisely quantified, while providing a check for normal instrument operation. The properties of ESPaDOnS and its data reduction using the Libre-ESpRIT \citep{d1997} package and the CFHT Upena pipeline were described in detail by \cite{2016MNRAS.456....2W} in the context of the MiMeS Large Programs. 

We observed $\rho$\,Oph\,C 9 times with ESPaDOnS between 18/05 and 28/05/2021, again in circular polarization mode. These observations used a subexposure time of 203 s, and were timed in order to evenly sample the 0.86~d rotation period reported by \cite{2018AJ....155..196R}. Exposure times were guided by the rotational broadening and longitudinal magnetic field strength reported by \cite{alecian2014}. The mean, minimum, and maximum peak $S/N$ of the dataset are respectively 533, 450, and 590. 

All observations were normalized to the continuum using standard order-by-order polynomial normalization \citep[e.g.][]{2018MNRAS.475.5144S}. 

\subsection{Time Domain Photometry}

\cite{2018AJ....155..196R} determined the rotational period of $\rho$\,Oph\,C using the {\em K2} light curve. The {\em K2} mission was an extension of the NASA {\em Kepler} mission \citep{2014PASP..126..398H}. The {\em Kepler} space telescope was a $\mu$mag precision photometer launched in 2009 with the goal of obtaining high-cadence, long-duration time series photometry in order to detect exoplanets via the transit method. When two of the reaction wheels failed, the {\em K2} mission extension was developed. {\em K2} used solar wind pressure to stabilize the satellite on a given field of view along the ecliptic for periods of about 3 months. $\rho$\,Oph\,C was observed during campaign 2, between 24/08 and 10/11/2014. We obtained these data from the Mikulski Archive for Space Telescopes (MAST). 

$\rho$\,Oph\,C was also observed by the {\em Hipparcos} mission, a European Space Agency (ESA) orbital observatory that provided the first high-precision parallax survey, as an element of which time series photometry was collected for a large number of stars \citep{perry1997,vanleeuwen2007}. The {\em Hipparcos} dataset is not as precise or as high-cadence as the {\em K2} dataset, however it covered a longer time baseline (23/01/1990 to 18/02/1993); by combining the {\em Hipparcos} and {\em K2} data, we hope to improve the precision of the rotational period. 

{\em TESS} uses four cameras with a total field of view of $24^\circ\times 96^\circ$, with a bandpass covering 600 to 1050 nm \citep{2015JATIS...1a4003R}. The initial two-year {\em TESS} mission began in 2018, during which it completed coverage of almost the entire sky. During each year, 13 sectors were observed for 27 days each, with a nominal precision of 60 ppm hr$^{-1}$ (although this varies between fields and targets). High-priority targets are observed with a two-minute cadence, and the processed light curves made available on the MAST archive immediately following reduction. $\rho$\,Oph\,C was observed for the first and so far only time in {\em TESS} Sector 91, from 04/09/2025 to 05/06/2025, with a gap of about six days and a cadence of 2 minutes. We obtained these data from the MAST archive.

{Kepler}'s pixels were $3.98 \times 3.98$ arcseconds, much smaller than the $21 \times 21$ arcsec pixels of {\em TESS}. Contamination by nearby stars is therefore a distinct possibility, especially given the close proximity of bright stars such as $\rho$\,Oph\,A, B, and D. As discussed at length by \cite{2025A&A...700A..14S}, the {\em K2} light curve was contaminated by the $\rho$\,Oph\,A light curve; indeed, this is how $\rho$\,Oph\,A's photometric variability was first detected. This was not a function of pixel size, but of the mask originally used for light curve extraction. Despite the larger {\em TESS} pixels, there is no indication in the frequency spectrum of $\rho$\,Oph\,A's variability signature, suggesting that the {\em TESS} data are relatively clean.

\section{Stellar parameters}\label{sect:stellarpars}

\begin{table}
\centering
\caption[]{Stellar, rotational, magnetic, and magnetospheric parameters of $\rho$\,Oph\,C. Reference key: $^a$\protect\cite{2021A&A...649A...6G}, $^b$\protect\cite{2002yCat.2237....0D}, $^c$\protect\cite{2016A&A...592A..88P}.}
\label{table:stellar_pars}
\begin{tabular}{l c c}
\hline\hline
Parameter & Value & Uncertainty \\
\hline
$\pi / {\rm mas}$ & $8.12^a$ & $0.07$ \\
$d / {\rm pc}$ & $139$ & $3$ \\
$E(B-V) / {\rm mag}$ & $0.59$ & $0.02$ \\
$V / {\rm mag}$ & 7.27$^b$ & -- \\
$BC / {\rm mag}$ & $-1.5$ & $0.7$ \\
$M_{\rm bol} / {\rm mag}$ & $-1.3$ & $0.8$ \\
$\log{L_{\rm bol} / {\rm L_\odot}}$ & $2.6$ & $0.3$ \\
$T_{\rm eff} / {\rm kK}$ & $16.8$ & $1.5$ \\
$\log{g / {\rm cm~s^{-2}}}$ & $4.01$ & $0.15$\\
\vsini~$ / {\rm km~s^{-1}}$ & $145$ & $5$ \\
$R_* / {\rm R_\odot}$ & 2.6 & 0.1 \\
$M_* / {\rm M_\odot}$ & 4.8 & 0.4 \\
${\rm Age / Myr}$ & 7$^c$ & $+3/-1$ \\
\hline
$P_{\rm rot} / {\rm d}$ & 0.86389657 & 0.00000004 \\ 
$T_0 / {\rm HJD}$ & 2447914.73 & 0.01 \\ 
$i_{\rm rot} / ^\circ$ & 64 & $+6/-4$ \\ 
$v_{\rm eq} / {\rm km~s^{-1}}$ & 156 & $+7/-4$ \\ 
$R_{\rm pol}/R_{\rm eq}$ & 0.968 & $+0.001/-0.003$ \\ 
$R_{\rm K} / R_*$ & 2.43 & $+0.03/-0.06$ \\ 
\hline
$B0 / {\rm kG}$ & $-1.49$ & 0.07 \\ 
$B1 / {\rm kG}$ & $-0.45$ & 0.09 \\ 
$\Phi_1 / {\rm rad}$ & 3.1 & 0.2 \\ 
$B2 / {\rm kG}$ & $-0.27$ & 0.12 \\ 
$\Phi_2 / {\rm rad}$ & 1.9 & 0.3 \\ 
$\beta / ^\circ$ & 7 & $+2/-2$ \\ 
$B_{\rm d} / {\rm kG}$ & 10.4 & $+13.1/-0.8$ \\ 
\hline
$B_{\rm K} / {\rm kG}$ & 0.42 & $+0.19/-0.03$ \\ 
$\log{\dot{M} / {\rm M_\odot~yr^{-1}}}$ & $-10.6$ & 0.2 \\ 
$v_\infty / {\rm km~s^{-1}}$ & 1100 & 10 \\ 
$\log{\eta_*}$ & 6.6 & $+0.5/-0.2$ \\ 
$R_{\rm A} / R_*$ & 43 & $+20/-4$ \\ 
$\log{R_{\rm A}/R_{\rm K}}$ & 1.29 & $+0.10/-0.04$ \\ 
$\tau_{\rm J} / {\rm Myr}$ & 5 & 1 \\ 
$t_{\rm S,max} / {\rm Myr}$ & 7 & 1 \\ 
\hline\hline
\end{tabular}
\end{table}

\subsection{Atmospheric parameters}\label{subsec:atmos}

   \begin{figure*}
   \centering
   \includegraphics[width=0.95\textwidth]{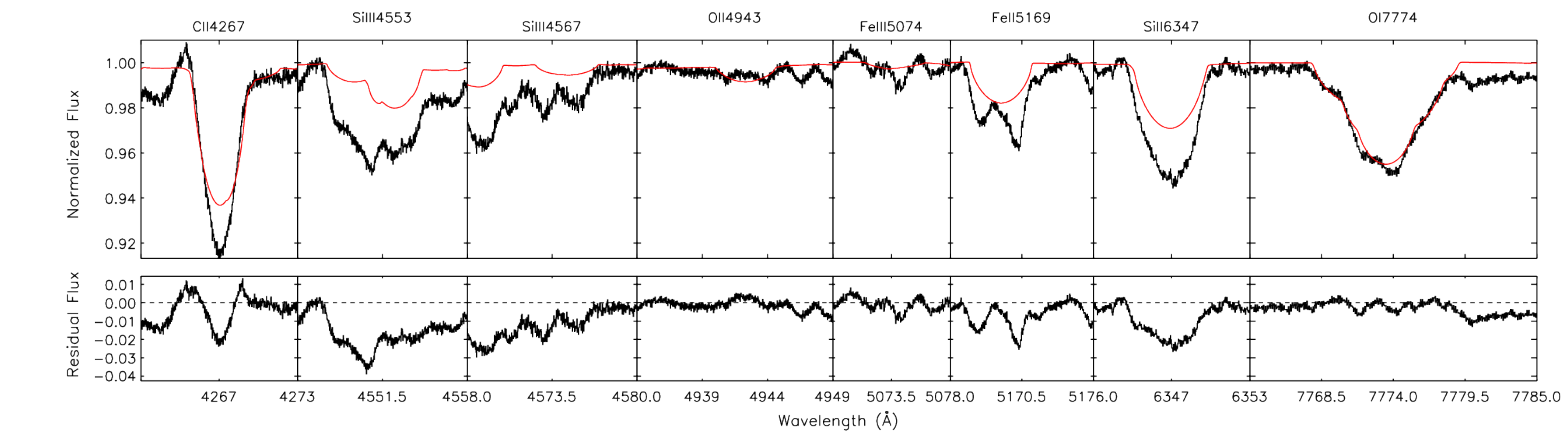} 
      \caption[]{Best-fit synthetic {\sc tlusty} spectrum (red) to the mean ESPaDOnS spectrum of $\rho$ Oph C.}
         \label{rhoOphC_met_line_fit}
   \end{figure*}

   \begin{figure}
   \centering
   \includegraphics[width=0.45\textwidth]{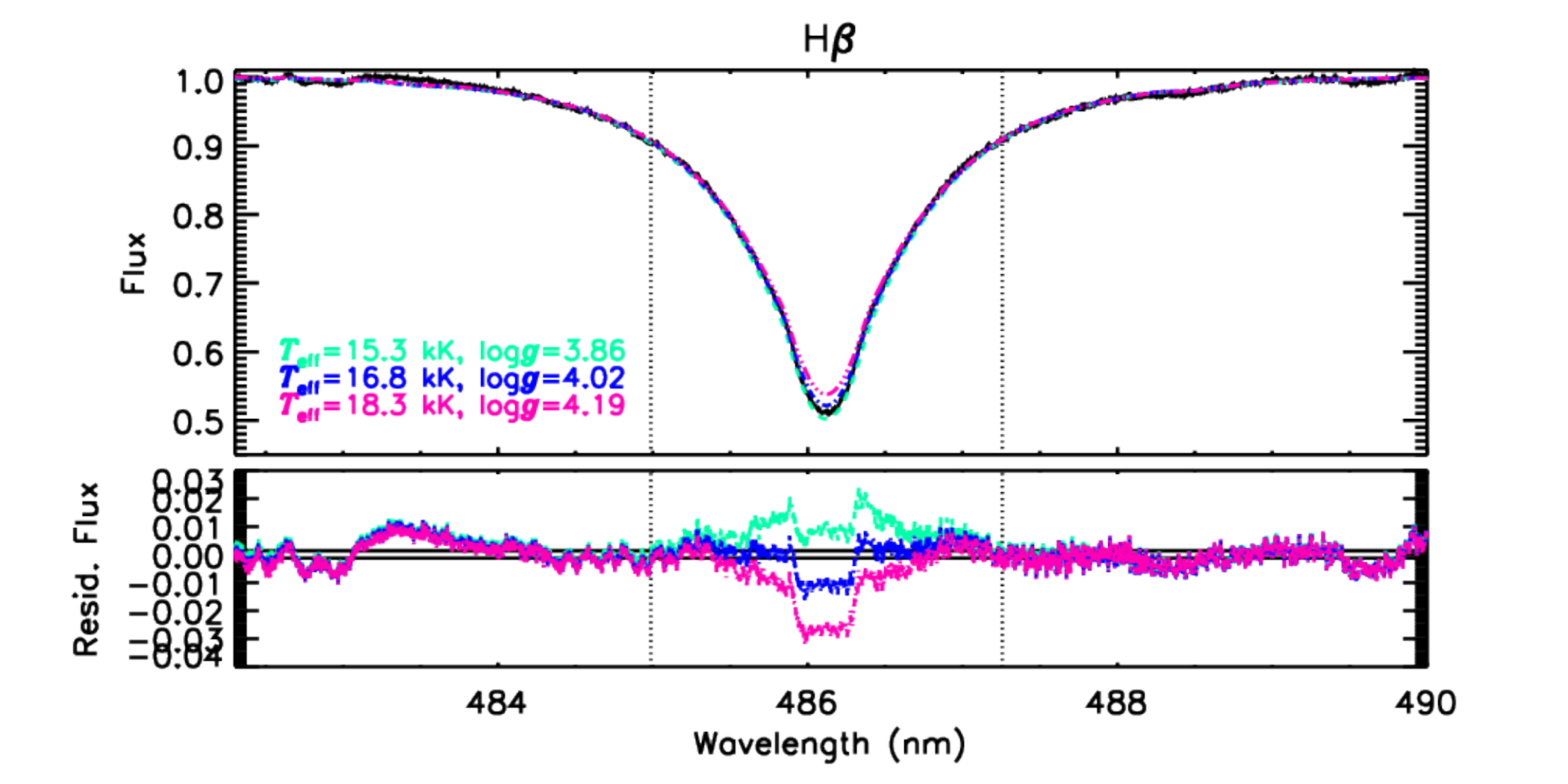}
   \includegraphics[width=0.45\textwidth]{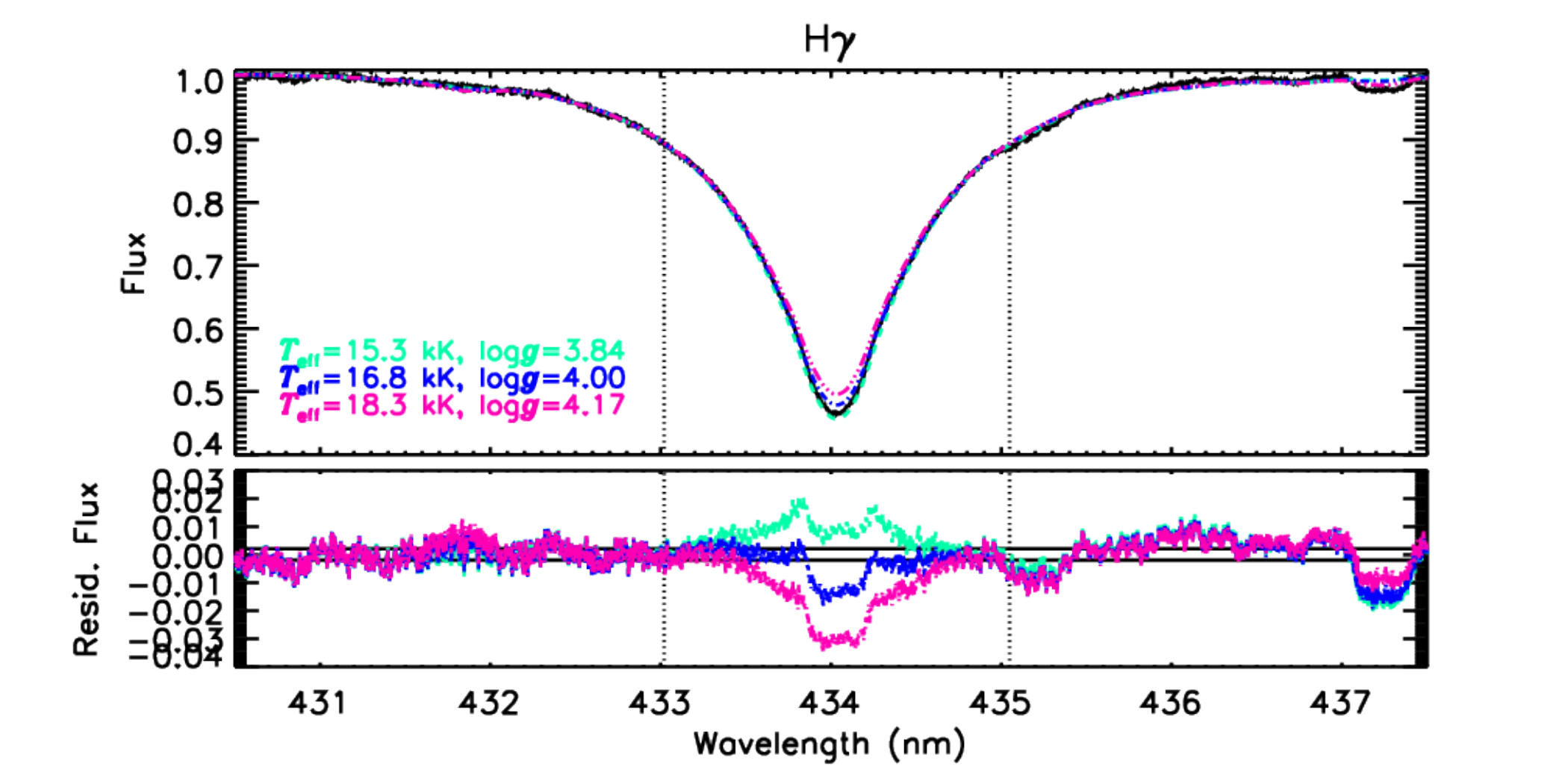} 
      \caption[]{Fits of {\sc tlusty} synthetic spectra to the H$\beta$ ({\em top}) and H$\gamma$ ({\em bottom}) line of $\rho$\,Oph\,C.}
         \label{rhoOphC_logg}
   \end{figure}

\cite{alecian2014} determined the stellar parameters of $\rho$\,Oph\,C using Johnson, Geneva, and Str\"omgren photometry, as well as via examination of the HARPSpol spectrum. They found the photometry to be consistent with \teff~$\sim 18$~kK, although the spectrum was more consistent with 17~kK, noting that the values found in the literature have a large spread between 12 kK and 18 kK. This is almost certainly due to the strongly chemically peculiar nature of the star, which exhibits weak He lines and strong and variable Si and Fe lines, as expected for a magnetic CP star in this \teff~range. Standard solar metallicity synthetic spectra therefore cannot achieve a good fit to the spectrum, as is plainly apparent from Fig.\ \ref{rhoOphC_met_line_fit} where a 17 kK {\sc tlusty} model is shown \citep{lanzhubeny2007}. Comparison of observed to synthetic spectra resulted in $v\sin{i} = 145 \pm 5$~\kms, consistent with the rotational broadening found by \cite{alecian2014}.

While the absolute depths of spectral lines are strongly affected by their abundances, the equivalent width (EW) ratios of lines formed by the same elements but with different ionizations are, to first order, independent of this affect (under the assumption that vertical abundance stratification can be ignored, and that there are no surface temperature variations in magnetic CP stars). We therefore adopted the same approach as used by \cite{2019MNRAS.485.1508S} in their analysis of the population of magnetic early B-type stars to which $\rho$\,Oph\,C belongs, i.e.\ identifying relatively isolated, strong spectral lines (in this case, Si~{\sc ii}~505.6, Si~{\sc ii}~634.7, Si~{\sc ii}~637.1, Si~{\sc iii}~456.8, P~{\sc ii}~604.3, P~{\sc iii}~422.2, S~{\sc ii}~564.0, and S~{\sc iii}~425.4~nm), measuring their EWs after renormalizing to the continuum on both sides of the line, and comparing to a grid of EWs measured from the BSTAR2006 library \citep{lanzhubeny2007}. This yielded \teff~$=16.8 \pm 1.5$~kK, with the Si lines being consistent with a value closer to 16.5 kK and the S lines more consistent with 20 kK. 

To determine the surface gravity we analyzed the H$\beta$ and H$\gamma$ lines, using the same method as used for single stars\footnote{As discussed below in Sect.\ \ref{subsubsect:rhooph_dist}, $\rho$\,Oph\,C is actually a multiple system, however the low-mass companions are much too dim to be detectable in the spectrum.} by \cite{2019MNRAS.485.1508S} i.e.\ fitting interpolated BSTAR2006 models over the range of \teff~uncertainty and identifying the minimum $\chi^2$ at each \teff. In this fit the region around $\pm 700$~\kms~of line centre was excluded in order to avoid biasing of the results by the weak H$\alpha$ emission reported by \cite{alecian2014} and examined in greater detail below in Sect.~\ref{sect:magnetospheres}. The fits are shown in Fig.\ \ref{rhoOphC_logg}; individually they yield identical results within uncertainties, and combining results from both lines yield $\log{g} = 4.01 \pm 0.15$. The surface gravity and effective temperature determined here are both consistent with the values reported by \cite{alecian2014}.

\subsection{Distance to the $\rho$ Ophichus group}\label{subsubsect:rhooph_dist}

   \begin{figure}
   \centering
   \includegraphics[width=0.45\textwidth]{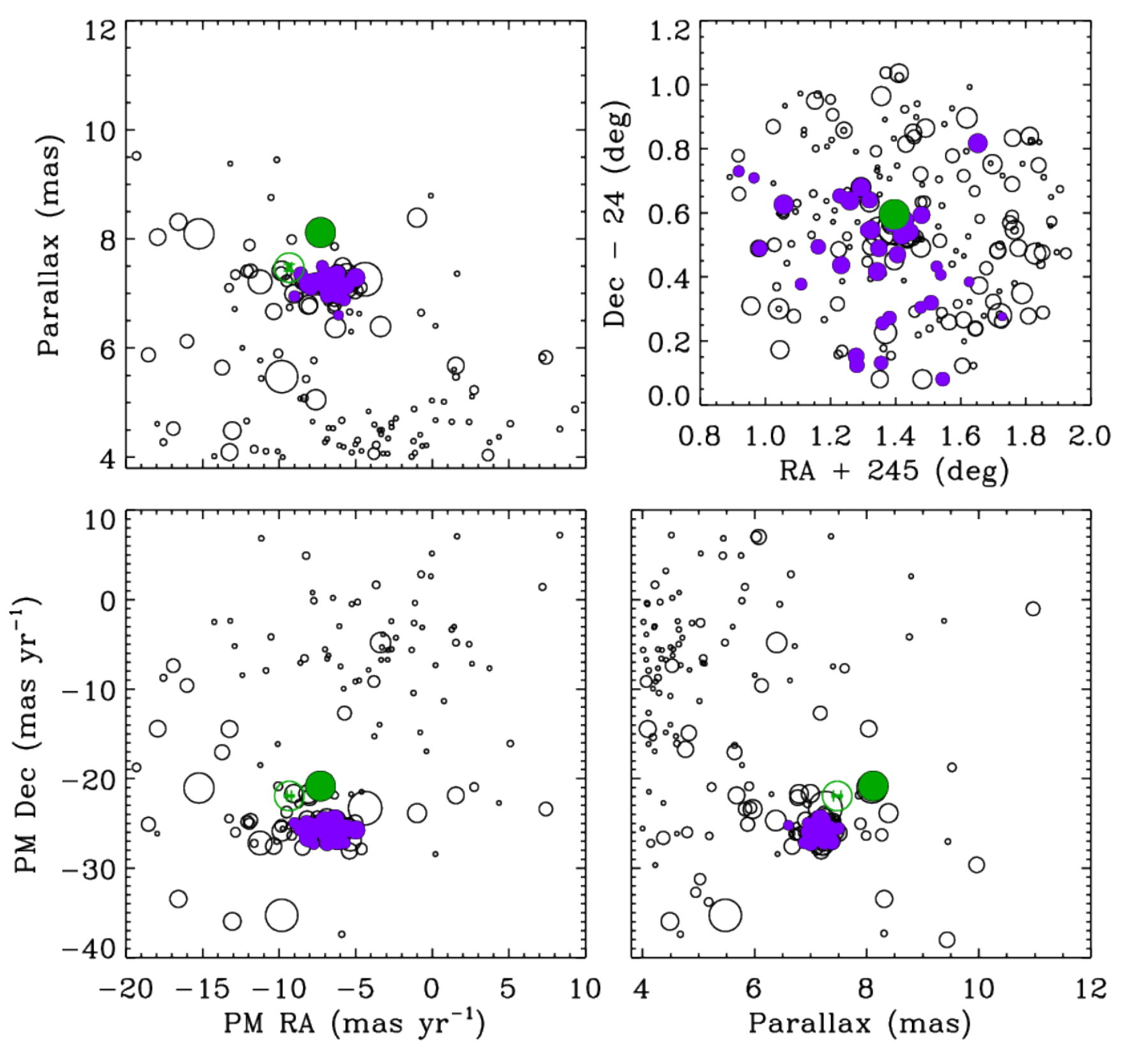}
      \caption[]{{\em Gaia} DR3 data proper motions and parallaxes for stars in the vicinity of $\rho$\,Oph\,C (top left and bottom panels). Purple circles indicate high-probability members of the $\rho$\,Oph cluster. Green circles indicate $\rho$\,Oph\,C, with open circles indicating DR2 measurements. The top right panel shows the distribution on the sky. Symbol size is inversely proportional to $G$ magnitude.}
         \label{rhooph_gaia}
   \end{figure}

As a first step in determining $\rho$\,Oph\,C's fundamental parameters it is necessary to know its bolometric luminosity $L_{\rm bol}$. While {\em Gaia} Data Release 2 \citep[DR2;][]{2018A&A...616A...1G} and Data Release 3 \citep[DR3][]{2021A&A...649A...6G} parallaxes are available, these are not necessarily reliable. There is a discrepancy of several sigma between the DR2 and DR3 parallaxes (see Fig.\ \ref{rhooph_gaia}), possibly attributable to the high renormalized unit weight error (RUWE) of 2.8: values greater than one can indicate that the astrometry is affected by multiplicity. 

Indeed, \cite{2025A&A...701A.171F} reported that $\rho$\,Oph\,C is a triple system, with 0.4 and 1.03 \msun~companions at distances of 6.7 and 83 mas, respectively. These correspond to physical separations of about 0.9 and 11.5 AU, orbital periods of around 0.4 and 16 years (assuming primary mass of 4.8 \msun), and maximum radial velocity amplitudes of 12.8 and 1.4 \kms. Given that the \vsini~of $\rho$\,Oph\,C is much higher than the reflex motion, it is unsurprising that radial velocity variations have not been detected in the star's spectrum. 

Since the parallax measurements of $\rho$\,Oph\,C are unreliable, we adopted the distance to $\rho$\,Oph determined by \cite{2025A&A...700A..14S} in their analysis of the nearby magnetic B star $\rho$\,Oph\,A, which yielded a distance of $d = 139 \pm 3$~pc. This analysis is reproduced in Fig.\ \ref{rhooph_gaia} for $\rho$\,Oph\,C. The cluster distance was found via an ensemble analysis of stars in the Gaia DR3 catalogue within 30$\arcmin$, with cluster membership determined via iteratively discarding stars with parallaxes and proper motions differing from the mean values by more than 1$\sigma$.

Whether $\rho$\,Oph\,C is a true cluster member is a matter of some dispute. \cite{2019A&A...626A..80C} considered it to be a member on the basis of DR2 data; \cite{2022A&A...667A.163M} instead classified it as a member of $\sigma$\,Sco, though this may have been an error due to the large RUWE. The star's proper motions are consistent with cluster means, as its DR2 parallax; only its DR3 parallax is inconsistent. The balance of evidence seems to suggest that the star is probably a member of $\rho$\,Oph, and we therefore adopt the cluster distance.

\subsection{Reddening}

\cite{1994ApJ...424..772F} estimated $E(B-V) = 0.48$ in the region by observing interstellar C$_2$ and CN absorption. \cite{2016A&A...592A..88P} used the dust map developed by \cite{2011ApJ...737..103S} based on IRAS (Infrared Astronomical Satellite) and COBE (Cosmic Background Explorer) data\footnote{Available at https://irsa.ipac.caltech.edu/applications/DUST/}, to derive the reddening $E(B-V)$ in the vicinity of $\rho$ Ophiuchus, which gives $E(B-V) \sim 1$~mag and thus an extinction $A_{\rm V} \sim 3$~mag assuming a usual reddening law $R_{\rm V} = 3.1$. However, the 3D STILISM tomographic dust map \citep[STructuring by Inversion the Local Interstellar Medium;][]{2014A&A...561A..91L,2017A&A...606A..65C} gives a much lower value of about $E(B-V) = 0.26$~mag at the distance to the cluster, with an asymptotic maximum of around 0.4 mag. 

$\rho$\,Oph\,C has measured UBVJHK magnitudes. Comparing the observed colours to the empirically determined intrinsic colours of young stars determined by \cite{2013ApJS..208....9P} for $\rho$\,Oph\,C's \teff~yields $E(B-V) = 0.59 \pm 0.02$~mag, almost exactly in the middle of the values retrieved from the DUST and STILISM models. We therefore adopt this as the colour excess, yielding $A_{\rm V} = 1.83 \pm 0.06$~mag. 

\subsection{Fundamental parameters}\label{subsubsect:rhoophc_fundpars}

Since $\rho$\,Oph\,C is a single star, at least insofar as its spectrum suggests, its fundamental parameters can be determined more easily. As the star is clearly strongly chemically peculiar, the \cite{2008AA...491..545N} bolometric correction for Bp stars was used, with the uncertainty increased by 0.15 mag since $\rho$\,Oph\,C is slightly hotter than the upper extent of their BC calibration. Fundamental parameters are given in Table \ref{table:stellar_pars}. The resulting bolometric luminosity $\log{L_{\rm bol}/L_\odot} = 2.6 \pm 0.3$ is slightly below the ZAMS for the \cite{ekstrom2012} evolutionary tracks for rotating stars, although it overlaps within uncertainty. Comparison to evolutionary tracks yields a mass of $4.8 \pm 0.4~{\rm M_\odot}$, while isochrones require an age less than 10 Myr, consistent with the cluster age determined by \cite{2016A&A...592A..88P}.

\section{Rotational Period}\label{sect:rot}

   \begin{figure*}
   \centering
   \includegraphics[width=0.95\textwidth]{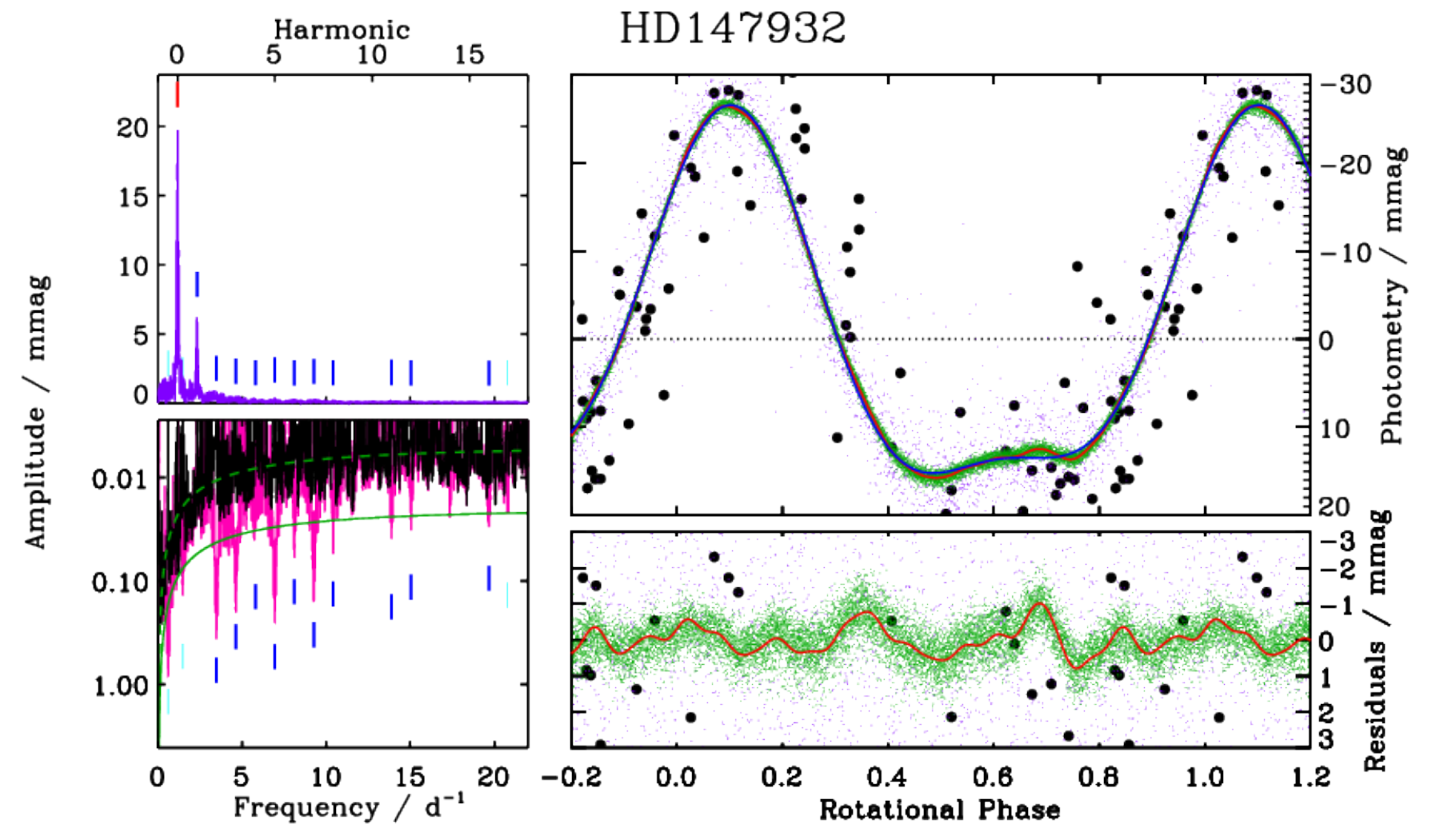}
      \caption[]{{\em Top left}: frequency spectrum for the {\em TESS} dataset. The rotational frequency $f_{\rm rot}$ is marked with a red dash; harmonics of $f_{\rm rot}$ with blue dashes; independent frequencies with cyan dashes. {\em Botton left}: Frequency spectrum after pre-whitening with $f_{\rm rot}$ and the first harmonic (magenta), and after pre-whitening with all significant frequencies (black). The dashed green curve indicates a noise model derived from the full pre-whitened frequency spectrum; the solid green curve indicates the S/N threshold for significance. {\em Top right}: {\em TESS} (green), {\em K2} (purple), and {\em Hipparcos} (black) light curves folded with the rotation frequency. The blue curve is a harmonic model for $f_{\rm rot}$ and its first harmonic; the red curve is the full harmonic model. {\em Bottom right}: residuals after subtraction of the simplified harmonic model, demonstrating substructure in the light curve.}
         \label{HD147932_periods_lc}
   \end{figure*}

\begin{table}
\centering
\caption[]{Photometric frequencies recovered from {\em TESS} data for $\rho$\,Oph\,C.}
\label{freqtab}
\begin{tabular}{l l r r}
\hline\hline
Freq. & ident & amp.   & $S/N$ \\
(c/d) &       & (mmag) &       \\
\hline
 0.609900(1) &  & 0.76(2) &  18.3 \\
 1.15754598(5) & $f_{\rm rot}$ & 20.16(2) & 871.6 \\
 1.462923(4) &  & 0.28(2) &  14.6 \\
 2.3150922(1) & 2$f_{\rm rot}$ & 6.47(2) & 472.6 \\
 3.472903(2) & 3$f_{\rm rot}$ & 0.38(2) &  35.7 \\
 4.630712(6) & 4$f_{\rm rot}$ & 0.18(2) &  19.8 \\
 5.79004(1) & 5$f_{\rm rot}$ & 0.07(2) &   9.0 \\
 6.944235(3) & 6$f_{\rm rot}$ & 0.28(2) &  37.2 \\
 8.10142(1) & 7$f_{\rm rot}$ & 0.07(2) &   9.3 \\
 9.259840(6) & 8$f_{\rm rot}$ & 0.17(2) &  25.5 \\
 10.42156(1) & 9$f_{\rm rot}$ & 0.07(2) &  10.7 \\
 13.88752(1) & 12$f_{\rm rot}$ & 0.09(2) &  15.4 \\
 15.05172(1) & 13$f_{\rm rot}$ & 0.06(2) &  10.1 \\
 19.68689(2) & 17$f_{\rm rot}$ & 0.05(2) &   8.8 \\
 20.77505(1) &  & 0.07(2) &  12.9 \\
\hline\hline
\end{tabular}
\end{table}

\cite{2018AJ....155..196R} determined a rotational period of 0.86404(7)~d for $\rho$\,Oph\,C from its {\em K2} light curve, where the number in parentheses indicates the uncertainty in the last significant digit. In order to ensure an accurate phasing of the ESPaDOnS and HARPSpol magnetometry, we revisit this rotational period using the {\em K2}, {\em Hipparcos}, and {\em TESS} light curves in an attempt to attain the highest possible precision. These light curves are shown folded with the new rotational period in Fig.\ \ref{HD147932_periods_lc}.

As discussed by \cite{2025A&A...700A..14S}, the {\em K2} light curve is heavily contaminated by light from $\rho$\,Oph\,A. We therefore use the light curve extracted by \citeauthor{2025A&A...700A..14S} in preference to the one available on MAST, as the newly extracted light curve used a mask that separated the light curves of the two stars. 

Frequency analysis was performed using {\sc period04}\footnote{\url{https://www.period04.net/}} \citep{2005CoAst.146...53L}. A first-pass analysis of the {\em TESS} light curve obtained from MAST was used to calculate a harmonic model, which was then used to detrend the {\em TESS} data by fitting a low order polynomial independently to the first and second chunks of the sector data (since these exhibit different trends on either side of the gap). The detrended light curve was then used to determine the final frequency spectrum. Frequency analysis of the {\em TESS} light curve reveals multiple significant frequencies, where a S/N threshold of 4 times the noise floor was used as the criterion for significance. Significant frequencies are provided in Table \ref{freqtab}. In addition to the rotational frequency $f_{\rm rot}$, there are 12 harmonics of $f_{\rm rot}$, and 2 independent frequencies of unknown origin.

Once significant frequencies had been determined from the {\em TESS} data, they were refined by appending the {\em K2} and {\em Hipparcos} datasets. While in principle there may be differences in photometric amplitude or light curve morphology in different bandpasses, as can be seen from Fig.\ \ref{HD147932_periods_lc} there is no discernable difference in this case. This resulted in a rotational period of 0.86389657(4)~d. This is within 2$\sigma$ of the period derived by \cite{2018AJ....155..196R} (using their larger uncertainty), but represents an improvement of about three orders of magnitude in precision. 

As can be seen from the top right panel Fig.\ \ref{HD147932_periods_lc}, the gross variations of the light curve can be reproduced quite well with a harmonic model derived from $f_{\rm rot}$ and its first harmonic. The higher harmonics arise from low-amplitude ($\sim$1 mmag) substructure in the light curve. Similar harmonic substructure has been observed in {\em TESS} data from a number of magnetic Ap/Bp stars, and is likely of magnetospheric origin \citep[e.g.][]{2022A&A...659A..37K}.

One of the 3 independent frequencies is rather high at 20.77~d$^{-1}$; while formally not a harmonic, this is close to the 18$^{\rm th}$ harmonic expected at 20.83$^{\rm -1}$. Two of the 3 independent frequencies are at low frequencies, around 0.6~d$^{-1}$ and 1.5~d$^{-1}$. Neither of these frequencies are consistent with $\rho$\,Oph\,A's rotational frequency of about 1.3~d$^{-1}$ \citep{2020MNRAS.493.4657L}. One possible explanation for these is that they are $g$-mode pulsations, in which case $\rho$\,Oph\,C might be a low-amplitude Slowly Pulsating B-type (SPB) star. Another possibility is that the frequency originates with the brighter of the two companions reported by \cite{2025A&A...701A.171F}. A parallel analysis of the {\em TESS} light curve was carried out by \cite{Gunderson2026}, who identified the same 0.6~d$^{-1}$ variation, and found that it can be produced with a 20\% flux variation of the G star, which is well within the range of expectations for a young, magnetically active pre-main sequence object.


\section{Magnetometry}\label{sect:magfld}

   \begin{figure}
   \centering
   \includegraphics[width=0.45\textwidth]{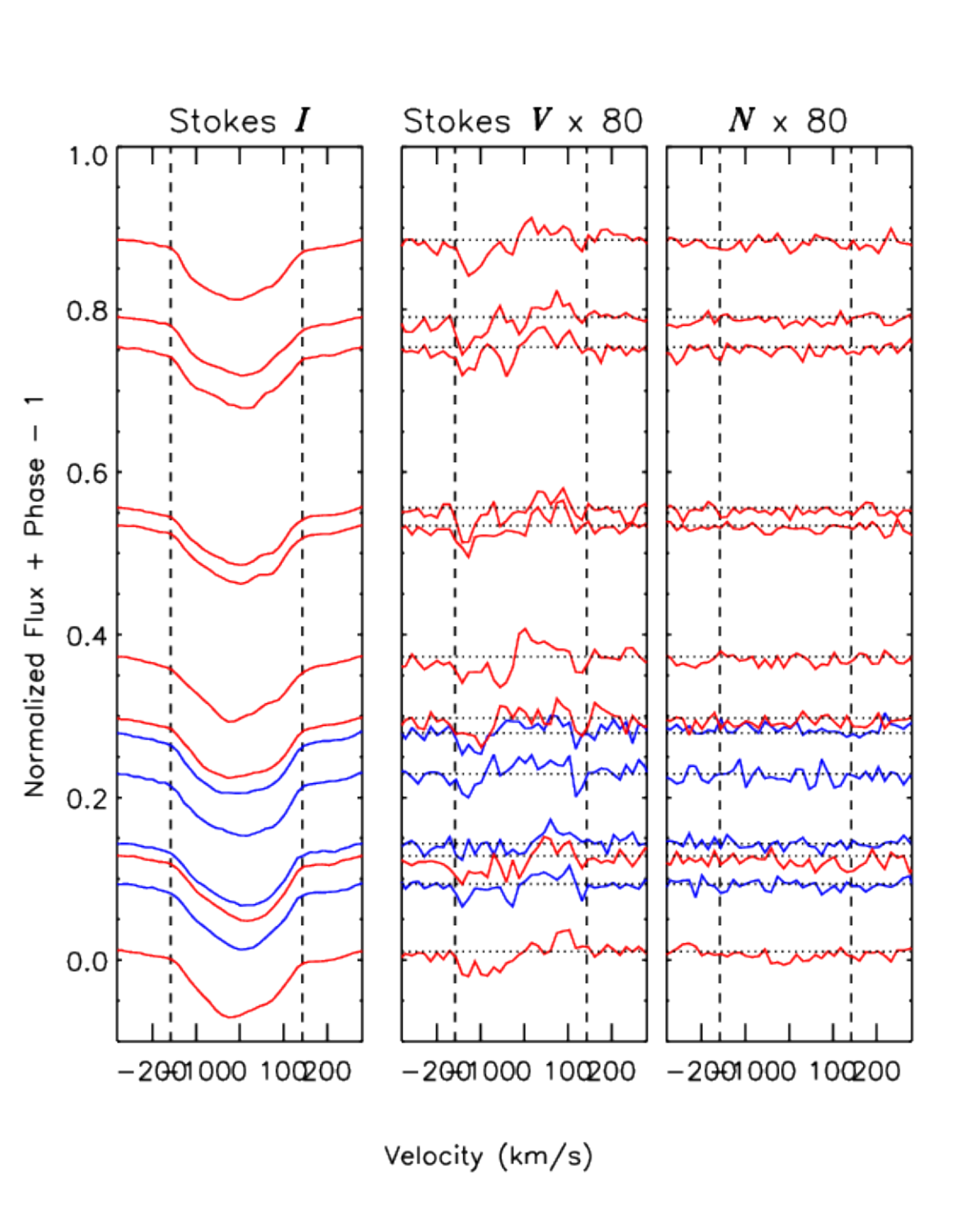}
      \caption[]{LSD Stokes $I$ (left), Stokes $V$ (centre), and $N$  (right) profiles of $\rho$\,Oph\,C, arranged in order of rotational phase, with the vertical offset equal to phase. Dashed lines indicate integration ranges. Blue and red profiles indicate HARPSpol and ESPaDOnS observations, respectively.}
         \label{rhoOphC_LSD_phaseplot}
   \end{figure}

   \begin{figure}
   \centering
   \includegraphics[width=0.45\textwidth]{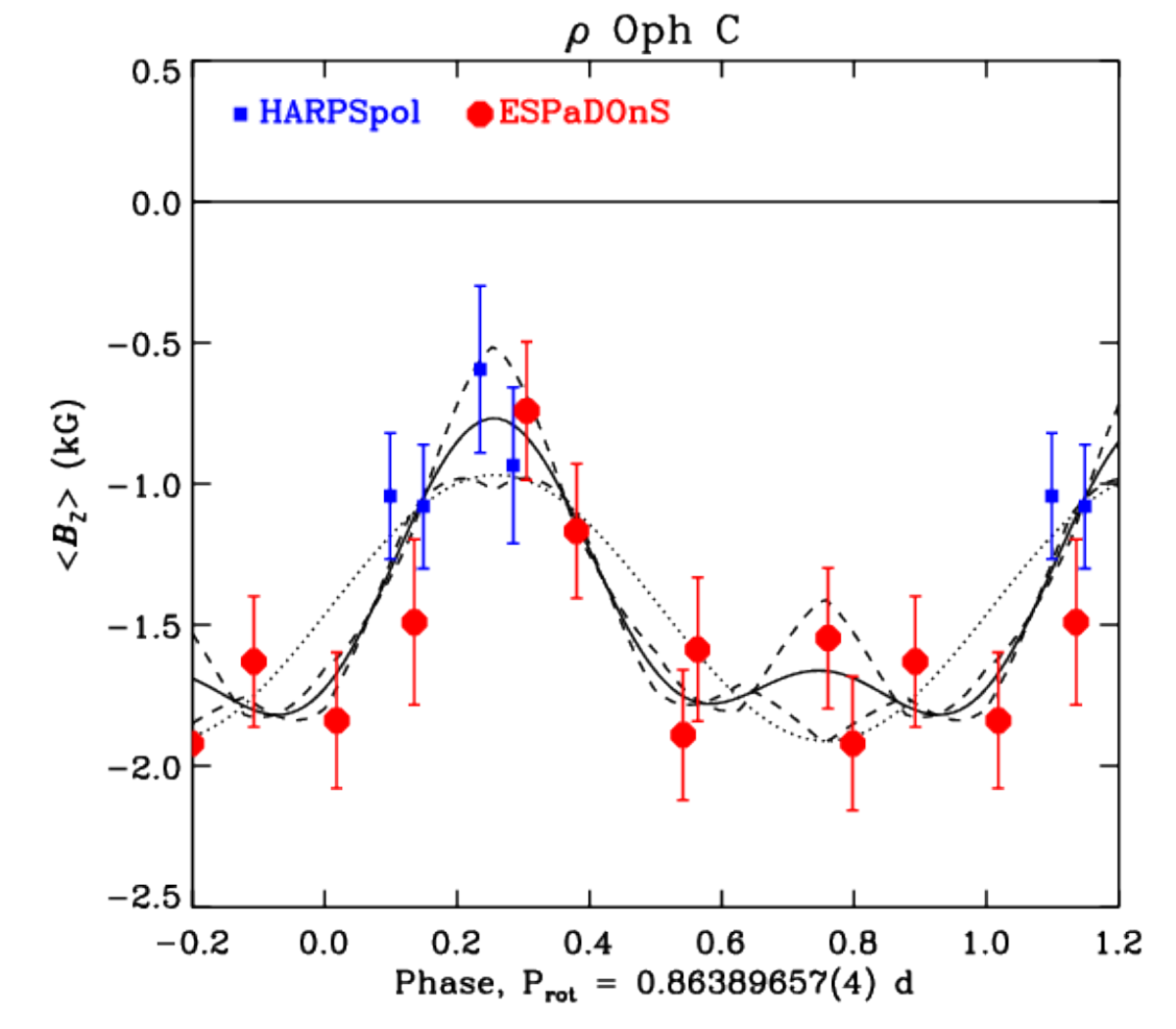}
      \caption[]{Longitudinal magnetic field \bz~of $\rho$\,Oph\,C measured from the LSD profiles shown in Fig.\ \ref{rhoOphC_LSD_phaseplot}, folded with the rotation period. Solid and dashed curves show the best harmonic fit and the fit uncertainty respectively. The dotted curve shows a sinsuoidal fit for comparison.}
         \label{rhoOphC_bz}
   \end{figure}

In order to evaluate the magnetic fields with the highest possible precision, mean line profiles were extracted from the spectra using least squares deconvolution \citep[LSD;][]{d1997}, specifically with the iLSD package developed by \cite{koch2010}. Since $\rho$\,Oph\,C is a chemically peculiar Bp star, a Vienna Atomic Line Database \citep[VALD3;][]{piskunov1995, ryabchikova1997, kupka1999, kupka2000,2015PhyS...90e4005R} line list was extracted via an `extract stellar' request using abundances matching those determined for HD\,61556, which is a He-weak, Si-enhanced star with a similar \teff~\citep{2015MNRAS.449.3945S}. The line list was then cleaned by eye in the usual way to remove H lines, lines strongly blended with the wings of H lines, and lines contaminated by telluric and interstellar lines, with the remaining lines then adjusted to match the depths of the lines in the observed spectrum \citep[the process is described in greater detail by][]{2018MNRAS.475.5144S}. After cleaning, 991 of the original 1610 lines remainded for analysis. Lines outside the HARPSpol spectral window were excluded in order to extract consistent LSD profiles simultaneously from ESPaDOnS and HARPSpol observations. A 14.4 \kms~velocity pixel was used for the LSD profiles, in order to boost the $S/N$; smaller velocity pixels resulted in some of the ESPaDOnS observations yielding NDs. 

Fig.\ \ref{rhoOphC_LSD_phaseplot} shows the LSD Stokes $I$, Stokes $V$, and $N$ profiles arranged in ascending order of rotational phase, with a vertical offset equal to phase. 

The presence of a magnetic field was evaluated using false alarm probabilities (FAPs), and classified as definite, marginal, or non-detections (DD, MD, ND) according to the usual criteria of, respectively, ${\rm FAP} < 10^{-5}$, $10^{-5} < {\rm FAP} < 10^{-3}$, and ${\rm FAP} > 10^{-3}$ \citep{d1997}, with FAPs measured using integration ranges set to the width of the Stokes $I$ spectral lines. The strength of the line-of-sight (longitudinal) magnetic field averaged over the visible stellar hemisphere \bz~was then measured using the centre-of-gravity method \citep[e.g.][]{mat1989}, with the same integration range as used for measuring FAPs; the null field \nz~was evaluated using $N$ profiles using the same method, as an additional check. In addition to evaluation of the Stokes $V$ profiles, the same measurements were performed on the null $N$ profiles in order to check for spurious signals. The \bz~and \nz~measurements, and the corresponding detection flags from the FAPs, are given in Table \ref{table:bz}. 

For the ESPaDOnS data, all $N$ profiles are NDs, while all Stokes $V$ profiles are DDs. Two of the HARPSpol Stokes $V$ LSD profiles yield MDs, although when these are co-added they yield a DD; the other two are DDs; all $N$ profiles are NDs. 

The \bz~variation was characterized by folding \bz~with the rotational period $P_{\rm rot}$ and fitting a harmonic function of the form 

\begin{equation}\label{eqn:bz}
\langle B_z\rangle(\phi) = \sum_{k=0}^{n} B_k \sin{(k\phi + \Phi_k)},
\end{equation}

\noindent where $\phi$ is the rotational phase, and $B_k$ and $\Phi_k$ are respectively the amplitude and phase offset of the $k^{\rm th}$ term, with phase terms in radians. Optimal fits were determined by calculating the reduced $\chi^2$ of the fit $\chi^2/\nu$ (where $\nu$ is the number of degrees of freedom): if $\chi^2/\nu$ is approximately 1, the fit is considered good; significantly greater than 1, poor; and below 1, over-fit. Expansion of Eq.\ \ref{eqn:bz} to $n=2$ is consistent with a dipolar magnetic field; if higher values of $n$ are required to achieve a reasonable fit, \bz~may be affected by higher-order contributions to the surface magnetic field, or the polarization signatures might be weighted by the variable surface brightness arising from chemical spots. 

\bz~is shown folded with the rotational period in Fig.\ \ref{rhoOphC_bz}. Fitting with Eq.\ \ref{eqn:bz} and $n=2$ yields $\chi^2/\nu = 1.8$, while $n=3$ gives $\chi^2/\nu = 1.2$. Both curves are shown in Fig.\ \ref{rhoOphC_bz}, and the amplitudes of the fitting terms are provided in Table \ref{table:stellar_pars} (the amplitude of $B_2$ is negligble compared to $B_0$ and $B_1$). Both HARPSpol observations were acquired in the phase window encompassing a dip in \bz~towards the magnetic equator, with the majority of ESPaDOnS observations yielding \bz~about twice as strong. The good agreement in \bz~between HARPSpol and ESPaDOnS observations obtained at similar rotational phases, as well as the agreement in the morphologies of the Stokes $V$ profiles obtained at similar phases (Fig.\ \ref{rhoOphC_LSD_phaseplot}), indicates that this discrepancy between the HARPSpol and ESPaDOnS datasets is not a result of systematic differences between the instruments, but reflects a real rotational modulation of the visible magnetic field.

\begin{table}
\centering
\caption[]{Longitudinal magnetic field (\bz) measurements of $\rho$\,Oph\,C. DF indicates detection flag (see text). Inst(rument) is either H(ARPSpol) or E(SPaDOnS). All null spectra yielded non-detections.}
\label{table:bz}
\begin{tabular}{l l r r r }
\hline\hline
HJD        & Inst & \bz   & DF$_V$ & \nz \\
$- 2450000$ & & (G) & & (G) \\
\hline
\\
6343.85434 & H & $-1043 \pm   223$ &  DD & $  41 \pm   223$ \\
6343.89756 & H & $-1079 \pm   219$ &  DD & $  21 \pm   219$ \\
6344.83557 & H & $ -594 \pm   294$ &  MD & $  310 \pm   294$ \\
6344.87880 & H & $ -933 \pm   277$ &  MD & $  375 \pm   277$ \\
9352.98448 & E & $ -788 \pm   249$ &  DD & $ -160 \pm   249$ \\
9353.91332 & E & $-1307 \pm   247$ &  DD & $  283 \pm   247$ \\
9354.93534 & E & $-1595 \pm   254$ &  DD & $  107 \pm   253$ \\
9355.96980 & E & $-1663 \pm   250$ &  DD & $  243 \pm   250$ \\
9357.91976 & E & $-1941 \pm   238$ &  DD & $  116 \pm   238$ \\
9358.88512 & E & $-1492 \pm   288$ &  DD & $  570 \pm   288$ \\
9361.82750 & E & $-1873 \pm   231$ &  DD & $   41 \pm   231$ \\
9362.91299 & E & $-1916 \pm   235$ &  DD & $  317 \pm   235$ \\
9362.99477 & E & $-1674 \pm   230$ &  DD & $   51 \pm   229$ \\
\hline\hline
\end{tabular}
\end{table}

The terms $B_0$ and $B_1$ from the n=2 fit were then used to solve the geometrical relationship developed by \cite{preston1967,preston1974} for the relationship between the inclination angle of the rotational axis from the line of sight $i_{\rm rot}$ and the obliquity angle $\beta$ between the rotational and magnetic axes for a tilted dipole:

\begin{equation}\label{preston_r}
r = \frac{|B_0| - B_1}{|B_0| + B_1} = \frac{\cos{(i_{\rm rot} + \beta)}}{\cos{(i_{\rm rot} - \beta)}},
\end{equation}

\noindent with $i_{\rm rot}$ independently constrained via

\begin{equation}\label{eqn:irot}
\sin{i_{\rm rot}} = \frac{v\sin{i_{\rm rot}}}{v_{\rm eq}} = \frac{P_{\rm rot}v\sin{i_{\rm rot}}}{2\pi R_*},
\end{equation}

\noindent where $v_{\rm eq}$ is the equatorial rotational velocity. The surface strength of the magnetic dipole $B_{\rm d}$ was then found via \citet{preston1967,preston1974}:

\begin{equation}\label{eqn:bd}
B_{\rm d} = \langle B_z \rangle_{\rm max} \frac{20(3 - \epsilon)(\cos{\beta}\cos{i_{\rm rot}} + \sin{\beta}\sin{i_{\rm rot}})}{15 + \epsilon},
\end{equation}

\noindent where $\epsilon$ is the limb-darkening coefficient, obtained from the tables calculated by \cite{2016MNRAS.456.1294R}, using the $B$ band values as those closest to the wavelength region containing the majority of spectral lines used in the analysis. 

In practice $i_{\rm rot}$, $\beta$, and $B_{\rm d}$ were calculated simultaneously using the Monte Carlo Hertzsprung Russell Diagram (MCHRD) sampler described by \cite{2019MNRAS.490..274S}, which combines all available atmospheric, rotational, and magnetic data with evolutionary models to obtain fully self-consistent fundamental, rotational, magnetic, wind, and magnetospheric parameters satisfying physical constraints such as under-critical rotation and oblateness (expressed here as the ratio $R_{\rm pol}/R_{\rm eq}$ of the polar to equatorial radii). Asymmetric uncertainties are obtained directly from posterior distributions, and automatically account for correlations between variables. The results of this section are summarized in Table \ref{table:stellar_pars}.

Magnetic and rotational parameters from the MCHRD sampler are given in Table \ref{table:stellar_pars}. The star is slightly oblate ($R_{\rm pol}/R_{\rm eq} = 0.968$), with a moderate inclination ($i_{\rm rot} = 64^\circ$), a magnetic axis nearly aligned with the rotational axis ($\beta = 7^\circ$), and a very strong magnetic field ($B_{\rm d} \sim 10$~kG, with a 1$\sigma$ upper uncertainty range extending to 23~kG). 

\cite{2020MNRAS.499L..72L} estimated $\rho$\,Oph\,C's rotational and magnetic parameters using the properties of its incoherent and coherent radio emission. They adopted a larger radius and \vsini~than those found here, differences which mutually compensate and lead to a similarly high $i_{\rm rot} \sim 74^\circ$. By varying $\beta$ between 0 and 10$^\circ$ (a range chosen based on the small level of variation of the two existing HARPSpol \bz~measurements) and $B_{\rm d}$ between 12 and 16 kG, they found a best-fit to the auroral emission properties with $\beta = 5^\circ$ and $B_{\rm d} = 13.2$~kG. The results found here are entirely consistent with these predictions. 

\section{Magnetospheric Variability}\label{sect:magnetospheres}

The magnetosphere of $\rho$\,Oph\,C has been thoroughly studied in X-ray and radio \citep{2016A&A...592A..88P,2020MNRAS.499L..72L,Gunderson2026}, but while its H$\alpha$ emission has been previously reported \citep{alecian2014,2020MNRAS.499.5379S}, the small number of observations precluded a detailed analysis of its characteristics. While the variable X-ray and radio emission of $\rho$\,Oph\,C are certainly dominated by the stellar magnetosphere, \cite{Gunderson2026} have shown that these diagnostics are also strongly affected by the low-mass, pre-main sequence companions detected by \cite{2025A&A...701A.171F}, which significantly complicates their analysis. This makes H$\alpha$ important as a magnetospheric diagnostic which is potentially less affected by the much dimmer companion stars. In this section we examine the variability of the star's H$\alpha$ emission. We also look at the star's photometric variability in an attempt to determine whether magnetospheric eclipsing or chemical spots are the primary source of variability.

The magnetospheric parameters inferred from the MCHRD sampler are reported in Table \ref{table:stellar_pars}. The mass-loss rate $\dot{M}$ and wind terminal velocity $v_\infty$ were evaluated using the \cite{vink2001} recipe.The magnetospheres of early-type stars are primarily characterized by two radii, the Kepler corotation radius \rk~and the Alfv\'en radius \ra~\citep[e.g.][]{petit2013,2019MNRAS.490..274S}. The Alfv\'en radius gives the maximum extent of magnetic confinement in the magnetic equatorial plane, and scales with the dimensionless wind magnetic confinement parameter $\eta_* \equiv B_{\rm eq}^2R_*^2\dot{M}^{-1}v_\infty^{-1}$ as $R_{\rm A} / R_* \sim \eta_*^{1/4}$, where $\eta_*$ expresses the ratio of magnetic to wind kinetic energy density at the surface in the vicinity of the magnetic equator, where the wind momentum and magnetic field are perpendicular to one another \citep{ud2002,ud2008}. If $\eta_* > 1$ the wind is considered to be magnetically confined, i.e.\ the star has a magnetosphere. In this case $\eta_* > 10^6$, i.e.\ the wind is strongly magnetically confined\footnote{If the \cite{krticka2014} mass-loss rates are used instead, $\eta_*$ is even larger since the \citeauthor{krticka2014} $\dot{M}$ is much lower.}. The corresponding Alfv\'en radius is $R_{\rm A} = 40^{20}_{-4}~R_*$, where the large positive uncertainty is a consequence of the corresponding uncertainty in $B_{\rm d}$. Systematic uncertainty in the reference spectrum due to uncertainty in stellar parameters $T_{\rm eff}$ and $\log{g}$ are incorporated into the EW uncertainty by measuring the EWs of synthetic spectra across the range of good fits; as can be seen from Fig. \ref{rhoOphC_logg}, these differences are in practice negligible.

The Kepler corotation radius \rk~is the distance from the rotational axis at which the centrifugal and gravitational forces acting upon corotating magnetically confined plasma are equivalent, and is given by $R_{\rm K} = (GM/\omega^2)^{1/3}$, where $G$ is the gravitational constant, $M_*$ is the stellar mass, and $\omega$ is the rotational frequency. For $\rho$\,Oph\,C, $R_{\rm K} = 2.43^{+0.03}_{-0.06}~R_*$. Since $R_{\rm A} > R_{\rm K}$, the star has a centrifugal magnetosphere, and indeed $\log{R_{\rm A}/R_{\rm K}} \sim 1.3$ is in the regime in which H$\alpha$ emission is generally expected for CM host stars \citep{2019MNRAS.490..274S}. A second numerical indication that H$\alpha$ emission should be present is provided by the strength $B_{\rm K} = B_{\rm eq}/R_{\rm K}^3$ of the equatorial magnetic field at $R_{\rm K}$. \cite{2020MNRAS.499.5379S} showed that all magnetic B-type stars with $B_{\rm K} > 100$~G display H$\alpha$ emission. $\rho$\,Oph\,C is well above this threshold ($B_{\rm K} = 420$~G; Table \ref{table:stellar_pars}).



The star's H$\alpha$ variability was evaluated using the same method as described by \cite{2020MNRAS.499.5379S}. First, synthetic spectra were subtracted from the observed profiles, with the equivalent width measured as ${\rm EW} = {\rm EW}_{\rm obs} - {\rm EW}_{\rm synth}$, yielding positive values for emission. Plasma projected in front of the stellar disk will yield enhanced absorption in the rotationally broadened core due to eclipsing of the star, whereas plasma projected to the side of the star will result in emission. In order to isolate these components, equivalent widths were measured in three integration ranges: from the furthest blue extent of emission (about $-900$~\kms) to $-$\vsini, between $\pm$\vsini, and from $+$\vsini~to the furthest red extent of emission (about $+700$~\kms). It is likely that the true red emission edge extends to similar velocities as the blue edge, however a shorter integration range was chosen to avoid contamination by the C~{\sc ii} doublet in the red wing of H$\alpha$. 

   \begin{figure}
   \centering
   \includegraphics[width=0.5\textwidth]{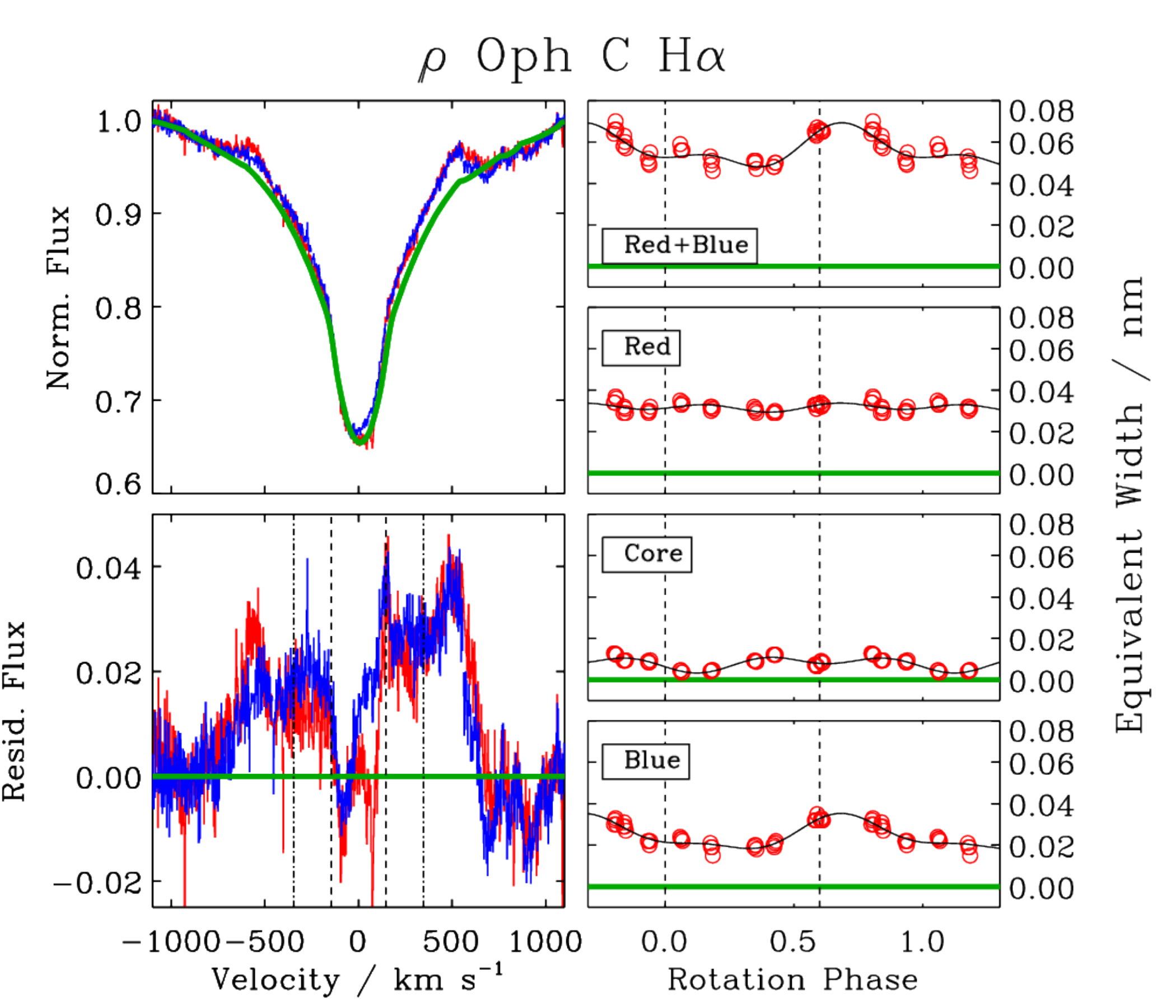}
      \caption[]{{\em Top left}: H$\alpha$ profiles of $\rho$\,Oph\,C at maximum emission (red) and minimum emission (blue), shown with a synthetic spectrum (thick green). {\em Bottom left}: residual flux after subtracting the synthetic spectrum from the observed spectra. The horizontal green line indicates the continuum level. Vertical dashed and dot-dashed lines show $\pm$\vsini~and $\pm (R_{\rm K}/R_*)$\vsini, respectively. {\em Right panels}: equivalent widths (EWs) measured from (bottom to top) the blue wing (i.e.\ from the negative extent of zero emission excess to $-$\vsini), the line core (i.e.\ within $\pm$\vsini), the red wing (i.e.\ from $+$\vsini~to the positive extent of zero emission excess), and the combined sum of the blue and red EWs. Dashed vertical lines show \bz~extrema from the fit in Fig.\ \ref{rhoOphC_bz}. Horizontal thick green lines show the continuum level. Curves show harmomic fits.}
         \label{rhoOphC_halpha_multiplot}
   \end{figure}

H$\alpha$ spectra of $\rho$\,Oph\,C at minimum and maximum emission are shown in the top left panel of Fig.\ \ref{rhoOphC_halpha_multiplot}, together with the reference synthetic spectrum. Residual flux is shown in the bottom left panel, where dashed and dot-dashed lines indicate $\pm$\vsini~and $\pm R_{\rm K}/$\vsini. There is very little difference between spectra at minimum and maximum emission, indicating a low level of variability, as expected given the small obliquity of $\rho$\,Oph\,C's magnetic field. The EW curves (right panels of Fig.\ \ref{rhoOphC_halpha_multiplot}) are also only weakly variable, and exhibit single-wave variations consistent with the Rigidly Rotating Magnetosphere \citep[RRM;][]{town2005c} expectations for a star for which the same magnetic pole is always in view \citep[][]{2020MNRAS.499.5379S}. Variability is dominated by the blue wing, with nearly constant emission in the red wing; however, the overall emission level is higher in the red wing. This may be indicative of asymmetry in the circumstellar plasma distribution arising from a slightly non-dipolar magnetic field. There is some suggestion of eclipsing in the core, however it also seems to be partly infilled by emission.  

   \begin{figure}
   \centering
   \includegraphics[width=0.45\textwidth,trim=10 10 50 0]{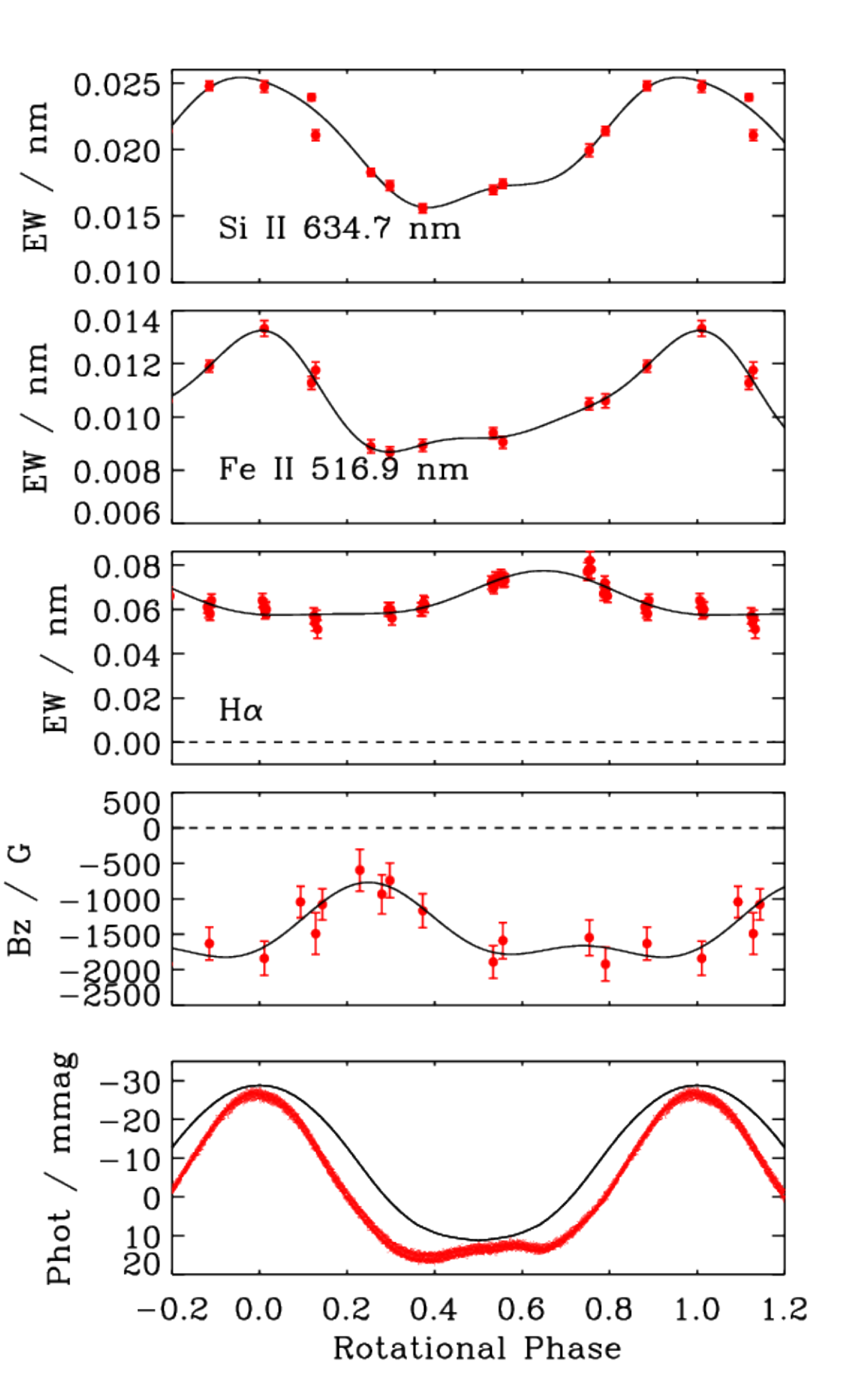}
      \caption[]{{\em Top panel}: Si~{\sc ii}~634.7~nm EWs folded with the rotation period. The solid curve shows a harmonic fit. {\em Second panel}: The same, for Fe~{\sc ii}~516.9~nm. {\em Third panel}: The same, for H$\alpha$ emission EWs (core plus wings) measured across the full line. The dashed line shows the continuum. {\em Fourth panel}: \bz~measurements. The dashed line shows the magnetic null. {\em Bottom panel}: {\em TESS} photometry (red points), compared to a CBO-RRM model with $\tau_{\rm K} = 0.5$ (black curve).}
         \label{rhoOphC_K2}
   \end{figure}

$\rho$\,Oph\,C's light curve is shown in Fig.\ \ref{rhoOphC_K2}, together with \bz, and EW curves for H$\alpha$ and two representative metallic lines.

A CBO-RRM model lightcurve of the kind developed by \cite{2022MNRAS.511.4815B,2023MNRAS.523.6371B} synthesized for $\rho$\,Oph\,C's angular and rotational parameters provides a reasonable match to the data with $i$, $\beta$, and $W$ as determined from Sect.\ \ref{sect:magfld} (see Table \ref{table:stellar_pars}) and an optical depth at the Kepler radius $\tau_{\rm K} = 0.5$. $W \equiv v_{\rm eq}/v_{\rm orb}$ is the dimensionless rotation parameter, defined as the ratio of the surface equatorial rotational velocity $v_{\rm eq}$ to the velocity $v_{\rm orb}$ necessary to maintain a Keplerian orbit at the stellar surface. The expected value of $\tau_{\rm K}$ is found from Eq.\ 7, \cite{2022MNRAS.511.4815B}:

\begin{equation}\label{tauk}
\tau_{\rm K} = 0.11 \frac{c_{\rm f}}{0.3}\left (\frac{B_{\rm d}}{\rm kG} \right )^2 W^{8/3},
\end{equation}

\noindent where $c_{\rm f}$ is a `correction factor' inferred by \cite{2020MNRAS.499.5366O} from magnetohydrodynamic models and taken to be about 0.3. Eq.\ \ref{tauk} predicts $\tau_{\rm K} = 0.34^{+1.6}_{-0.07}$, consistent with the best fit. 

While the correspondence between the light curve and the CBO-RRM model is promising, there is no obvious correlation between the light curve and either \bz~or H$\alpha$, as would be expected if the photometric variability was primarily produced in the magnetosphere. By contrast, EWs measured from the Fe~{\sc ii}~516.9~nm and Si~{\sc ii}~634.7~nm lines both correlate closely to the light curve, with EW maxima at maximum light, and broad minima corresponding to the broad photometric minimum. This is just as would be expected if $\rho$\,Oph\,C's photometric variability is primarily a consequence of chemical abundance spots, rather than its magnetosphere. The photometric variation produced by spots originates due to flux redistribution from the UV to the visible via bound-bound and bound-free transitions. Line blanketing due to abundance enhancements leads to light being absorbed in the UV, and re-emitted at longer wavelengths. Thus, regions with higher abundances of He and Fe-group elements result in darkening in the UV, and brightening in the visible. Higher EWs indicate higher surface abundances, which in turn should result in brighter visible light due to redistribution of flux from the UV to the visible. For a visible light curve we therefore expect that stronger EWs should correlate to brightening of the star, and vice versa. This has been shown in detail for several individual mCP stars covering a wide range of spectral types \citep[e.g.][]{2007A&A...470.1089K,2009A&A...499..567K,2010A&A...524A..66S,2012A&A...537A..14K,2013A&A...556A..18K,2015A&A...576A..82K}.

While a CBO-RRM model can reproduce the light curve with reasonable parameters, the correlation between metallic line variability and the light curve on the one hand, and the lack of correlation between H$\alpha$ and \bz~as compared to the light curve on the other, are reasonable empirical grounds for suspecting that the photometric variation is mainly a consequence of chemical spots rather than the magnetosphere. While the gross variations of the {\em TESS} light curve are almost certainly a result of chemical spots, the harmonic substructure of the light curve, as shown in Fig.\ \ref{HD147932_periods_lc}, is difficult to reproduce using surface features. Similar substructure has been seen in other stars, for instance HD\,37776 \citep{2020pase.conf...46M}, and has been reproduced with multipolar RRM models \citep{2022A&A...659A..37K}. If the substructure in $\rho$\,Oph\,C's light curve originates in its magnetosphere, the $\sim$1~mmag amplitude implies a much smaller $\tau_{\rm K}$ than the value inferred from Eq.\ \ref{tauk}, probably around 0.02 rather 0.5 (if we simply scale by the ratio of the full photometric amplitude to the residual photometric amplitude in Fig.\ \ref{HD147932_periods_lc}). This much smaller $\tau_{\rm K}$ is easier to reconcile with the star's extremely weak H$\alpha$ emission.


\section{Discussion}\label{sect:discuss}

   \begin{figure}
   \centering
   \includegraphics[width=0.45\textwidth]{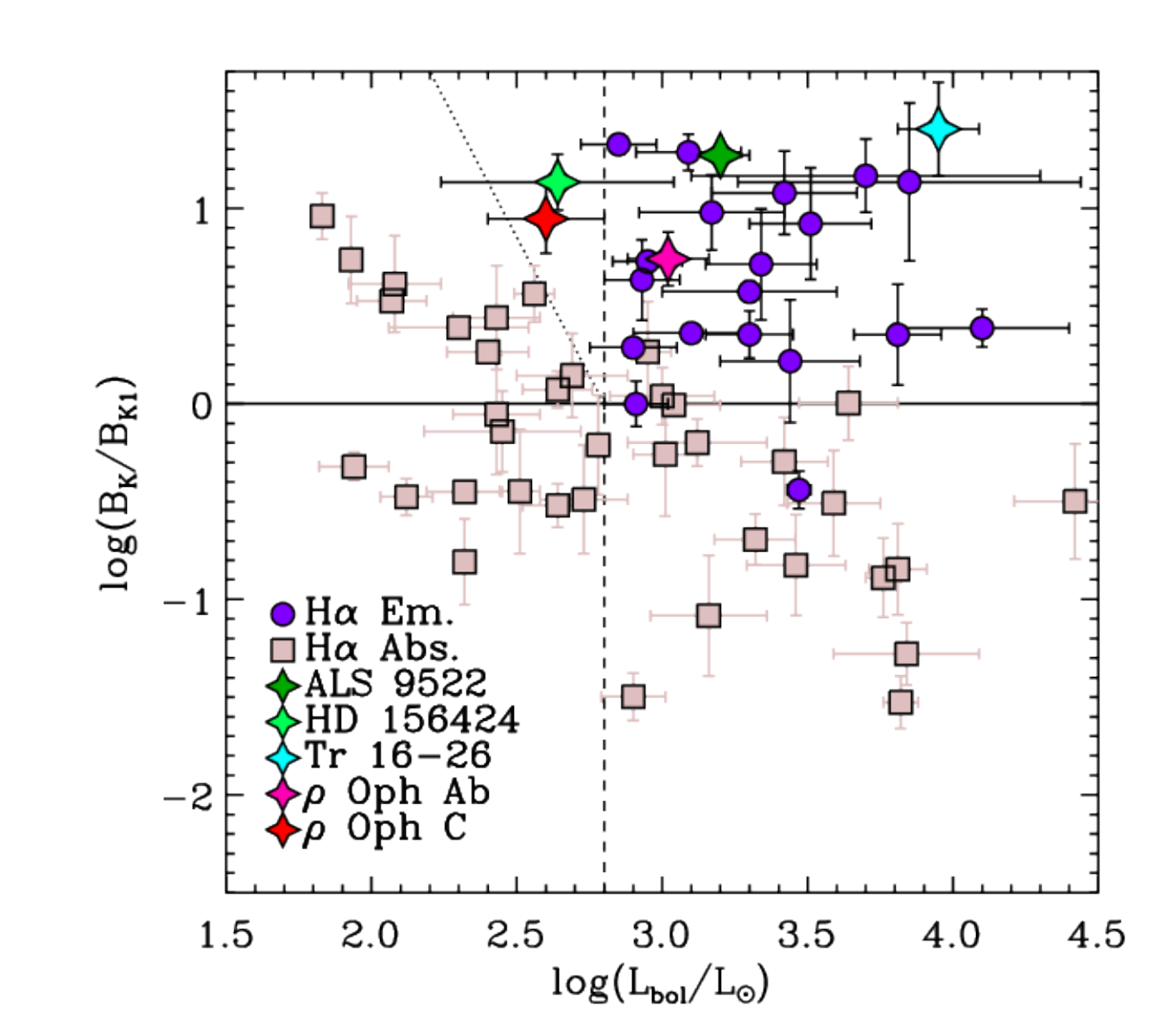}
      \caption[]{Ratio between the magnetic field at the Kepler radius $B_{\rm K}$ and the magnetic field strength necessary to attain an optical depth of unity $B_{\rm K1}$ as a function of bolometric luminosity $L_{\rm bol}$. Newly added H$\alpha$-bright stars are highlighted. The solid horizontal line indicates an emission threshold based purely on $B_{\rm K}$. The dashed vertical line indicates an emission threshold based on $\log{L}$. The dotted diagonal line indicates a variable, luminosity-dependendent emission threshold as might be expected in the case of magnetospheric leakage.}
         \label{rhooph_lum_bk1}
   \end{figure}

   \begin{figure}
   \centering
   \includegraphics[width=0.45\textwidth]{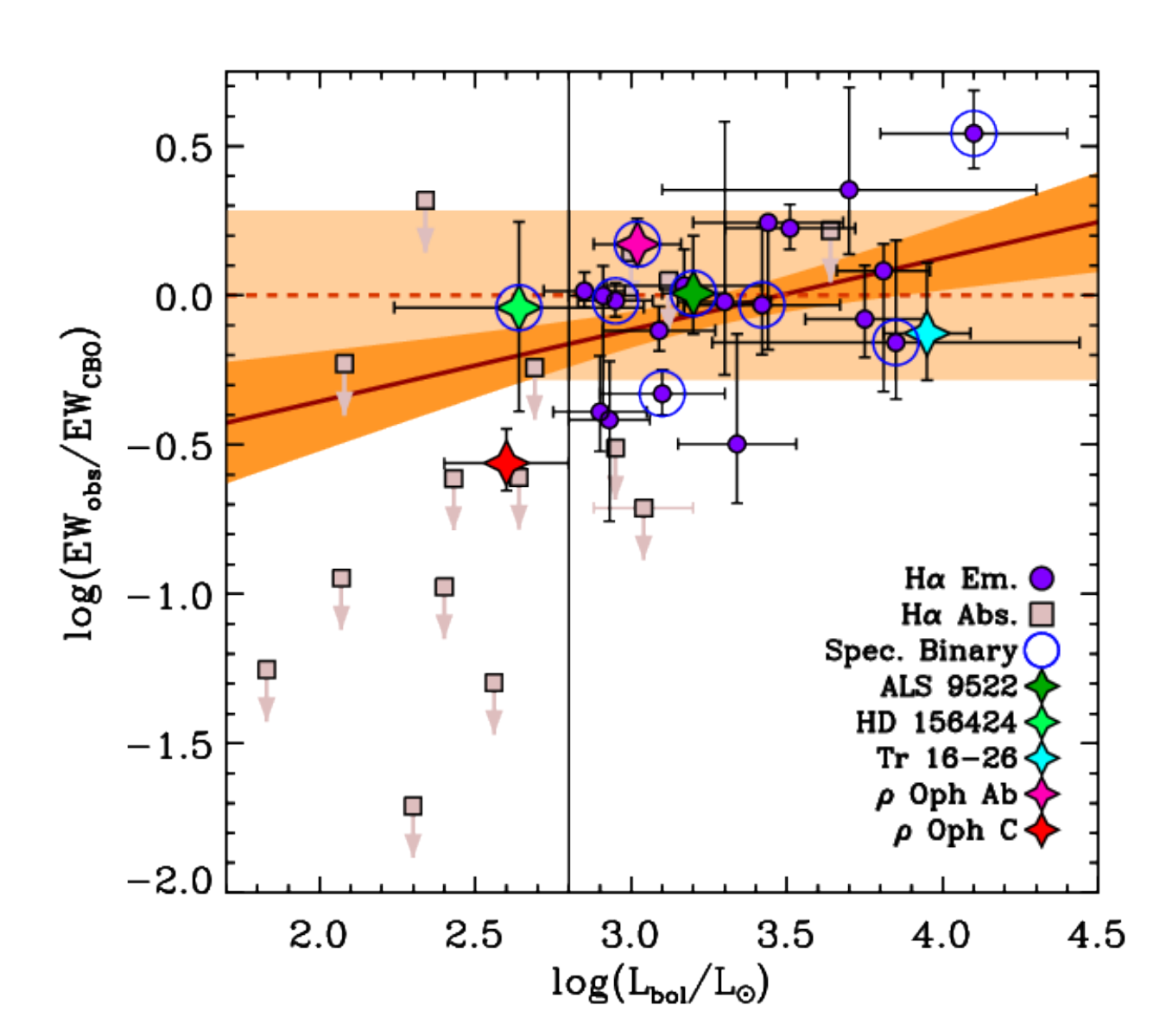}
      \caption[]{Logarithmic ratio of the observed emission EW to the value predicted by the CBO scaling relationship as a function of bolometric luminosity $L_{\rm bol}$. Horizontal dashed line shows $x=y$, with the shaded region indicating the dispersion in the scaling relationship. The vertical line shows the previously identified luminosity cutoff for H$\alpha$ emission. The solid orange line shows a linear regression, with shaded regions indicating the standard deviation in regressions.}
         \label{halpha_population}
   \end{figure}

\cite{2020MNRAS.499.5379S} and \cite{2020MNRAS.499.5366O} demonstrated that a centrifugal breakout (CBO) plasma transport mechanism matches the observed characteristics of the H$\alpha$ emission of magnetic early B-type stars. Since their analyses were published 5 additional H$\alpha$-bright stars have been examined in detail\footnote{\cite{2021MNRAS.507.1283S} also reported CM-style H$\alpha$ emission from a sixth star, HD\,144941, which with its nearly 14-day rotation period is by far the most slowly rotating star with CM-type emission. As \citeauthor{2021MNRAS.507.1283S} discussed in detail, given the slow rotation, this emission only makes sense if HD\,144941 is a post-main sequence extreme helium star with an extremely low mass. In this case its emission properties are in line with the CBO scaling relationship. If HD\,144941 is instead a main sequence He-strong star, as argued by \cite{2021A&A...654A.119P}, its emission strength is a significant outlier on the scaling relationship. Given the substantial systematic ambiguity in the star's fundamental parameters, we elected to exclude it from this discussion.}: HD\,156424, \citep{2021MNRAS.504.4850S}, ALS\,9522 \citep{2021MNRAS.504.3203S}, Tr\,16-26 \citep{2022MNRAS.516.2812C}, $\rho$\,Oph\,Ab \citep{2025A&A...700A..14S}, and the subject of the present work, $\rho$\,Oph\,C. Here, we revisit the original CBO scaling relationship in light of the expanded sample. H$\alpha$ measurements, together with stellar, rotational, and magnetic parameters necessary for calculation of the theoretical H$\alpha$ emission properties of the sample, can be found in the cited works. 

Fig.\ \ref{rhooph_lum_bk1} shows, as a function of the bolometric luminosity, the ratio of the strength of the magnetic field at the Kepler radius $B_{\rm K}$ to the magnetic field strength $B_{\rm K1}$ required to achieve an optical depth of unity at $R_{\rm K}$, assuming a CM in which the mass budget is dominated by CBO. Stars with $\log{L_{\rm bol}} > 2.7$ and $\log{(B_{\rm K}/B_{\rm K1})} > 0$ all have H$\alpha$ in emission, whereas stars outside these ranges generally do not. \cite{2020MNRAS.499.5379S} and \cite{2020MNRAS.499.5366O} discussed two possibilities to explain the apparent absence of emission at low luminosities: 

\noindent 1) a leakage mechanism operating via some combination of steady diffusion and drift across magnetic field lines, as described by \cite{2018MNRAS.474.3090O}, might become relevant in the magnetospheres of low-luminosity stars within which the wind-feeding rate slows to a relative trickle. As discussed by \cite{petit2013}, in this scenario a higher threshold in $B_{\rm K}$ for emission would be expected with decreasing luminosity;

\noindent 2) cool, low-luminosity stars may have runaway metallic winds with very small or zero H content \citep[e.g.][]{2002ApJ...568..965O}. In this case, the weak H$\alpha$ emission is a consequence of the magnetosphere being filled with elements other than H, which themselves never reach a sufficient density to become optically thick in any given spectral line (at least at visible wavelengths). This scenario is explored in detail by \cite{Owocki2026}.

As can be seen in Fig.\ \ref{rhooph_lum_bk1}, both HD\,156424 and $\rho$\,Oph\,C probe the region of high $B_{\rm K}$ below the luminosity cutoff. HD\,156424 is actually a double-lined spectroscopic binary with two magnetic stars, with the H$\alpha$ emission originating in the much cooler secondary, HD\,156424\,B. The spectrum of HD\,156424\,B is heavily blended with the much brighter early B-type star HD\,156424\,A; as a result of this, its stellar, magnetic, and rotational parameters are not precisely determined. As the spectrum of $\rho$\,Oph\,C is completely unaffected by its much dimmer M-type companions, there is no systematic uncertainty in its parameters, and it is firmly below the luminosity threshold. 

Fig.\ \ref{halpha_population} shows, as a function of bolometric luminosity, a comparison between the emission equivalent width EW$_{\rm obs}$ measured directly from the H$\alpha$ profile and the values EW$_{\rm CBO}$ predicted by the scaling relationship determined by \cite{2020MNRAS.499.5366O}, using the measurements presented here as well as those provided by \cite{2020MNRAS.499.5379S,2021MNRAS.504.4850S,2021MNRAS.504.3203S,2025A&A...700A..14S} and \cite{2022MNRAS.516.2812C}. Following \cite{2020MNRAS.499.5379S}, spectroscopic binaries (indicated in the figure) have had corrections applied to account for dilution of the spectrum by the non-magnetic companion(s), following the method described by \cite{2020MNRAS.499.5379S}. Also following \cite{2020MNRAS.499.5379S}, Fig.\ \ref{halpha_population} further shows single-cloud emission EWs (i.e.\ the highest value measured in either the blue or the red wing individually), in order to avoid biasing the comparison with anomalously low values for some spectroscopic binaries with highly asymmetric magnetospheres \citep[e.g. HD\,37017, HD\,156324, and $\rho$\,Oph\,Ab;][]{2018MNRAS.475..839S,2020MNRAS.499.5379S,2025A&A...700A..14S}. 

As can be seen in Fig.\ \ref{halpha_population}, ALS\,9522\,B, HD\,156424\,B, $\rho$\,Oph\,Ab, and Tr\,16-26 are all consistent within uncertainty with their expected values, validating the \cite{2020MNRAS.499.5366O} CBO scaling relationship. However, $\rho$\,Oph\,C is a notable outlier, with H$\alpha$ emission almost 4 times weaker than expected. The other outliers are mostly objects with much larger uncertainties, agreeing with the scaling relationship at about the 1$\sigma$ level; however, $\rho$\,Oph\,C's parameters are much more precisely determined. 

 The highest-luminosity H$\alpha$-bright star in the sample \citep[the SB3 HD\,164492\,C;][]{2014A&A...564L..10H,2017MNRAS.465.2517W} is the only star to lie as far off of the scaling relationship as $\rho$\,Oph\,C, albeit in the other direction i.e.\ its magnetosphere is much brighter than expected. This is suggestive of a possible residual trend with luminosity, as illustrated with the linear regression in Fig.\ \ref{halpha_population}. The Pearson's correlation coefficient of the regression is 0.5, which is suggestive of a real relationship but not close enough to 1 to be absolutely certain. Dropping either $\rho$\,Oph\,C or HD\,164492\,C from the regression reduces the correlation, and dropping both reduces it to 0.4, i.e.\ probably not significant. It is therefore very sensitive to the two most extreme stars in the population. A similar test was attempted by \cite{2020MNRAS.499.5379S}, but yielded even more ambiguous results due to the use of a coarser regression of EW vs. $B_{\rm K}$ \citep[rather than the scaling relationship developed by][]{2020MNRAS.499.5366O}, as well as of course the absence of $\rho$\,Oph\,C in their dataset. 

Stars with H$\alpha$ in absorption almost all have $\log{(\rm EW_{obs}/EW_{CBO})}$ below 0, with the discrepancy between the measured upper limit and the expected value increasing with decreasing luminosity. In this context, $\rho$\,Oph\,C's relatively weak H$\alpha$ emission is consistent with the discrepancies between upper limits and predicted emission strengths that are seen in stars with comparable luminosities. 

   \begin{figure}
   \centering
   \includegraphics[width=0.45\textwidth]{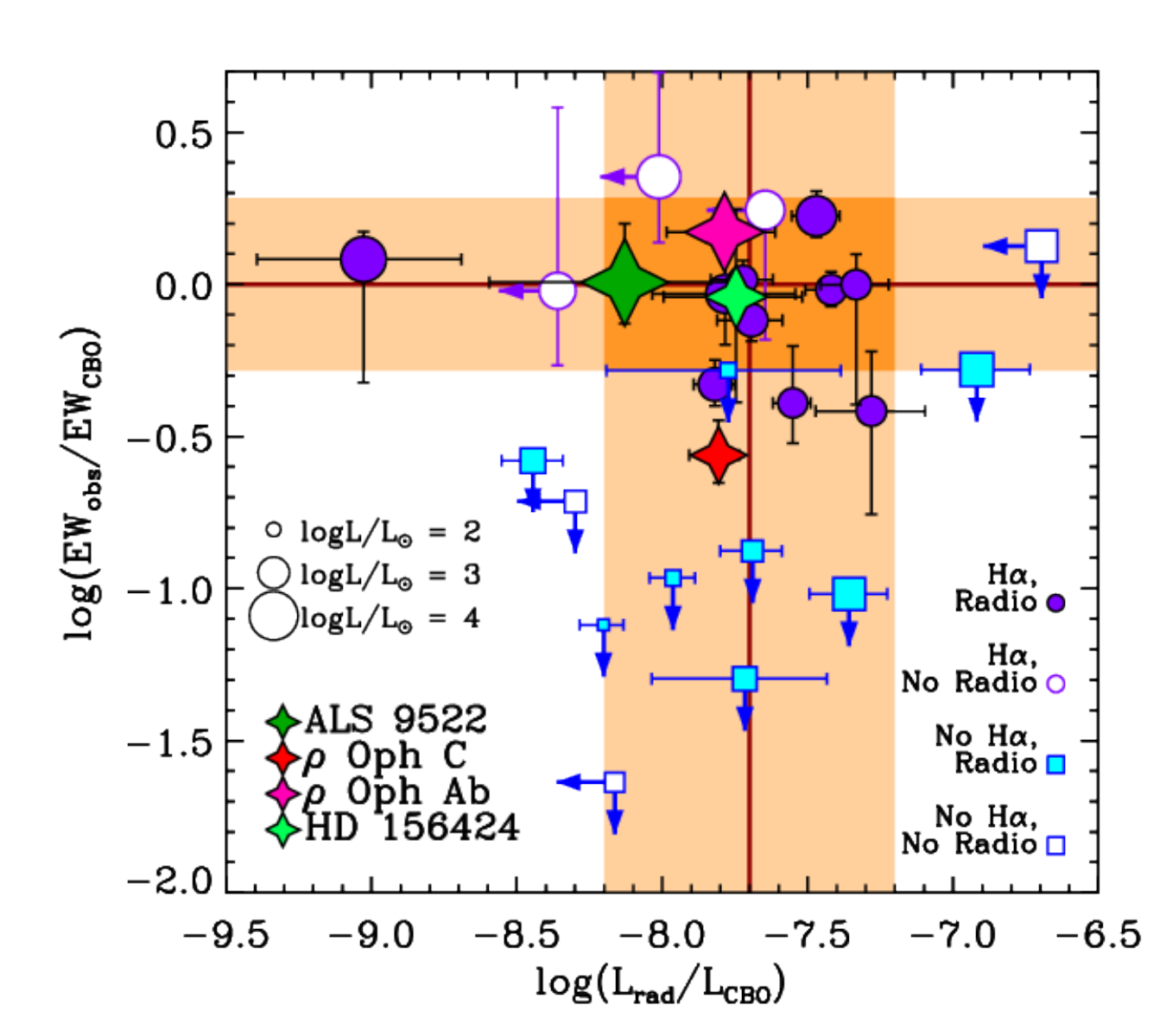}
      \caption[]{Residual EW as a function of residual radio luminosity. Symbol size is proportional to bolometric luminosity. Highlighted stars are as in Figs.\ \ref{rhooph_lum_bk1} and \ref{halpha_population} (Tr 16-26 is not shown because there is no radio data for that star). The horizontal shaded region indicates the nominal CBO EW and the mean measured error bar, as in Fig.\ \ref{halpha_population}. The vertical shaded region indicates the nominal CBO radio luminosity adjusted for the efficiency factor, and the dispersion in the radio scaling relationship. The dark shaded box indicates stars with good agreement between observation and theory for both H$\alpha$ and radio.}
         \label{delta_halpha_delta_radio}
   \end{figure}

In Fig.\ \ref{delta_halpha_delta_radio} we show the ratio of the observed to predicted EW as a function of the ratio of the observed to predicted radio luminosity, with symbol size scaled by bolometric luminosity. Residual radio luminosities were found using the radio luminosities $L_{\rm rad}$ collected by \cite{2022MNRAS.513.1429S}  and the breakout radio luminosity $L_{\rm CBO}$ calculated using the prescription given by \cite{2022MNRAS.513.1449O}, adjusted for obliquity angle as discussed by \cite{2022MNRAS.513.1429S}. The mean ratio $L_{\rm rad}/L_{\rm CBO} \sim 10^{-7.7}$ is consistent with the efficiency factor discussed by \cite{2022MNRAS.513.1449O}. 

For stars with detected H$\alpha$ emission and detected radio emission, all but 2 cluster tightly at the centre of the diagram. The exceptions are $\rho$\,Oph\,C, which is underluminous in H$\alpha$ but has the expected radio luminosity, and HD\,64740, which is underluminous in radio but has the expected H$\alpha$ emission strength. As discussed by \cite{2022MNRAS.513.1429S}, HD\,64740 is the most luminous star with detected radio emission, and due to its strong wind has a radio photosphere larger than its Alfv\'en radius, which attenuates the star's gyrosynchrotron radiation. 

Turning our attention to stars with detected radio but no H$\alpha$, the majority of them have upper limits far below the expected H$\alpha$ emission strength, but radio luminosities consistent with the values expected from CBO. Conversely, there are no stars with both H$\alpha$ and radio far below the expected values, nor are there any stars with both diagnostics far stronger than the expected values. 

As demonstrated in Figs.\ \ref{HD147932_periods_lc} and \ref{rhoOphC_K2}, $\rho$\,Oph\,C's light curve is broadly consistent with the pattern of variability of its metallic absorption lines, indicating that the photometric variation is primarily a result of chemical spots. However, the {\em TESS} light curve also demonstrates complex substructure at the mmag level which is also modulated on the rotational period, which probably has an origin in occlusion of the photosphere by the magnetosphere. The amplitude of the magnetospheric variation is considerably lower than the amplitude that would be expected from a CBO model, which to the contrary predicts a light curve with prominent magnetospheric eclipses at the tens of mmag level (Fig.\ \ref{rhoOphC_K2}). Weaker than expected photometric amplitude is consistent with weaker than expected H$\alpha$ emission strength. 

In a leakage scenario, the total mass in the magnetosphere should be lower than the value predicted by a CBO model. If the leakage rate is relatively independent of stellar, magnetic, or rotational parameters, this discrepancy should increase as the mass-loss rates decrease. Of course, if the leakage mechanism is sensitive to one or more of these properties, there may be complex dependencies that make any such decline difficult to discern. The assumption of a steady-state leakage rate would be consistent with a simultaneous decline in both H$\alpha$ emission and photometric eclipse amplitude in a transitional object like $\rho$\,Oph\,C, in which leakage is presumably just beginning to become important. 

A real decline in magnetospheric density should also lead to a decline in radio luminosity in cooler, less luminous stars, which should be less radio bright than predicted by the CBO model. To the contrary, there are no systematic trends of $L_{\rm rad}$ with decreasing $L_{\rm bol}$: the radio luminosity scaling relationship holds even at very low luminosities. 

In a metallic wind scenario, H$\alpha$ disappears for the simple reason that there is no H in the magnetosphere, but the total mass trapped by the magnetic field remains the same. However, the mass of each ion is much higher than the mass of an individual proton. Since metallic ions will each contribute only one or two electrons per particle, whereas in a H dominated plasma there is a 1:1 ratio of ionized H to electrons, this means that the electron density of metallic magnetospheres would be much lower. Thus, in addition to H emission disappearing, the photometric amplitude of eclipses (which depends on electron scattering) should also decrease significantly (but should not disappear entirely). Radio emission is produced by high-energy electrons accelerated by centrifugally driven magnetic reconnection, so we might expect that radio luminosity should also decrease. However, only a very small fraction of electrons need to be excited to produce the necessary emission levels, so a reduced baseline electron density in the cool magnetosphere may well not affect gyrosynchrotron levels. 

As explored by \cite{Owocki2026}, wind models indicate that H decoupling is expected around $\rho$\,Oph\,C's effective temperature, suggesting that the star may be a transitional object. Furthermore, runaway metallic winds are not expected to appear isotropically; instead, they begin at the highest magnetic colatitudes and work their way down. Thus, the outermost magnetospheric regions are denuded of H first, with the metallic magnetosphere appearing at large distances and working its way inward to the Kepler radius. Consistent with this scenario, \cite{Owocki2026} show that the outer wings of $\rho$\,Oph\,C's emission line profile (Fig. \ref{rhoOphC_halpha_multiplot}) are truncated in comparison to the averaged emission profile wing observed by \cite{2020MNRAS.499.5379S} in other CM host stars \citep[which][demonstrated could be explained as a result of optically thick emission in a CBO-regulated CM]{2020MNRAS.499.5366O}.

A runaway metallic wind is the preferred hypothesis to explain the disappearance of H$\alpha$ emission, the near-disappearance of photometric eclipses, and the persistence of radio emission at low luminosities. If so, this opens low-luminosity magnetic AB stars as probes for the exploration of a wind regime which has heretofore been practically ignored. 

\section{Summary}\label{sect:summary}

$\rho$\,Oph\,C's magnetic field was previously estimated from only two high-resolution HARPSpol observations. A preliminary magnetic model was based on those data, and on the behaviour of the star's auroral radio emission, such that the star must be seen at high inclination, with a small magnetic obliquity angle and a surface field on the order of 10 kG. Our dataset confirms these expectations. 

H$\alpha$ emission was previously reported for $\rho$\,Oph\,C, consistent with its status as a rapidly rotating, strongly magnetized hot star with a large centrifugal magnetosphere. Our dataset shows that this emission is present at all phases and exhibits a low level of variability, consistent with the small obliquity angle of the magnetic field and the presence of persistent auroral radio emission. 

The major fraction of the variability seen in the star's {\em TESS} light curve is consistent with an origin in surface chemical abundance patches, with photometric behaviour approximately matching the patterns seen in the equivalent widths of metallic lines. However, there are complex rotationally modulated low amplitude variations which are difficult to explain as a result of surface features, and seem more consistent with weak eclipsing by magnetospheric plasma. 

The H$\alpha$ emission of $\rho$\,Oph\,C is much weaker than expected from the CBO-derived scaling law, and the amplitude of the light curve variations that can plausibly be attributed to circumstellar material is similarly much weaker than expected from the CBO model. However, the star's radio luminosity is perfectly consistent with the CBO scaling law. This suggests that $\rho$\,Oph\,C is probably a transitional object between regimes with H-dominated winds and runaway metallic winds. This scenario is explored in greater depth in a companion paper by \cite{Owocki2026}.

\section*{Acknowledgements}

This work is based on data obtained as part of the Canada-France Imaging Survey, a CFHT large program of the National Research Council of Canada and the French Centre National de la Recherche Scientifique. Based on observations obtained with MegaPrime/MegaCam, a joint project of CFHT and CEA Saclay, at the Canada-France-Hawai'i Telescope (CFHT) which is operated by the National Research Council (NRC) of Canada, the Institut National des Science de l'Univers (INSU) of the Centre National de la Recherche Scientifique (CNRS) of France, and the University of Hawai'i; at the La Silla Observatory, ESO Chile with the 3.6 m telescope; and based on observations obtained at the Dominion Astrophysical Observatory, National Research Council Herzberg Astronomy and Astrophysics Research Centre, National Research Council of Canada. Based on observations made with ESO Telescopes at the La Silla and Paranal Observatories under programme IDs 60.A-9800(D), 71.C-0068(A), 073.C-0138(A), 082.C-0005(B), and 187.D-0917(D). This work has made use of the VALD database, operated at Uppsala University, the Institute of Astronomy RAS in Moscow, and the University of Vienna. This research has made use of the SIMBAD database, operated at CDS, Strasbourg, France. MES acknowledges the financial support provided by the Annie Jump Cannon Fellowship, supported by the University of Delaware and endowed by the Mount Cuba Astronomical Observatory. AuD acknowledges NASA ATP grant number 80NSSC22K0628 and support by NASA through Chandra Award number TM1-22001B and GO2-23003X issued by the Chandra X-ray Observatory 27 Center, which is operated by the Smithsonian Astrophysical Observatory for and on behalf of NASA under contract NAS8-03060. GAW acknowledges Discovery Grant support from the Natural Sciences and Engineering Research Council (NSERC) of Canada. JK was supported by grant GA \v{C}R 25-15910S. Funded/Co-funded by the European Union (ERC, MAGNIFY, Project 101126182 ). Views and opinions expressed are however those of the author(s) only and do not necessarily reflect those of the European Union or the European Research Council. Neither the European Union nor the granting authority can be held responsible for them. We would like to thank the Canada-France-Hawai'i Telescope (CFHT) Operations, Instrumentation, and Software Groups for their contributions and diligence in maintaining observatory operations; the CFHT Astronomy Group for their observation coordination and data acquisition efforts; and the CFHT Finance \& Administration Group for their contributions to the management and administration of the observatory. CFHT is located on Maunakea on Hawai'i Island, a mountain of considerable cultural, natural, and ecological significance. Maunakea is a sacred site to Native Hawaiians, also known as Kanaka'Oiwi. This paper includes data collected with the {\em TESS} mission, obtained from the MAST data archive at the Space Telescope Science Institute (STScI). Funding for US Institutions for the {\em TESS} mission is provided by the NASA Explorer Program. STScI is operated by the Association of Universities for Research in Astronomy, Inc., under NASA contract NAS 5–26555. This paper includes data collected by the Kepler mission and obtained from the MAST data archive at the Space Telescope Science Institute (STScI). Funding to US Institutions for the Kepler mission was provided by the NASA Science Mission Directorate. STScI is operated by the Association of Universities for Research in Astronomy, Inc., under NASA contract NAS 5–26555.

\section*{Data Availability Statement}

Reduced ESPaDOnS spectra are available at the CFHT archive maintained by the CADC\footnote{\url{https://www.cadc-ccda.hia-iha.nrc-cnrc.gc.ca/en/}}, where they can be found via standard stellar designations. ESPaDOnS and Narval data can also be obtained at the PolarBase archive\footnote{\url{http://polarbase.irap.omp.eu/}}. {\em K2} and {\em TESS} data are available at the MAST archive\footnote{\url{https://mast.stsci.edu/portal/Mashup/Clients/Mast/Portal.html}}. {\em Hipparcos} data are available through SIMBAD and Vizier\footnote{\url{https://vizier.u-strasbg.fr/viz-bin/VizieR}}. HARPSpol spectra, along with the reduced data products used in this paper, are available from the authors on request.



\bibliographystyle{mnras}
\bibliography{bib_dat} 

@ARTICLE{bm1997,
   author = {{Babel}, J. and {Montmerle}, T.},
    title = "{On the Periodic X-Ray Emission from the O7 V Star theta 1 Orionis C}",
  journal = {\apjl},
     year = 1997,
    month = aug,
   volume = 485,
    pages = {L29},
      doi = {10.1086/310806},
   adsurl = {http://cdsads.u-strasbg.fr/abs/1997ApJ...485L..29B}
}

@ARTICLE{d1997,
   author = {{Donati}, J.-F. and {Semel}, M. and {Carter}, B.~D. and {Rees}, D.~E. and 
	{Collier Cameron}, A.},
    title = "{Spectropolarimetric observations of active stars}",
  journal = {MNRAS},
     year = 1997,
    month = nov,
   volume = 291,
    pages = {658},
   adsurl = {http://adsabs.harvard.edu/abs/1997MNRAS.291..658D}
}

@ARTICLE{grun2012,
   author = {{Grunhut}, J.~H. and {Rivinius}, T. and {Wade}, G.~A. and {Townsend}, R.~H.~D. and 
	{Marcolino}, W.~L.~F. and {Bohlender}, D.~A. and {Szeifert}, T. and 
	{Petit}, V. and {Matthews}, J.~M. and {Rowe}, J.~F. and {Moffat}, A.~F.~J. and 
	{Kallinger}, T. and {Kuschnig}, R. and {Guenther}, D.~B. and 
	{Rucinski}, S.~M. and {Sasselov}, D. and {Weiss}, W.~W.},
    title = "{HR 5907: Discovery of the most rapidly rotating magnetic early B-type star by the MiMeS Collaboration}",
  journal = {\mnras},
archivePrefix = "arXiv",
   eprint = {1109.3157},
 primaryClass = "astro-ph.SR",
     year = 2012,
    month = jan,
   volume = 419,
    pages = {1610-1627},
      doi = {10.1111/j.1365-2966.2011.19824.x},
   adsurl = {http://cdsads.u-strasbg.fr/abs/2012MNRAS.419.1610G}
}

@ARTICLE{koch2010,
   author = {{Kochukhov}, O. and {Makaganiuk}, V. and {Piskunov}, N.},
    title = "{Least-squares deconvolution of the stellar intensity and polarization spectra}",
  journal = {\aap},
archivePrefix = "arXiv",
   eprint = {1008.5115},
 primaryClass = "astro-ph.SR",
     year = 2010,
    month = dec,
   volume = 524,
      eid = {A5},
    pages = {A5},
      doi = {10.1051/0004-6361/201015429},
   adsurl = {http://cdsads.u-strasbg.fr/abs/2010A26A...524A...5K}
}

@ARTICLE{lb1978,
   author = {{Landstreet}, J.~D. and {Borra}, E.~F.},
    title = "{The magnetic field of Sigma Orionis E}",
  journal = {\apjl},
     year = 1978,
    month = aug,
   volume = 224,
    pages = {L5-L8},
      doi = {10.1086/182746},
   adsurl = {http://cdsads.u-strasbg.fr/abs/1978ApJ...224L...5L}
}

@ARTICLE{mat1989,
   author = {{Mathys}, G.},
    title = "{The Observation of Magnetic Fields in Nondegenerate Stars}",
  journal = {FCPh},
     year = 1989,
   volume = 13,
    pages = {143-308},
   adsurl = {http://adsabs.harvard.edu/abs/1989FCPh...13..143M}
}

@ARTICLE{perry1997,
   author = {{Perryman}, M.~A.~C. and {Lindegren}, L. and {Kovalevsky}, J. and 
	{Hoeg}, E. and {Bastian}, U. and {Bernacca}, P.~L. and {Cr{\'e}z{\'e}}, M. and 
	{Donati}, F. and {Grenon}, M. and {Grewing}, M. and {van Leeuwen}, F. and 
	{van der Marel}, H. and {Mignard}, F. and {Murray}, C.~A. and 
	{Le Poole}, R.~S. and {Schrijver}, H. and {Turon}, C. and {Arenou}, F. and 
	{Froeschl{\'e}}, M. and {Petersen}, C.~S.},
    title = "{The HIPPARCOS Catalogue}",
  journal = {\aap},
     year = 1997,
    month = jul,
   volume = 323,
    pages = {L49-L52},
   adsurl = {http://adsabs.harvard.edu/abs/1997A26A...323L..49P}
}

@ARTICLE{petit2013,
   author = {{Petit}, V. and {Owocki}, S.~P. and {Wade}, G.~A. and {Cohen}, D.~H. and 
	{Sundqvist}, J.~O. and {Gagn{\'e}}, M. and {Ma{\'{\i}}z Apell{\'a}niz}, J. and 
	{Oksala}, M.~E. and {Bohlender}, D.~A. and {Rivinius}, T. and 
	{Henrichs}, H.~F. and {Alecian}, E. and {Townsend}, R.~H.~D. and 
	{ud-Doula}, A. and {MiMeS Collaboration}},
    title = "{{\bf P13:} A magnetic confinement versus rotation classification of massive-star magnetospheres}",
  journal = {\mnras},
archivePrefix = "arXiv",
   eprint = {1211.0282},
 primaryClass = "astro-ph.SR",
     year = 2013,
    month = feb,
   volume = 429,
    pages = {398-422},
      doi = {10.1093/mnras/sts344},
   adsurl = {http://adsabs.harvard.edu/abs/2013MNRAS.429..398P}
}

@ARTICLE{town2005c,
   author = {{Townsend}, R.~H.~D. and {Owocki}, S.~P.},
    title = "{A rigidly rotating magnetosphere model for circumstellar emission from magnetic OB stars}",
  journal = {\mnras},
   eprint = {0408565},
     year = 2005,
    month = feb,
   volume = 357,
    pages = {251-264},
      doi = {10.1111/j.1365-2966.2005.08642.x},
   adsurl = {http://adsabs.harvard.edu/abs/2005MNRAS.357..251T}
}

@ARTICLE{ud2002,
   author = {{ud-Doula}, A. and {Owocki}, S.~P.},
    title = "{Dynamical Simulations of Magnetically Channeled Line-driven Stellar Winds. I. Isothermal, Nonrotating, Radially Driven Flow}",
  journal = {ApJ},
   eprint = {0201195},
     year = 2002,
    month = sep,
   volume = 576,
    pages = {413-428},
      doi = {10.1086/341543},
   adsurl = {http://adsabs.harvard.edu/abs/2002ApJ...576..413U}
}

@ARTICLE{ud2008,
   author = {{ud-Doula}, A. and {Owocki}, S.~P. and {Townsend}, R.~H.~D.},
    title = "{Dynamical simulations of magnetically channelled line-driven stellar winds - II. The effects of field-aligned rotation}",
  journal = {MNRAS},
archivePrefix = "arXiv",
   eprint = {0712.2780},
     year = 2008,
    month = mar,
   volume = 385,
    pages = {97-108},
      doi = {10.1111/j.1365-2966.2008.12840.x},
   adsurl = {http://adsabs.harvard.edu/abs/2008MNRAS.385...97U}
}

@ARTICLE{ud2009,
   author = {{ud-Doula}, A. and {Owocki}, S.~P. and {Townsend}, R.~H.~D.},
    title = "{Dynamical simulations of magnetically channelled line-driven stellar winds - III. Angular momentum loss and rotational spin-down}",
  journal = {MNRAS},
archivePrefix = "arXiv",
   eprint = {0810.4247},
     year = 2009,
    month = jan,
   volume = 392,
    pages = {1022-1033},
      doi = {10.1111/j.1365-2966.2008.14134.x},
   adsurl = {http://adsabs.harvard.edu/abs/2009MNRAS.392.1022U}
}

@ARTICLE{vink2001,
   author = {{Vink}, J.~S. and {de Koter}, A. and {Lamers}, H.~J.~G.~L.~M.
	},
    title = "{Mass-loss predictions for O and B stars as a function of metallicity}",
  journal = {\aap},
   eprint = {0101509},
     year = 2001,
    month = apr,
   volume = 369,
    pages = {574-588},
      doi = {10.1051/0004-6361:20010127},
   adsurl = {http://adsabs.harvard.edu/abs/2001A26A...369..574V}
}

@ARTICLE{lanzhubeny2007,
   author = {{Lanz}, T. and {Hubeny}, I.},
    title = "{A Grid of NLTE Line-blanketed Model Atmospheres of Early B-Type Stars}",
  journal = {\apjs},
   eprint = {0611891},
     year = 2007,
    month = mar,
   volume = 169,
    pages = {83-104},
      doi = {10.1086/511270},
   adsurl = {http://adsabs.harvard.edu/abs/2007ApJS..169...83L}
}

@ARTICLE{henrichs2013,
   author = {{Henrichs}, H.~F. and {de Jong}, J.~A. and {Verdugo}, E. and 
	{Schnerr}, R.~S. and {Neiner}, C. and {Donati}, J.-F. and {Catala}, C. and 
	{Shorlin}, S.~L.~S. and {Wade}, G.~A. and {Veen}, P.~M. and 
	{Nichols}, J.~S. and {Damen}, E.~M.~F. and {Talavera}, A. and 
	{Hill}, G.~M. and {Kaper}, L. and {Tijani}, A.~M. and {Geers}, V.~C. and 
	{Wiersema}, K. and {Plaggenborg}, B. and {Rygl}, K.~L.~J.},
    title = "{Discovery of the magnetic field in the pulsating B star {$\beta$} Cephei}",
  journal = {\aap},
archivePrefix = "arXiv",
   eprint = {1305.2601},
 primaryClass = "astro-ph.SR",
     year = 2013,
    month = jul,
   volume = 555,
      eid = {A46},
    pages = {A46},
      doi = {10.1051/0004-6361/201321584},
   adsurl = {http://adsabs.harvard.edu/abs/2013A26A...555A..46H}
}

@ARTICLE{krticka2014,
   author = {{Krti{\v c}ka}, J.},
    title = "{Mass loss in main-sequence B stars}",
  journal = {\aap},
archivePrefix = "arXiv",
   eprint = {1401.5511},
 primaryClass = "astro-ph.SR",
     year = 2014,
    month = apr,
   volume = 564,
      eid = {A70},
    pages = {A70},
      doi = {10.1051/0004-6361/201321980},
   adsurl = {http://cdsads.u-strasbg.fr/abs/2014A26A...564A..70K}
}

@ARTICLE{vanleeuwen2007,
   author = {{van Leeuwen}, F.},
    title = "{Validation of the new Hipparcos reduction}",
  journal = {\aap},
archivePrefix = "arXiv",
   eprint = {0708.1752},
     year = 2007,
    month = nov,
   volume = 474,
    pages = {653-664},
      doi = {10.1051/0004-6361:20078357},
   adsurl = {http://adsabs.harvard.edu/abs/2007A26A...474..653V}
}

@ARTICLE{preston1967,
   author = {{Preston}, G.~W.},
    title = "{A Statistical Investigation of the Orientation of Magnetic Axes in the Periodic Magnetic Variables}",
  journal = {\apj},
     year = 1967,
    month = nov,
   volume = 150,
    pages = {547},
      doi = {10.1086/149358},
   adsurl = {http://adsabs.harvard.edu/abs/1967ApJ...150..547P}
}

@ARTICLE{alecian2014,
   author = {{Alecian}, E. and {Kochukhov}, O. and {Petit}, V. and {Grunhut}, J. and 
	{Landstreet}, J. and {Oksala}, M.~E. and {Wade}, G.~A. and {Hussain}, G. and 
	{Neiner}, C. and {Bohlender}, D. and {MiMeS Collaboration}},
    title = "{Discovery of new magnetic early-B stars within the MiMeS HARPSpol survey}",
  journal = {\aap},
archivePrefix = "arXiv",
   eprint = {1404.5508},
 primaryClass = "astro-ph.SR",
     year = 2014,
    month = jul,
   volume = 567,
      eid = {A28},
    pages = {A28},
      doi = {10.1051/0004-6361/201323286},
   adsurl = {http://adsabs.harvard.edu/abs/2014A26A...567A..28A}
}

@ARTICLE{kupka2000,
   author = {{Kupka}, F.~G. and {Ryabchikova}, T.~A. and {Piskunov}, N.~E. and 
	{Stempels}, H.~C. and {Weiss}, W.~W.},
    title = "{VALD-2 -- The New Vienna Atomic Line Database}",
  journal = {Balt.\ Astron.},
     year = 2000,
   volume = 9,
    pages = {590-594},
   adsurl = {http://adsabs.harvard.edu/abs/2000BaltA...9..590K}
}

@ARTICLE{kupka1999,
   author = {{Kupka}, F.~G. and {Piskunov}, N. and {Ryabchikova}, T.~A. and 
	{Stempels}, H.~C. and {Weiss}, W.~W.},
    title = "{VALD-2: Progress of the Vienna Atomic Line Data Base}",
  journal = {\aaps},
     year = 1999,
    month = jul,
   volume = 138,
    pages = {119-133},
      doi = {10.1051/aas:1999267},
   adsurl = {http://adsabs.harvard.edu/abs/1999A26AS..138..119K}
}

@ARTICLE{ryabchikova1997,
   author = {{Ryabchikova}, T.~A. and {Piskunov}, N.~E. and {Kupka}, F. and 
	{Weiss}, W.~W.},
    title = "{The Vienna Atomic Line Database : Present State and Future Development}",
  journal = {Balt.\ Astron.},
     year = 1997,
    month = mar,
   volume = 6,
    pages = {244-247},
   adsurl = {http://cdsads.u-strasbg.fr/abs/1997BaltA...6..244R}
}

@ARTICLE{piskunov1995,
   author = {{Piskunov}, N.~E. and {Kupka}, F. and {Ryabchikova}, T.~A. and 
	{Weiss}, W.~W. and {Jeffery}, C.~S.},
    title = "{VALD: The Vienna Atomic Line Data Base.}",
  journal = {\aaps},
     year = 1995,
    month = sep,
   volume = 112,
    pages = {525},
   adsurl = {http://adsabs.harvard.edu/abs/1995A26AS..112..525P}
}

@ARTICLE{ekstrom2012,
   author = {{Ekstr{\"o}m}, S. and {Georgy}, C. and {Eggenberger}, P. and 
	{Meynet}, G. and {Mowlavi}, N. and {Wyttenbach}, A. and {Granada}, A. and 
	{Decressin}, T. and {Hirschi}, R. and {Frischknecht}, U. and 
	{Charbonnel}, C. and {Maeder}, A.},
    title = "{Grids of stellar models with rotation. I. Models from 0.8 to 120 M$_{⊙}$ at solar metallicity (Z = 0.014)}",
  journal = {\aap},
archivePrefix = "arXiv",
   eprint = {1110.5049},
 primaryClass = "astro-ph.SR",
     year = 2012,
    month = jan,
   volume = 537,
      eid = {A146},
    pages = {A146},
      doi = {10.1051/0004-6361/201117751},
   adsurl = {http://cdsads.u-strasbg.fr/abs/2012A26A...537A.146E}
}

@ARTICLE{preston1974,
   author = {{Preston}, G.~W.},
    title = "{The chemically peculiar stars of the upper main sequence}",
  journal = {\araa},
     year = 1974,
   volume = 12,
    pages = {257-277},
      doi = {10.1146/annurev.aa.12.090174.001353},
   adsurl = {http://cdsads.u-strasbg.fr/abs/1974ARA26A..12..257P}
}

@ARTICLE{2002yCat.2237....0D,
   author = {{Ducati}, J.~R.},
    title = "{VizieR Online Data Catalog: Catalogue of Stellar Photometry in Johnson's 11-color system.}",
  journal = {VizieR Online Data Catalog},
     year = 2002,
   volume = 2237,
    pages = {0},
   adsurl = {http://cdsads.u-strasbg.fr/abs/2002yCat.2237....0D}
}

@ARTICLE{2014ApJS..215...10N,
   author = {{Naz{\'e}}, Y. and {Petit}, V. and {Rinbrand}, M. and {Cohen}, D. and 
	{Owocki}, S. and {ud-Doula}, A. and {Wade}, G.~A.},
    title = "{X-Ray Emission from Magnetic Massive Stars}",
  journal = {\apjs},
archivePrefix = "arXiv",
   eprint = {1409.1690},
 primaryClass = "astro-ph.SR",
     year = 2014,
    month = nov,
   volume = 215,
      eid = {10},
    pages = {10},
      doi = {10.1088/0067-0049/215/1/10},
   adsurl = {http://adsabs.harvard.edu/abs/2014ApJS..215...10N}
}

@ARTICLE{2015MNRAS.449.3945S,
   author = {{Shultz}, M.~E. and {Rivinius}, T. and {Folsom}, C.~P. and {Wade}, G.~A. and 
	{Townsend}, R.~H.~D. and {Sikora}, J. and {Grunhut}, J. and 
	{Stahl}, O. and {MiMeS Collaboration}},
    title = "{The magnetic field and spectral variability of the He-weak star HR 2949}",
  journal = {\mnras},
archivePrefix = "arXiv",
   eprint = {1504.03289},
 primaryClass = "astro-ph.SR",
     year = 2015,
    month = jun,
   volume = 449,
    pages = {3945-3965},
      doi = {10.1093/mnras/stv564},
   adsurl = {http://adsabs.harvard.edu/abs/2015MNRAS.449.3945S}
}

@ARTICLE{2016MNRAS.456....2W,
   author = {{Wade}, G.~A. and {Neiner}, C. and {Alecian}, E. and {Grunhut}, J.~H. and 
	{Petit}, V. and {Batz}, B.~d. and {Bohlender}, D.~A. and {Cohen}, D.~H. and 
	{Henrichs}, H.~F. and {Kochukhov}, O. and {Landstreet}, J.~D. and 
	{Manset}, N. and {Martins}, F. and {Mathis}, S. and {Oksala}, M.~E. and 
	{Owocki}, S.~P. and {Rivinius}, T. and {Shultz}, M.~E. and {Sundqvist}, J.~O. and 
	{Townsend}, R.~H.~D. and {ud-Doula}, A. and {Bouret}, J.-C. and 
	{Braithwaite}, J. and {Briquet}, M. and {Carciofi}, A.~C. and 
	{David-Uraz}, A. and {Folsom}, C.~P. and {Fullerton}, A.~W. and 
	{Leroy}, B. and {Marcolino}, W.~L.~F. and {Moffat}, A.~F.~J. and 
	{Naz{\'e}}, Y. and {Louis}, N.~S. and {Auri{\`e}re}, M. and 
	{Bagnulo}, S. and {Bailey}, J.~D. and {Barb{\'a}}, R.~H. and 
	{Blaz{\`e}re}, A. and {B{\"o}hm}, T. and {Catala}, C. and {Donati}, J.-F. and 
	{Ferrario}, L. and {Harrington}, D. and {Howarth}, I.~D. and 
	{Ignace}, R. and {Kaper}, L. and {L{\"u}ftinger}, T. and {Prinja}, R. and 
	{Vink}, J.~S. and {Weiss}, W.~W. and {Yakunin}, I.},
    title = "{The MiMeS survey of magnetism in massive stars: introduction and overview}",
  journal = {\mnras},
archivePrefix = "arXiv",
   eprint = {1511.08425},
 primaryClass = "astro-ph.SR",
     year = 2016,
    month = feb,
   volume = 456,
    pages = {2-22},
      doi = {10.1093/mnras/stv2568},
   adsurl = {http://adsabs.harvard.edu/abs/2016MNRAS.456....2W}
}

@ARTICLE{2014A&A...564L..10H,
   author = {{Hubrig}, S. and {Fossati}, L. and {Carroll}, T.~A. and {Castro}, N. and 
	{Gonz{\'a}lez}, J.~F. and {Ilyin}, I. and {Przybilla}, N. and 
	{Sch{\"o}ller}, M. and {Oskinova}, L.~M. and {Morel}, T. and 
	{Langer}, N. and {Scholz}, R.~D. and {Kharchenko}, N.~V. and 
	{Nieva}, M.-F.},
    title = "{B fields in OB stars (BOB): The discovery of a magnetic field in a multiple system in the Trifid nebula, one of the youngest star forming regions}",
  journal = {\aap},
archivePrefix = "arXiv",
   eprint = {1403.0491},
 primaryClass = "astro-ph.SR",
     year = 2014,
    month = apr,
   volume = 564,
      eid = {L10},
    pages = {L10},
      doi = {10.1051/0004-6361/201423490},
   adsurl = {http://adsabs.harvard.edu/abs/2014A26A...564L..10H}
}

@ARTICLE{1987ApJ...322..902D,
   author = {{Drake}, S.~A. and {Abbott}, D.~C. and {Bastian}, T.~S. and 
	{Bieging}, J.~H. and {Churchwell}, E. and {Dulk}, G. and {Linsky}, J.~L.
	},
    title = "{The discovery of nonthermal radio emission from magnetic Bp-Ap stars}",
  journal = {\apj},
     year = 1987,
    month = nov,
   volume = 322,
    pages = {902-908},
      doi = {10.1086/165784},
   adsurl = {http://adsabs.harvard.edu/abs/1987ApJ...322..902D}
}

@ARTICLE{2000A&A...362..281T,
   author = {{Trigilio}, C. and {Leto}, P. and {Leone}, F. and {Umana}, G. and 
	{Buemi}, C.},
    title = "{Coherent radio emission from the magnetic chemically peculiar star CU Virginis}",
  journal = {\aap},
   eprint = {0007097},
     year = 2000,
    month = oct,
   volume = 362,
    pages = {281-288},
   adsurl = {http://adsabs.harvard.edu/abs/2000A%26A...362..281T}
}

@ARTICLE{2011Msngr.143....7P,
   author = {{Piskunov}, N. and {Snik}, F. and {Dolgopolov}, A. and {Kochukhov}, O. and 
	{Rodenhuis}, M. and {Valenti}, J. and {Jeffers}, S. and {Makaganiuk}, V. and 
	{Johns-Krull}, C. and {Stempels}, E. and {Keller}, C.},
    title = "{HARPSpol {\mdash} The New Polarimetric Mode for HARPS}",
  journal = {The Messenger},
     year = 2011,
    month = mar,
   volume = 143,
    pages = {7-10},
   adsurl = {http://adsabs.harvard.edu/abs/2011Msngr.143....7P}
}

@ARTICLE{2008AA...491..545N,
   author = {{Netopil}, M. and {Paunzen}, E. and {Maitzen}, H.~M. and {North}, P. and 
	{Hubrig}, S.},
    title = "{Chemically peculiar stars and their temperature calibration}",
  journal = {\aap},
archivePrefix = "arXiv",
   eprint = {0809.5131},
     year = 2008,
    month = nov,
   volume = 491,
    pages = {545-554},
      doi = {10.1051/0004-6361:200810325},
   adsurl = {http://cdsads.u-strasbg.fr/abs/2008A26A...491..545N}
}

@ARTICLE{2015A&A...578A.112O,
   author = {{Oksala}, M.~E. and {Grunhut}, J.~H. and {Kraus}, M. and {Borges Fernandes}, M. and 
	{Neiner}, C. and {Condori}, C.~A.~H. and {Campagnolo}, J.~C.~N. and 
	{Souza}, T.~B.},
    title = "{An infrared diagnostic for magnetism in hot stars}",
  journal = {\aap},
archivePrefix = "arXiv",
   eprint = {1505.01739},
 primaryClass = "astro-ph.SR",
     year = 2015,
    month = jun,
   volume = 578,
      eid = {A112},
    pages = {A112},
      doi = {10.1051/0004-6361/201525987},
   adsurl = {http://adsabs.harvard.edu/abs/2015A%26A...578A.112O}
}

@ARTICLE{2010MNRAS.407.1423M,
   author = {{Martins}, F. and {Donati}, J.-F. and {Marcolino}, W.~L.~F. and 
	{Bouret}, J.-C. and {Wade}, G.~A. and {Escolano}, C. and {Howarth}, I.~D. and 
	{Mimes Collaboration}},
    title = "{Detection of a magnetic field on HD108: clues to extreme magnetic braking and the Of?p phenomenon}",
  journal = {\mnras},
archivePrefix = "arXiv",
   eprint = {1005.1854},
 primaryClass = "astro-ph.SR",
     year = 2010,
    month = sep,
   volume = 407,
    pages = {1423-1432},
      doi = {10.1111/j.1365-2966.2010.17005.x},
   adsurl = {http://cdsads.u-strasbg.fr/abs/2010MNRAS.407.1423M}
}

@ARTICLE{2017MNRAS.465.2517W,
   author = {{Wade}, G.~A. and {Shultz}, M. and {Sikora}, J. and {Bernier}, M.-{\'E}. and 
	{Rivinius}, T. and {Alecian}, E. and {Petit}, V. and {Grunhut}, J.~H. and 
	{BinaMIcS Collaboration}},
    title = "{HD 164492C: a rapidly rotating, H{$\alpha$}-bright, magnetic early B star associated with a 12.5 d spectroscopic binary}",
  journal = {\mnras},
archivePrefix = "arXiv",
   eprint = {1610.02585},
 primaryClass = "astro-ph.SR",
     year = 2017,
    month = mar,
   volume = 465,
    pages = {2517-2530},
      doi = {10.1093/mnras/stw2799},
   adsurl = {http://adsabs.harvard.edu/abs/2017MNRAS.465.2517W}
}

@ARTICLE{2017MNRAS.471.2286S,
   author = {{Shultz}, M. and {Wade}, G.~A. and {Rivinius}, T. and {Neiner}, C. and 
	{Henrichs}, H. and {Marcolino}, W. and {MiMeS Collaboration}
	},
    title = "{The pulsating magnetosphere of the extremely slowly rotating magnetic {$\beta$} Cep star {$\xi$}$^{1}$ CMa}",
  journal = {\mnras},
archivePrefix = "arXiv",
   eprint = {1706.08820},
 primaryClass = "astro-ph.SR",
     year = 2017,
    month = oct,
   volume = 471,
    pages = {2286-2310},
      doi = {10.1093/mnras/stx1632},
   adsurl = {http://adsabs.harvard.edu/abs/2017MNRAS.471.2286S}
}

@ARTICLE{2018MNRAS.475.5144S,
   author = {{Shultz}, M.~E. and {Wade}, G.~A. and {Rivinius}, T. and {Neiner}, C. and 
	{Alecian}, E. and {Bohlender}, D. and {Monin}, D. and {Sikora}, J. and 
	{MiMeS Collaboration} and {BinaMIcS Collaboration}},
    title = "{The magnetic early B-type stars I: magnetometry and rotation}",
  journal = {\mnras},
archivePrefix = "arXiv",
   eprint = {1801.02924},
 primaryClass = "astro-ph.SR",
     year = 2018,
    month = apr,
   volume = 475,
    pages = {5144-5178},
      doi = {10.1093/mnras/sty103},
   adsurl = {http://adsabs.harvard.edu/abs/2018MNRAS.475.5144S}
}

@ARTICLE{2018MNRAS.474.3090O,
   author = {{Owocki}, S.~P. and {Cranmer}, S.~R.},
    title = "{Diffusion-plus-drift models for the mass leakage from centrifugal magnetospheres of magnetic hot-stars}",
  journal = {\mnras},
archivePrefix = "arXiv",
   eprint = {1711.05414},
 primaryClass = "astro-ph.SR",
     year = 2018,
    month = mar,
   volume = 474,
    pages = {3090-3100},
      doi = {10.1093/mnras/stx2989},
   adsurl = {http://adsabs.harvard.edu/abs/2018MNRAS.474.3090O}
}

@ARTICLE{2018MNRAS.475..839S,
   author = {{Shultz}, M.~E. and {Rivinius}, T. and {Wade}, G.~A. and {Alecian}, E. and 
	{Petit}, V.},
    title = "{HD 156324: a tidally locked magnetic triple spectroscopic binary with a disrupted magnetosphere}",
  journal = {\mnras},
archivePrefix = "arXiv",
   eprint = {1712.04280},
 primaryClass = "astro-ph.SR",
     year = 2018,
    month = mar,
   volume = 475,
    pages = {839-852},
      doi = {10.1093/mnras/stx3238},
   adsurl = {http://adsabs.harvard.edu/abs/2018MNRAS.475..839S}
}

@ARTICLE{2013ApJS..208....9P,
   author = {{Pecaut}, M.~J. and {Mamajek}, E.~E.},
    title = "{Intrinsic Colors, Temperatures, and Bolometric Corrections of Pre-main-sequence Stars}",
  journal = {\apjs},
archivePrefix = "arXiv",
   eprint = {1307.2657},
 primaryClass = "astro-ph.SR",
     year = 2013,
    month = sep,
   volume = 208,
      eid = {9},
    pages = {9},
      doi = {10.1088/0067-0049/208/1/9},
   adsurl = {http://adsabs.harvard.edu/abs/2013ApJS..208....9P}
}

@ARTICLE{2015PhyS...90e4005R,
   author = {{Ryabchikova}, T. and {Piskunov}, N. and {Kurucz}, R.~L. and 
	{Stempels}, H.~C. and {Heiter}, U. and {Pakhomov}, Y. and {Barklem}, P.~S.
	},
    title = "{A major upgrade of the VALD database}",
  journal = {\physscr},
     year = 2015,
    month = may,
   volume = 90,
   number = 5,
      eid = {054005},
    pages = {054005},
      doi = {10.1088/0031-8949/90/5/054005},
   adsurl = {http://adsabs.harvard.edu/abs/2015PhyS...90e4005R}
}

@ARTICLE{2019MNRAS.485.1508S,
       author = {{Shultz}, M.~E. and {Wade}, G.~A. and {Rivinius}, Th and {Alecian}, E. and
         {Neiner}, C. and {Petit}, V. and {Wisniewski}, J.~P. and
         {MiMeS Collaboration} and {BinaMIcS Collaboration}},
        title = "{The magnetic early B-type Stars II: stellar atmospheric parameters in the era of Gaia}",
      journal = {\mnras},
         year = "2019",
        month = "May",
       volume = {485},
       number = {2},
        pages = {1508-1527},
          doi = {10.1093/mnras/stz416},
archivePrefix = {arXiv},
       eprint = {1902.02713},
 primaryClass = {astro-ph.SR},
       adsurl = {https://ui.adsabs.harvard.edu/abs/2019MNRAS.485.1508S}
}

@ARTICLE{2005CoAst.146...53L,
   author = {{Lenz}, P. and {Breger}, M.},
    title = "{Period04 User Guide}",
  journal = {Communications in Asteroseismology},
     year = 2005,
    month = jun,
   volume = 146,
    pages = {53-136},
      doi = {10.1553/cia146s53},
   adsurl = {https://ui.adsabs.harvard.edu/abs/2005CoAst.146...53L}
}

@ARTICLE{2019MNRAS.490..274S,
   author = {{Shultz}, M.~E. and {Wade}, G.~A. and {Rivinius}, T. and {Alecian}, E. and 
	{Neiner}, C. and {Petit}, V. and {Owocki}, S. and {ud-Doula}, A. and 
	{Kochukhov}, O. and {Bohlender}, D. and {Keszthelyi}, Z. and 
	{MiMeS Collaboration} and {BinaMIcS Collaboration}},
    title = "{The magnetic early B-type stars - III. A main-sequence magnetic, rotational, and magnetospheric biography}",
  journal = {\mnras},
archivePrefix = "arXiv",
   eprint = {1909.02530},
 primaryClass = "astro-ph.SR",
     year = 2019,
    month = nov,
   volume = 490,
    pages = {274-295},
      doi = {10.1093/mnras/stz2551},
   adsurl = {https://ui.adsabs.harvard.edu/abs/2019MNRAS.490..274S}
}

@ARTICLE{2016MNRAS.456.1294R,
   author = {{Reeve}, D.~C. and {Howarth}, I.~D.},
    title = "{Limb-darkening coefficients from line-blanketed non-LTE hot-star model atmospheres}",
  journal = {\mnras},
archivePrefix = "arXiv",
   eprint = {1511.02029},
 primaryClass = "astro-ph.SR",
     year = 2016,
    month = feb,
   volume = 456,
    pages = {1294-1298},
      doi = {10.1093/mnras/stv2631},
   adsurl = {http://adsabs.harvard.edu/abs/2016MNRAS.456.1294R}
}

@ARTICLE{2018A&A...616A...1G,
       author = {{Gaia Collaboration} and {Brown}, A.~G.~A. and {Vallenari}, A. and
         {Prusti}, T. and {de Bruijne}, J.~H.~J. and {Babusiaux}, C. and
         {Bailer-Jones}, C.~A.~L. and {Biermann}, M. and {Evans}, D.~W. and
         {Eyer}, L. and {Jansen}, F. and {Jordi}, C. and {Klioner}, S.~A. and
         {Lammers}, U. and {Lindegren}, L. and {Luri}, X. and {Mignard}, F. and
         {Panem}, C. and {Pourbaix}, D. and {Randich}, S. and {Sartoretti}, P. and
         {Siddiqui}, H.~I. and {Soubiran}, C. and {van Leeuwen}, F. and
         {Walton}, N.~A. and {Arenou}, F. and {Bastian}, U. and {Cropper}, M. and
         {Drimmel}, R. and {Katz}, D. and {Lattanzi}, M.~G. and {Bakker}, J. and
         {Cacciari}, C. and {Casta{\~n}eda}, J. and {Chaoul}, L. and
         {Cheek}, N. and {De Angeli}, F. and {Fabricius}, C. and {Guerra}, R. and
         {Holl}, B. and {Masana}, E. and {Messineo}, R. and {Mowlavi}, N. and
         {Nienartowicz}, K. and {Panuzzo}, P. and {Portell}, J. and
         {Riello}, M. and {Seabroke}, G.~M. and {Tanga}, P. and
         {Th{\'e}venin}, F. and {Gracia-Abril}, G. and {Comoretto}, G. and
         {Garcia-Reinaldos}, M. and {Teyssier}, D. and {Altmann}, M. and
         {Andrae}, R. and {Audard}, M. and {Bellas-Velidis}, I. and
         {Benson}, K. and {Berthier}, J. and {Blomme}, R. and {Burgess}, P. and
         {Busso}, G. and {Carry}, B. and {Cellino}, A. and {Clementini}, G. and
         {Clotet}, M. and {Creevey}, O. and {Davidson}, M. and {De Ridder}, J. and
         {Delchambre}, L. and {Dell'Oro}, A. and {Ducourant}, C. and
         {Fern{\'a}ndez-Hern{\'a}ndez}, J. and {Fouesneau}, M. and
         {Fr{\'e}mat}, Y. and {Galluccio}, L. and {Garc{\'\i}a-Torres}, M. and
         {Gonz{\'a}lez-N{\'u}{\~n}ez}, J. and {Gonz{\'a}lez-Vidal}, J.~J. and
         {Gosset}, E. and {Guy}, L.~P. and {Halbwachs}, J. -L. and
         {Hambly}, N.~C. and {Harrison}, D.~L. and {Hern{\'a}ndez}, J. and
         {Hestroffer}, D. and {Hodgkin}, S.~T. and {Hutton}, A. and
         {Jasniewicz}, G. and {Jean-Antoine-Piccolo}, A. and {Jordan}, S. and
         {Korn}, A.~J. and {Krone-Martins}, A. and {Lanzafame}, A.~C. and
         {Lebzelter}, T. and {L{\"o}ffler}, W. and {Manteiga}, M. and
         {Marrese}, P.~M. and {Mart{\'\i}n-Fleitas}, J.~M. and {Moitinho}, A. and
         {Mora}, A. and {Muinonen}, K. and {Osinde}, J. and {Pancino}, E. and
         {Pauwels}, T. and {Petit}, J. -M. and {Recio-Blanco}, A. and
         {Richards}, P.~J. and {Rimoldini}, L. and {Robin}, A.~C. and
         {Sarro}, L.~M. and {Siopis}, C. and {Smith}, M. and {Sozzetti}, A. and
         {S{\"u}veges}, M. and {Torra}, J. and {van Reeven}, W. and {Abbas}, U. and
         {Abreu Aramburu}, A. and {Accart}, S. and {Aerts}, C. and
         {Altavilla}, G. and {{\'A}lvarez}, M.~A. and {Alvarez}, R. and
         {Alves}, J. and {Anderson}, R.~I. and {Andrei}, A.~H. and
         {Anglada Varela}, E. and {Antiche}, E. and {Antoja}, T. and
         {Arcay}, B. and {Astraatmadja}, T.~L. and {Bach}, N. and
         {Baker}, S.~G. and {Balaguer-N{\'u}{\~n}ez}, L. and {Balm}, P. and
         {Barache}, C. and {Barata}, C. and {Barbato}, D. and {Barblan}, F. and
         {Barklem}, P.~S. and {Barrado}, D. and {Barros}, M. and
         {Barstow}, M.~A. and {Bartholom{\'e} Mu{\~n}oz}, S. and
         {Bassilana}, J. -L. and {Becciani}, U. and {Bellazzini}, M. and
         {Berihuete}, A. and {Bertone}, S. and {Bianchi}, L. and
         {Bienaym{\'e}}, O. and {Blanco-Cuaresma}, S. and {Boch}, T. and
         {Boeche}, C. and {Bombrun}, A. and {Borrachero}, R. and {Bossini}, D. and
         {Bouquillon}, S. and {Bourda}, G. and {Bragaglia}, A. and
         {Bramante}, L. and {Breddels}, M.~A. and {Bressan}, A. and
         {Brouillet}, N. and {Br{\"u}semeister}, T. and {Brugaletta}, E. and
         {Bucciarelli}, B. and {Burlacu}, A. and {Busonero}, D. and
         {Butkevich}, A.~G. and {Buzzi}, R. and {Caffau}, E. and
         {Cancelliere}, R. and {Cannizzaro}, G. and {Cantat-Gaudin}, T. and
         {Carballo}, R. and {Carlucci}, T. and {Carrasco}, J.~M. and
         {Casamiquela}, L. and {Castellani}, M. and {Castro-Ginard}, A. and
         {Charlot}, P. and {Chemin}, L. and {Chiavassa}, A. and {Cocozza}, G. and
         {Costigan}, G. and {Cowell}, S. and {Crifo}, F. and {Crosta}, M. and
         {Crowley}, C. and {Cuypers}, J. and {Dafonte}, C. and {Damerdji}, Y. and
         {Dapergolas}, A. and {David}, P. and {David}, M. and {de Laverny}, P. and
         {De Luise}, F. and {De March}, R. and {de Martino}, D. and
         {de Souza}, R. and {de Torres}, A. and {Debosscher}, J. and
         {del Pozo}, E. and {Delbo}, M. and {Delgado}, A. and {Delgado}, H.~E. and
         {Di Matteo}, P. and {Diakite}, S. and {Diener}, C. and {Distefano}, E. and
         {Dolding}, C. and {Drazinos}, P. and {Dur{\'a}n}, J. and
         {Edvardsson}, B. and {Enke}, H. and {Eriksson}, K. and {Esquej}, P. and
         {Eynard Bontemps}, G. and {Fabre}, C. and {Fabrizio}, M. and
         {Faigler}, S. and {Falc{\~a}o}, A.~J. and {Farr{\`a}s Casas}, M. and
         {Federici}, L. and {Fedorets}, G. and {Fernique}, P. and
         {Figueras}, F. and {Filippi}, F. and {Findeisen}, K. and {Fonti}, A. and
         {Fraile}, E. and {Fraser}, M. and {Fr{\'e}zouls}, B. and {Gai}, M. and
         {Galleti}, S. and {Garabato}, D. and {Garc{\'\i}a-Sedano}, F. and
         {Garofalo}, A. and {Garralda}, N. and {Gavel}, A. and {Gavras}, P. and
         {Gerssen}, J. and {Geyer}, R. and {Giacobbe}, P. and {Gilmore}, G. and
         {Girona}, S. and {Giuffrida}, G. and {Glass}, F. and {Gomes}, M. and
         {Granvik}, M. and {Gueguen}, A. and {Guerrier}, A. and {Guiraud}, J. and
         {Guti{\'e}rrez-S{\'a}nchez}, R. and {Haigron}, R. and
         {Hatzidimitriou}, D. and {Hauser}, M. and {Haywood}, M. and
         {Heiter}, U. and {Helmi}, A. and {Heu}, J. and {Hilger}, T. and
         {Hobbs}, D. and {Hofmann}, W. and {Holland}, G. and {Huckle}, H.~E. and
         {Hypki}, A. and {Icardi}, V. and {Jan{\ss}en}, K. and
         {Jevardat de Fombelle}, G. and {Jonker}, P.~G. and
         {Juh{\'a}sz}, {\'A}. L. and {Julbe}, F. and {Karampelas}, A. and
         {Kewley}, A. and {Klar}, J. and {Kochoska}, A. and {Kohley}, R. and
         {Kolenberg}, K. and {Kontizas}, M. and {Kontizas}, E. and
         {Koposov}, S.~E. and {Kordopatis}, G. and {Kostrzewa-Rutkowska}, Z. and
         {Koubsky}, P. and {Lambert}, S. and {Lanza}, A.~F. and {Lasne}, Y. and
         {Lavigne}, J. -B. and {Le Fustec}, Y. and {Le Poncin-Lafitte}, C. and
         {Lebreton}, Y. and {Leccia}, S. and {Leclerc}, N. and
         {Lecoeur-Taibi}, I. and {Lenhardt}, H. and {Leroux}, F. and {Liao}, S. and
         {Licata}, E. and {Lindstr{\o}m}, H.~E.~P. and {Lister}, T.~A. and
         {Livanou}, E. and {Lobel}, A. and {L{\'o}pez}, M. and {Managau}, S. and
         {Mann}, R.~G. and {Mantelet}, G. and {Marchal}, O. and
         {Marchant}, J.~M. and {Marconi}, M. and {Marinoni}, S. and
         {Marschalk{\'o}}, G. and {Marshall}, D.~J. and {Martino}, M. and
         {Marton}, G. and {Mary}, N. and {Massari}, D. and
         {Matijevi{\v{c}}}, G. and {Mazeh}, T. and {McMillan}, P.~J. and
         {Messina}, S. and {Michalik}, D. and {Millar}, N.~R. and {Molina}, D. and
         {Molinaro}, R. and {Moln{\'a}r}, L. and {Montegriffo}, P. and
         {Mor}, R. and {Morbidelli}, R. and {Morel}, T. and {Morris}, D. and
         {Mulone}, A.~F. and {Muraveva}, T. and {Musella}, I. and
         {Nelemans}, G. and {Nicastro}, L. and {Noval}, L. and {O'Mullane}, W. and
         {Ord{\'e}novic}, C. and {Ord{\'o}{\~n}ez-Blanco}, D. and {Osborne}, P. and
         {Pagani}, C. and {Pagano}, I. and {Pailler}, F. and {Palacin}, H. and
         {Palaversa}, L. and {Panahi}, A. and {Pawlak}, M. and
         {Piersimoni}, A.~M. and {Pineau}, F. -X. and {Plachy}, E. and
         {Plum}, G. and {Poggio}, E. and {Poujoulet}, E. and {Pr{\v{s}}a}, A. and
         {Pulone}, L. and {Racero}, E. and {Ragaini}, S. and {Rambaux}, N. and
         {Ramos-Lerate}, M. and {Regibo}, S. and {Reyl{\'e}}, C. and
         {Riclet}, F. and {Ripepi}, V. and {Riva}, A. and {Rivard}, A. and
         {Rixon}, G. and {Roegiers}, T. and {Roelens}, M. and
         {Romero-G{\'o}mez}, M. and {Rowell}, N. and {Royer}, F. and
         {Ruiz-Dern}, L. and {Sadowski}, G. and {Sagrist{\`a} Sell{\'e}s}, T. and
         {Sahlmann}, J. and {Salgado}, J. and {Salguero}, E. and {Sanna}, N. and
         {Santana-Ros}, T. and {Sarasso}, M. and {Savietto}, H. and
         {Schultheis}, M. and {Sciacca}, E. and {Segol}, M. and
         {Segovia}, J.~C. and {S{\'e}gransan}, D. and {Shih}, I. -C. and
         {Siltala}, L. and {Silva}, A.~F. and {Smart}, R.~L. and {Smith}, K.~W. and
         {Solano}, E. and {Solitro}, F. and {Sordo}, R. and {Soria Nieto}, S. and
         {Souchay}, J. and {Spagna}, A. and {Spoto}, F. and {Stampa}, U. and
         {Steele}, I.~A. and {Steidelm{\"u}ller}, H. and {Stephenson}, C.~A. and
         {Stoev}, H. and {Suess}, F.~F. and {Surdej}, J. and {Szabados}, L. and
         {Szegedi-Elek}, E. and {Tapiador}, D. and {Taris}, F. and {Tauran}, G. and
         {Taylor}, M.~B. and {Teixeira}, R. and {Terrett}, D. and {Teyssand
        ier}, P. and {Thuillot}, W. and {Titarenko}, A. and {Torra Clotet}, F. and
         {Turon}, C. and {Ulla}, A. and {Utrilla}, E. and {Uzzi}, S. and
         {Vaillant}, M. and {Valentini}, G. and {Valette}, V. and
         {van Elteren}, A. and {Van Hemelryck}, E. and {van Leeuwen}, M. and
         {Vaschetto}, M. and {Vecchiato}, A. and {Veljanoski}, J. and
         {Viala}, Y. and {Vicente}, D. and {Vogt}, S. and {von Essen}, C. and
         {Voss}, H. and {Votruba}, V. and {Voutsinas}, S. and {Walmsley}, G. and
         {Weiler}, M. and {Wertz}, O. and {Wevers}, T. and {Wyrzykowski}, {\L}. and
         {Yoldas}, A. and {{\v{Z}}erjal}, M. and {Ziaeepour}, H. and
         {Zorec}, J. and {Zschocke}, S. and {Zucker}, S. and {Zurbach}, C. and
         {Zwitter}, T.},
        title = "{Gaia Data Release 2. Summary of the contents and survey properties}",
      journal = {\aap},
         year = 2018,
        month = aug,
       volume = {616},
          eid = {A1},
        pages = {A1},
          doi = {10.1051/0004-6361/201833051},
archivePrefix = {arXiv},
       eprint = {1804.09365},
 primaryClass = {astro-ph.GA},
       adsurl = {https://ui.adsabs.harvard.edu/abs/2018A&A...616A...1G}
}

@ARTICLE{2020MNRAS.499.5366O,
       author = {{Owocki}, Stanley P. and {Shultz}, Matt E. and {ud-Doula}, Asif and {Sundqvist}, Jon O. and {Townsend}, Richard H.~D. and {Cranmer}, Steven R.},
        title = "{How the breakout-limited mass in B-star centrifugal magnetospheres controls their circumstellar H {\ensuremath{\alpha}} emission}",
      journal = {\mnras},
         year = 2020,
        month = oct,
       volume = {499},
       number = {4},
        pages = {5366-5378},
          doi = {10.1093/mnras/staa2325},
archivePrefix = {arXiv},
       eprint = {2009.12359},
 primaryClass = {astro-ph.SR},
       adsurl = {https://ui.adsabs.harvard.edu/abs/2020MNRAS.499.5366O}
}

@ARTICLE{2020MNRAS.499.5379S,
       author = {{Shultz}, M.~E. and {Owocki}, S. and {Rivinius}, Th and {Wade}, G.~A. and {Neiner}, C. and {Alecian}, E. and {Kochukhov}, O. and {Bohlender}, D. and {ud-Doula}, A. and {Landstreet}, J.~D. and {Sikora}, J. and {David-Uraz}, A. and {Petit}, V. and {Cerraho{\u{g}}lu}, P. and {Fine}, R. and {Henson}, G. and {MiMeS Collaboration} and {BinaMIcS Collaboration}},
        title = "{The magnetic early B-type stars - IV. Breakout or leakage? H {\ensuremath{\alpha}} emission as a diagnostic of plasma transport in centrifugal magnetospheres}",
      journal = {\mnras},
         year = 2020,
        month = oct,
       volume = {499},
       number = {4},
        pages = {5379-5395},
          doi = {10.1093/mnras/staa3102},
archivePrefix = {arXiv},
       eprint = {2009.12336},
 primaryClass = {astro-ph.SR},
       adsurl = {https://ui.adsabs.harvard.edu/abs/2020MNRAS.499.5379S}
}

@INPROCEEDINGS{2020pase.conf...46M,
       author = {{Mikul{\'a}{\v{s}}ek}, Z. and {Krti{\v{c}}ka}, J. and {Shultz}, M.~E. and {Henry}, G.~W. and {Prv{\'a}k}, M. and {David-Uraz}, A. and {Jan{\'\i}k}, J. and {Zejda}, M. and {Romanyuk}, I.~I.},
        title = "{What's New with Landstreet's Star HD 37776 (V901 Ori)?}",
    booktitle = {Stellar Magnetism: A Workshop in Honour of the Career and Contributions of John D. Landstreet},
         year = 2020,
       editor = {{Wade}, Gregg and {Alecian}, Evelyne and {Bohlender}, David and {Sigut}, Aaron},
       volume = {11},
        month = nov,
        pages = {46-53},
archivePrefix = {arXiv},
       eprint = {1912.04121},
 primaryClass = {astro-ph.SR},
       adsurl = {https://ui.adsabs.harvard.edu/abs/2020pase.conf...46M},
	series = {}
}

@ARTICLE{2020MNRAS.493.4657L,
       author = {{Leto}, P. and {Trigilio}, C. and {Leone}, F. and {Pillitteri}, I. and {Buemi}, C.~S. and {Fossati}, L. and {Cavallaro}, F. and {Oskinova}, L.~M. and {Ignace}, R. and {Krti{\v{c}}ka}, J. and {Umana}, G. and {Catanzaro}, G. and {Ingallinera}, A. and {Bufano}, F. and {Agliozzo}, C. and {Phillips}, N.~M. and {Cerrigone}, L. and {Riggi}, S. and {Loru}, S. and {Munari}, M. and {Gangi}, M. and {Giarrusso}, M. and {Robrade}, J.},
        title = "{Evidence for radio and X-ray auroral emissions from the magnetic B-type star {\ensuremath{\rho}} Oph A}",
      journal = {\mnras},
         year = 2020,
        month = apr,
       volume = {493},
       number = {4},
        pages = {4657-4676},
          doi = {10.1093/mnras/staa587},
archivePrefix = {arXiv},
       eprint = {2002.09251},
 primaryClass = {astro-ph.SR},
       adsurl = {https://ui.adsabs.harvard.edu/abs/2020MNRAS.493.4657L}
}

@ARTICLE{2020MNRAS.499L..72L,
       author = {{Leto}, P. and {Trigilio}, C. and {Buemi}, C.~S. and {Leone}, F. and {Pillitteri}, I. and {Fossati}, L. and {Cavallaro}, F. and {Oskinova}, L.~M. and {Ignace}, R. and {Krti{\v{c}}ka}, J. and {Umana}, G. and {Catanzaro}, G. and {Ingallinera}, A. and {Bufano}, F. and {Riggi}, S. and {Cerrigone}, L. and {Loru}, S. and {Schillir{\'o}}, F. and {Agliozzo}, C. and {Phillips}, N.~M. and {Giarrusso}, M. and {Robrade}, J.},
        title = "{The auroral radio emission of the magnetic B-type star {\ensuremath{\rho}}-=OphC}",
      journal = {\mnras},
         year = 2020,
        month = dec,
       volume = {499},
       number = {1},
        pages = {L72-L76},
          doi = {10.1093/mnrasl/slaa157},
archivePrefix = {arXiv},
       eprint = {2009.02363},
 primaryClass = {astro-ph.SR},
       adsurl = {https://ui.adsabs.harvard.edu/abs/2020MNRAS.499L..72L}
}

@ARTICLE{2016A&A...592A..88P,
       author = {{Pillitteri}, I. and {Wolk}, S.~J. and {Chen}, H.~H. and {Goodman}, A.},
        title = "{First stars of the {\ensuremath{\rho}} Ophiuchi dark cloud. XMM-Newton view of {\ensuremath{\rho}} Oph and its neighbors}",
      journal = {\aap},
         year = 2016,
        month = aug,
       volume = {592},
          eid = {A88},
        pages = {A88},
          doi = {10.1051/0004-6361/201628284},
archivePrefix = {arXiv},
       eprint = {1605.07365},
 primaryClass = {astro-ph.SR},
       adsurl = {https://ui.adsabs.harvard.edu/abs/2016A&A...592A..88P}
}

@ARTICLE{2018AJ....155..196R,
       author = {{Rebull}, L.~M. and {Stauffer}, J.~R. and {Cody}, A.~M. and {Hillenbrand}, L.~A. and {David}, T.~J. and {Pinsonneault}, M.},
        title = "{Rotation of Low-mass Stars in Upper Scorpius and {\ensuremath{\rho}} Ophiuchus with K2}",
      journal = {\aj},
         year = 2018,
        month = may,
       volume = {155},
       number = {5},
          eid = {196},
        pages = {196},
          doi = {10.3847/1538-3881/aab605},
archivePrefix = {arXiv},
       eprint = {1803.04440},
 primaryClass = {astro-ph.SR},
       adsurl = {https://ui.adsabs.harvard.edu/abs/2018AJ....155..196R}
}

@ARTICLE{1994ApJ...424..772F,
       author = {{Federman}, S.~R. and {Strom}, C.~J. and {Lambert}, D.~L. and {Cardelli}, Jason A. and {Smith}, V.~V. and {Joseph}, C.~L.},
        title = "{Chemical Transitions for Interstellar C 2 and CN in Cloud Envelopes}",
      journal = {\apj},
         year = 1994,
        month = apr,
       volume = {424},
        pages = {772},
          doi = {10.1086/173930},
       adsurl = {https://ui.adsabs.harvard.edu/abs/1994ApJ...424..772F}
}

@ARTICLE{2021MNRAS.504.3203S,
       author = {{Shultz}, M.~E. and {Alecian}, E. and {Petit}, V. and {Bagnulo}, S. and {B{\"o}hm}, T. and {Folsom}, C.~P. and {Wade}, G.~A. and {Wade}, G.~A. and {MiMeS Collaboration}},
        title = "{NGC 6611 601: a hot pre-main-sequence spectroscopic binary containing a centrifugal magnetosphere host star}",
      journal = {\mnras},
         year = 2021,
        month = jul,
       volume = {504},
       number = {3},
        pages = {3203-3220},
          doi = {10.1093/mnras/stab822},
archivePrefix = {arXiv},
       eprint = {2103.09670},
 primaryClass = {astro-ph.SR},
       adsurl = {https://ui.adsabs.harvard.edu/abs/2021MNRAS.504.3203S}
}

@ARTICLE{2021MNRAS.504.4850S,
       author = {{Shultz}, M.~E. and {Rivinius}, Th and {Wade}, G.~A. and {Kochukhov}, O. and {Alecian}, E. and {David-Uraz}, A. and {Sikora}, J.},
        title = "{MOBSTER - V. Discovery of a magnetic companion star to the magnetic {\ensuremath{\beta}} Cep pulsator HD 156424}",
      journal = {\mnras},
         year = 2021,
        month = jul,
       volume = {504},
       number = {4},
        pages = {4850-4864},
          doi = {10.1093/mnras/staa3158},
archivePrefix = {arXiv},
       eprint = {2010.04221},
 primaryClass = {astro-ph.SR},
       adsurl = {https://ui.adsabs.harvard.edu/abs/2021MNRAS.504.4850S}
}

@ARTICLE{2021A&A...649A...6G,
       author = {{Gaia Collaboration} and {Smart}, R.~L. and {Sarro}, L.~M. and {Rybizki}, J. and {Reyl{\'e}}, C. and {Robin}, A.~C. and {Hambly}, N.~C. and {Abbas}, U. and {Barstow}, M.~A. and {de Bruijne}, J.~H.~J. and {Bucciarelli}, B. and {Carrasco}, J.~M. and {Cooper}, W.~J. and {Hodgkin}, S.~T. and {Masana}, E. and {Michalik}, D. and {Sahlmann}, J. and {Sozzetti}, A. and {Brown}, A.~G.~A. and {Vallenari}, A. and {Prusti}, T. and {Babusiaux}, C. and {Biermann}, M. and {Creevey}, O.~L. and {Evans}, D.~W. and {Eyer}, L. and {Hutton}, A. and {Jansen}, F. and {Jordi}, C. and {Klioner}, S.~A. and {Lammers}, U. and {Lindegren}, L. and {Luri}, X. and {Mignard}, F. and {Panem}, C. and {Pourbaix}, D. and {Randich}, S. and {Sartoretti}, P. and {Soubiran}, C. and {Walton}, N.~A. and {Arenou}, F. and {Bailer-Jones}, C.~A.~L. and {Bastian}, U. and {Cropper}, M. and {Drimmel}, R. and {Katz}, D. and {Lattanzi}, M.~G. and {van Leeuwen}, F. and {Bakker}, J. and {Casta{\~n}eda}, J. and {De Angeli}, F. and {Ducourant}, C. and {Fabricius}, C. and {Fouesneau}, M. and {Fr{\'e}mat}, Y. and {Guerra}, R. and {Guerrier}, A. and {Guiraud}, J. and {Jean-Antoine Piccolo}, A. and {Messineo}, R. and {Mowlavi}, N. and {Nicolas}, C. and {Nienartowicz}, K. and {Pailler}, F. and {Panuzzo}, P. and {Riclet}, F. and {Roux}, W. and {Seabroke}, G.~M. and {Sordo}, R. and {Tanga}, P. and {Th{\'e}venin}, F. and {Gracia-Abril}, G. and {Portell}, J. and {Teyssier}, D. and {Altmann}, M. and {Andrae}, R. and {Bellas-Velidis}, I. and {Benson}, K. and {Berthier}, J. and {Blomme}, R. and {Brugaletta}, E. and {Burgess}, P.~W. and {Busso}, G. and {Carry}, B. and {Cellino}, A. and {Cheek}, N. and {Clementini}, G. and {Damerdji}, Y. and {Davidson}, M. and {Delchambre}, L. and {Dell'Oro}, A. and {Fern{\'a}ndez-Hern{\'a}ndez}, J. and {Galluccio}, L. and {Garc{\'\i}a-Lario}, P. and {Garcia-Reinaldos}, M. and {Gonz{\'a}lez-N{\'u}{\~n}ez}, J. and {Gosset}, E. and {Haigron}, R. and {Halbwachs}, J. -L. and {Harrison}, D.~L. and {Hatzidimitriou}, D. and {Heiter}, U. and {Hern{\'a}ndez}, J. and {Hestroffer}, D. and {Holl}, B. and {Jan{\ss}en}, K. and {Jevardat de Fombelle}, G. and {Jordan}, S. and {Krone-Martins}, A. and {Lanzafame}, A.~C. and {L{\"o}ffler}, W. and {Lorca}, A. and {Manteiga}, M. and {Marchal}, O. and {Marrese}, P.~M. and {Moitinho}, A. and {Mora}, A. and {Muinonen}, K. and {Osborne}, P. and {Pancino}, E. and {Pauwels}, T. and {Recio-Blanco}, A. and {Richards}, P.~J. and {Riello}, M. and {Rimoldini}, L. and {Roegiers}, T. and {Siopis}, C. and {Smith}, M. and {Ulla}, A. and {Utrilla}, E. and {van Leeuwen}, M. and {van Reeven}, W. and {Abreu Aramburu}, A. and {Accart}, S. and {Aerts}, C. and {Aguado}, J.~J. and {Ajaj}, M. and {Altavilla}, G. and {{\'A}lvarez}, M.~A. and {{\'A}lvarez Cid-Fuentes}, J. and {Alves}, J. and {Anderson}, R.~I. and {Anglada Varela}, E. and {Antoja}, T. and {Audard}, M. and {Baines}, D. and {Baker}, S.~G. and {Balaguer-N{\'u}{\~n}ez}, L. and {Balbinot}, E. and {Balog}, Z. and {Barache}, C. and {Barbato}, D. and {Barros}, M. and {Bartolom{\'e}}, S. and {Bassilana}, J. -L. and {Bauchet}, N. and {Baudesson-Stella}, A. and {Becciani}, U. and {Bellazzini}, M. and {Bernet}, M. and {Bertone}, S. and {Bianchi}, L. and {Blanco-Cuaresma}, S. and {Boch}, T. and {Bombrun}, A. and {Bossini}, D. and {Bouquillon}, S. and {Bragaglia}, A. and {Bramante}, L. and {Breedt}, E. and {Bressan}, A. and {Brouillet}, N. and {Burlacu}, A. and {Busonero}, D. and {Butkevich}, A.~G. and {Buzzi}, R. and {Caffau}, E. and {Cancelliere}, R. and {C{\'a}novas}, H. and {Cantat-Gaudin}, T. and {Carballo}, R. and {Carlucci}, T. and {Carnerero}, M.~I. and {Casamiquela}, L. and {Castellani}, M. and {Castro-Ginard}, A. and {Castro Sampol}, P. and {Chaoul}, L. and {Charlot}, P. and {Chemin}, L. and {Chiavassa}, A. and {Cioni}, M. -R.~L. and {Comoretto}, G. and {Cornez}, T. and {Cowell}, S. and {Crifo}, F. and {Crosta}, M. and {Crowley}, C. and {Dafonte}, C. and {Dapergolas}, A. and {David}, M. and {David}, P. and {de Laverny}, P. and {De Luise}, F. and {De March}, R. and {De Ridder}, J. and {de Souza}, R. and {de Teodoro}, P. and {de Torres}, A. and {del Peloso}, E.~F. and {del Pozo}, E. and {Delgado}, A. and {Delgado}, H.~E. and {Delisle}, J. -B. and {Di Matteo}, P. and {Diakite}, S. and {Diener}, C. and {Distefano}, E. and {Dolding}, C. and {Eappachen}, D. and {Edvardsson}, B. and {Enke}, H. and {Esquej}, P. and {Fabre}, C. and {Fabrizio}, M. and {Faigler}, S. and {Fedorets}, G. and {Fernique}, P. and {Fienga}, A. and {Figueras}, F. and {Fouron}, C. and {Fragkoudi}, F. and {Fraile}, E. and {Franke}, F. and {Gai}, M. and {Garabato}, D. and {Garcia-Gutierrez}, A. and {Garc{\'\i}a-Torres}, M. and {Garofalo}, A. and {Gavras}, P. and {Gerlach}, E. and {Geyer}, R. and {Giacobbe}, P. and {Gilmore}, G. and {Girona}, S. and {Giuffrida}, G. and {Gomel}, R. and {Gomez}, A. and {Gonzalez-Santamaria}, I. and {Gonz{\'a}lez-Vidal}, J.~J. and {Granvik}, M. and {Guti{\'e}rrez-S{\'a}nchez}, R. and {Guy}, L.~P. and {Hauser}, M. and {Haywood}, M. and {Helmi}, A. and {Hidalgo}, S.~L. and {Hilger}, T. and {H{\l}adczuk}, N. and {Hobbs}, D. and {Holland}, G. and {Huckle}, H.~E. and {Jasniewicz}, G. and {Jonker}, P.~G. and {Juaristi Campillo}, J. and {Julbe}, F. and {Karbevska}, L. and {Kervella}, P. and {Khanna}, S. and {Kochoska}, A. and {Kontizas}, M. and {Kordopatis}, G. and {Korn}, A.~J. and {Kostrzewa-Rutkowska}, Z. and {Kruszy{\'n}ska}, K. and {Lambert}, S. and {Lanza}, A.~F. and {Lasne}, Y. and {Le Campion}, J. -F. and {Le Fustec}, Y. and {Lebreton}, Y. and {Lebzelter}, T. and {Leccia}, S. and {Leclerc}, N. and {Lecoeur-Taibi}, I. and {Liao}, S. and {Licata}, E. and {Lindstr{\o}m}, H.~E.~P. and {Lister}, T.~A. and {Livanou}, E. and {Lobel}, A. and {Madrero Pardo}, P. and {Managau}, S. and {Mann}, R.~G. and {Marchant}, J.~M. and {Marconi}, M. and {Marcos Santos}, M.~M.~S. and {Marinoni}, S. and {Marocco}, F. and {Marshall}, D.~J. and {Martin Polo}, L. and {Mart{\'\i}n-Fleitas}, J.~M. and {Masip}, A. and {Massari}, D. and {Mastrobuono-Battisti}, A. and {Mazeh}, T. and {McMillan}, P.~J. and {Messina}, S. and {Millar}, N.~R. and {Mints}, A. and {Molina}, D. and {Molinaro}, R. and {Moln{\'a}r}, L. and {Montegriffo}, P. and {Mor}, R. and {Morbidelli}, R. and {Morel}, T. and {Morris}, D. and {Mulone}, A.~F. and {Munoz}, D. and {Muraveva}, T. and {Murphy}, C.~P. and {Musella}, I. and {Noval}, L. and {Ord{\'e}novic}, C. and {Orr{\`u}}, G. and {Osinde}, J. and {Pagani}, C. and {Pagano}, I. and {Palaversa}, L. and {Palicio}, P.~A. and {Panahi}, A. and {Pawlak}, M. and {Pe{\~n}alosa Esteller}, X. and {Penttil{\"a}}, A. and {Piersimoni}, A.~M. and {Pineau}, F. -X. and {Plachy}, E. and {Plum}, G. and {Poggio}, E. and {Poretti}, E. and {Poujoulet}, E. and {Pr{\v{s}}a}, A. and {Pulone}, L. and {Racero}, E. and {Ragaini}, S. and {Rainer}, M. and {Raiteri}, C.~M. and {Rambaux}, N. and {Ramos}, P. and {Ramos-Lerate}, M. and {Re Fiorentin}, P. and {Regibo}, S. and {Ripepi}, V. and {Riva}, A. and {Rixon}, G. and {Robichon}, N. and {Robin}, C. and {Roelens}, M. and {Rohrbasser}, L. and {Romero-G{\'o}mez}, M. and {Rowell}, N. and {Royer}, F. and {Rybicki}, K.~A. and {Sadowski}, G. and {Sagrist{\`a} Sell{\'e}s}, A. and {Salgado}, J. and {Salguero}, E. and {Samaras}, N. and {Sanchez Gimenez}, V. and {Sanna}, N. and {Santove{\~n}a}, R. and {Sarasso}, M. and {Schultheis}, M. and {Sciacca}, E. and {Segol}, M. and {Segovia}, J.~C. and {S{\'e}gransan}, D. and {Semeux}, D. and {Shahaf}, S. and {Siddiqui}, H.~I. and {Siebert}, A. and {Siltala}, L. and {Slezak}, E. and {Solano}, E. and {Solitro}, F. and {Souami}, D. and {Souchay}, J. and {Spagna}, A. and {Spoto}, F. and {Steele}, I.~A. and {Steidelm{\"u}ller}, H. and {Stephenson}, C.~A. and {S{\"u}veges}, M. and {Szabados}, L. and {Szegedi-Elek}, E. and {Taris}, F. and {Tauran}, G. and {Taylor}, M.~B. and {Teixeira}, R. and {Thuillot}, W. and {Tonello}, N. and {Torra}, F. and {Torra}, J. and {Turon}, C. and {Unger}, N. and {Vaillant}, M. and {van Dillen}, E. and {Vanel}, O. and {Vecchiato}, A. and {Viala}, Y. and {Vicente}, D. and {Voutsinas}, S. and {Weiler}, M. and {Wevers}, T. and {Wyrzykowski}, {\L}. and {Yoldas}, A. and {Yvard}, P. and {Zhao}, H. and {Zorec}, J. and {Zucker}, S. and {Zurbach}, C. and {Zwitter}, T.},
        title = "{Gaia Early Data Release 3. The Gaia Catalogue of Nearby Stars}",
      journal = {\aap},
         year = 2021,
        month = may,
       volume = {649},
          eid = {A6},
        pages = {A6},
          doi = {10.1051/0004-6361/202039498},
archivePrefix = {arXiv},
       eprint = {2012.02061},
 primaryClass = {astro-ph.SR},
       adsurl = {https://ui.adsabs.harvard.edu/abs/2021A&A...649A...6G}
}

@ARTICLE{2011ApJ...737..103S,
       author = {{Schlafly}, Edward F. and {Finkbeiner}, Douglas P.},
        title = "{Measuring Reddening with Sloan Digital Sky Survey Stellar Spectra and Recalibrating SFD}",
      journal = {\apj},
         year = 2011,
        month = aug,
       volume = {737},
       number = {2},
          eid = {103},
        pages = {103},
          doi = {10.1088/0004-637X/737/2/103},
archivePrefix = {arXiv},
       eprint = {1012.4804},
 primaryClass = {astro-ph.GA},
       adsurl = {https://ui.adsabs.harvard.edu/abs/2011ApJ...737..103S}
}

@ARTICLE{2014A&A...561A..91L,
       author = {{Lallement}, R. and {Vergely}, J. -L. and {Valette}, B. and {Puspitarini}, L. and {Eyer}, L. and {Casagrande}, L.},
        title = "{3D maps of the local ISM from inversion of individual color excess measurements}",
      journal = {\aap},
         year = 2014,
        month = jan,
       volume = {561},
          eid = {A91},
        pages = {A91},
          doi = {10.1051/0004-6361/201322032},
archivePrefix = {arXiv},
       eprint = {1309.6100},
 primaryClass = {astro-ph.GA},
       adsurl = {https://ui.adsabs.harvard.edu/abs/2014A&A...561A..91L}
}

@ARTICLE{2017A&A...606A..65C,
       author = {{Capitanio}, L. and {Lallement}, R. and {Vergely}, J.~L. and {Elyajouri}, M. and {Monreal-Ibero}, A.},
        title = "{Three-dimensional mapping of the local interstellar medium with composite data}",
      journal = {\aap},
         year = 2017,
        month = oct,
       volume = {606},
          eid = {A65},
        pages = {A65},
          doi = {10.1051/0004-6361/201730831},
archivePrefix = {arXiv},
       eprint = {1706.07711},
 primaryClass = {astro-ph.GA},
       adsurl = {https://ui.adsabs.harvard.edu/abs/2017A&A...606A..65C}
}

@ARTICLE{2007A&A...470.1089K,
       author = {{Krti{\v{c}}ka}, J. and {Mikul{\'a}{\v{s}}ek}, Z. and {Zverko}, J. and {{\v{Z}}i{\v{z}}{\'n}ovsk{\'y}}, J.},
        title = "{The light variability of the helium strong star HD 37776 as a result of its inhomogeneous elemental surface distribution}",
      journal = {\aap},
         year = 2007,
        month = aug,
       volume = {470},
       number = {3},
        pages = {1089-1098},
          doi = {10.1051/0004-6361:20066627},
archivePrefix = {arXiv},
       eprint = {0706.2338},
 primaryClass = {astro-ph},
       adsurl = {https://ui.adsabs.harvard.edu/abs/2007A&A...470.1089K}
}

@ARTICLE{2009A&A...499..567K,
       author = {{Krti{\v{c}}ka}, J. and {Mikul{\'a}{\v{s}}ek}, Z. and {Henry}, G.~W. and {Zverko}, J. and {{\v{Z}}i{\v{z}}ovsk{\'y}}, J. and {Skalick{\'y}}, J. and {Zv{\v{e}}{\v{r}}ina}, P.},
        title = "{The nature of the light variability of the silicon star HR 7224}",
      journal = {\aap},
         year = 2009,
        month = may,
       volume = {499},
       number = {2},
        pages = {567-577},
          doi = {10.1051/0004-6361/200811123},
archivePrefix = {arXiv},
       eprint = {0904.0231},
 primaryClass = {astro-ph.SR},
       adsurl = {https://ui.adsabs.harvard.edu/abs/2009A&A...499..567K}
}

@ARTICLE{2012A&A...537A..14K,
       author = {{Krti{\v{c}}ka}, J. and {Mikul{\'a}{\v{s}}ek}, Z. and {L{\"u}ftinger}, T. and {Shulyak}, D. and {Zverko}, J. and {{\v{Z}}i{\v{z}}{\v{n}}ovsk{\'y}}, J. and {Sokolov}, N.~A.},
        title = "{Modelling of the ultraviolet and visual SED variability in the hot magnetic Ap star CU Virginis}",
      journal = {\aap},
         year = 2012,
        month = jan,
       volume = {537},
          eid = {A14},
        pages = {A14},
          doi = {10.1051/0004-6361/201117490},
archivePrefix = {arXiv},
       eprint = {1111.2746},
 primaryClass = {astro-ph.SR},
       adsurl = {https://ui.adsabs.harvard.edu/abs/2012A&A...537A..14K}
}

@ARTICLE{2013A&A...556A..18K,
       author = {{Krti{\v{c}}ka}, J. and {Jan{\'\i}k}, J. and {Markov{\'a}}, H. and {Mikul{\'a}{\v{s}}ek}, Z. and {Zverko}, J. and {Prv{\'a}k}, M. and {Skarka}, M.},
        title = "{Ultraviolet and visual flux and line variations of one of the least variable Bp stars HD 64740}",
      journal = {\aap},
         year = 2013,
        month = aug,
       volume = {556},
          eid = {A18},
        pages = {A18},
          doi = {10.1051/0004-6361/201221018},
archivePrefix = {arXiv},
       eprint = {1306.2458},
 primaryClass = {astro-ph.SR},
       adsurl = {https://ui.adsabs.harvard.edu/abs/2013A&A...556A..18K}
}

@ARTICLE{2015A&A...576A..82K,
       author = {{Krti{\v{c}}ka}, J. and {Mikul{\'a}{\v{s}}ek}, Z. and {L{\"u}ftinger}, T. and {Jagelka}, M.},
        title = "{Visual and ultraviolet flux variability of the bright CP star {\ensuremath{\theta}} Aurigae}",
      journal = {\aap},
         year = 2015,
        month = apr,
       volume = {576},
          eid = {A82},
        pages = {A82},
          doi = {10.1051/0004-6361/201425097},
archivePrefix = {arXiv},
       eprint = {1502.01985},
 primaryClass = {astro-ph.SR},
       adsurl = {https://ui.adsabs.harvard.edu/abs/2015A&A...576A..82K}
}

@ARTICLE{2010A&A...524A..66S,
       author = {{Shulyak}, D. and {Krti{\v{c}}ka}, J. and {Mikul{\'a}{\v{s}}ek}, Z. and {Kochukhov}, O. and {L{\"u}ftinger}, T.},
        title = "{Modelling the light variability of the Ap star {\ensuremath{\in}} Ursae Majoris}",
      journal = {\aap},
         year = 2010,
        month = dec,
       volume = {524},
          eid = {A66},
        pages = {A66},
          doi = {10.1051/0004-6361/201015094},
archivePrefix = {arXiv},
       eprint = {1009.1297},
 primaryClass = {astro-ph.SR},
       adsurl = {https://ui.adsabs.harvard.edu/abs/2010A&A...524A..66S}
}

@ARTICLE{2019A&A...626A..80C,
       author = {{C{\'a}novas}, H. and {Cantero}, C. and {Cieza}, L. and {Bombrun}, A. and {Lammers}, U. and {Mer{\'\i}n}, B. and {Mora}, A. and {Ribas}, {\'A}. and {Ru{\'\i}z-Rodr{\'\i}guez}, D.},
        title = "{Census of {\ensuremath{\rho}} Ophiuchi candidate members from Gaia Data Release 2}",
      journal = {\aap},
         year = 2019,
        month = jun,
       volume = {626},
          eid = {A80},
        pages = {A80},
          doi = {10.1051/0004-6361/201935321},
archivePrefix = {arXiv},
       eprint = {1902.07600},
 primaryClass = {astro-ph.EP},
       adsurl = {https://ui.adsabs.harvard.edu/abs/2019A&A...626A..80C}
}

@ARTICLE{2014PASP..126..398H,
       author = {{Howell}, Steve B. and {Sobeck}, Charlie and {Haas}, Michael and {Still}, Martin and {Barclay}, Thomas and {Mullally}, Fergal and {Troeltzsch}, John and {Aigrain}, Suzanne and {Bryson}, Stephen T. and {Caldwell}, Doug and {Chaplin}, William J. and {Cochran}, William D. and {Huber}, Daniel and {Marcy}, Geoffrey W. and {Miglio}, Andrea and {Najita}, Joan R. and {Smith}, Marcie and {Twicken}, J.~D. and {Fortney}, Jonathan J.},
        title = "{The K2 Mission: Characterization and Early Results}",
      journal = {\pasp},
         year = 2014,
        month = apr,
       volume = {126},
       number = {938},
        pages = {398},
          doi = {10.1086/676406},
archivePrefix = {arXiv},
       eprint = {1402.5163},
 primaryClass = {astro-ph.IM},
       adsurl = {https://ui.adsabs.harvard.edu/abs/2014PASP..126..398H}
}

@ARTICLE{2021MNRAS.507.1979L,
       author = {{Leto}, P. and {Trigilio}, C. and {Krti{\v{c}}ka}, J. and {Fossati}, L. and {Ignace}, R. and {Shultz}, M.~E. and {Buemi}, C.~S. and {Cerrigone}, L. and {Umana}, G. and {Ingallinera}, A. and {Bordiu}, C. and {Pillitteri}, I. and {Bufano}, F. and {Oskinova}, L.~M. and {Agliozzo}, C. and {Cavallaro}, F. and {Riggi}, S. and {Loru}, S. and {Todt}, H. and {Giarrusso}, M. and {Phillips}, N.~M. and {Robrade}, J. and {Leone}, F.},
        title = "{A scaling relationship for non-thermal radio emission from ordered magnetospheres: from the top of the main sequence to planets}",
      journal = {\mnras},
         year = 2021,
        month = oct,
       volume = {507},
       number = {2},
        pages = {1979-1998},
          doi = {10.1093/mnras/stab2168},
archivePrefix = {arXiv},
       eprint = {2107.11995},
 primaryClass = {astro-ph.SR},
       adsurl = {https://ui.adsabs.harvard.edu/abs/2021MNRAS.507.1979L}
}

@ARTICLE{2022MNRAS.513.1429S,
       author = {{Shultz}, M.~E. and {Owocki}, S.~P. and {ud-Doula}, A. and {Biswas}, A. and {Bohlender}, D. and {Chandra}, P. and {Das}, B. and {David-Uraz}, A. and {Khalack}, V. and {Kochukhov}, O. and {Landstreet}, J.~D. and {Leto}, P. and {Monin}, D. and {Neiner}, C. and {Rivinius}, Th and {Wade}, G.~A.},
        title = "{MOBSTER - VI. The crucial influence of rotation on the radio magnetospheres of hot stars}",
      journal = {\mnras},
         year = 2022,
        month = jun,
       volume = {513},
       number = {1},
        pages = {1429-1448},
          doi = {10.1093/mnras/stac136},
archivePrefix = {arXiv},
       eprint = {2201.05512},
 primaryClass = {astro-ph.SR},
       adsurl = {https://ui.adsabs.harvard.edu/abs/2022MNRAS.513.1429S}
}

@ARTICLE{2022MNRAS.513.1449O,
       author = {{Owocki}, S.~P. and {Shultz}, M.~E. and {ud-Doula}, A. and {Chandra}, P. and {Das}, B. and {Leto}, P.},
        title = "{Centrifugal breakout reconnection as the electron acceleration mechanism powering the radio magnetospheres of early-type stars}",
      journal = {\mnras},
         year = 2022,
        month = jun,
       volume = {513},
       number = {1},
        pages = {1449-1458},
          doi = {10.1093/mnras/stac341},
archivePrefix = {arXiv},
       eprint = {2202.05449},
 primaryClass = {astro-ph.SR},
       adsurl = {https://ui.adsabs.harvard.edu/abs/2022MNRAS.513.1449O}
}

@ARTICLE{2022MNRAS.511.4815B,
       author = {{Berry}, I.~D. and {Owocki}, S.~P. and {Shultz}, M.~E. and {ud-Doula}, A.},
        title = "{Electron scattering emission in the light curves of stars with centrifugal magnetospheres}",
      journal = {\mnras},
         year = 2022,
        month = apr,
       volume = {511},
       number = {4},
        pages = {4815-4825},
          doi = {10.1093/mnras/stac322},
archivePrefix = {arXiv},
       eprint = {2202.00615},
 primaryClass = {astro-ph.SR},
       adsurl = {https://ui.adsabs.harvard.edu/abs/2022MNRAS.511.4815B}
}

@ARTICLE{2023MNRAS.523.6371B,
       author = {{Berry}, I.~D. and {Shultz}, M.~E. and {Owocki}, S.~P. and {ud-Doula}, A.},
        title = "{MOBSTER - VII. Using light curves to infer magnetic and rotational properties of stars with centrifugal magnetospheres}",
      journal = {\mnras},
         year = 2023,
        month = aug,
       volume = {523},
       number = {4},
        pages = {6371-6385},
          doi = {10.1093/mnras/stad1726},
archivePrefix = {arXiv},
       eprint = {2306.03990},
 primaryClass = {astro-ph.SR},
       adsurl = {https://ui.adsabs.harvard.edu/abs/2023MNRAS.523.6371B}
}

@ARTICLE{2022A&A...659A..37K,
       author = {{Krti{\v{c}}ka}, J. and {Mikul{\'a}{\v{s}}ek}, Z. and {Kurf{\"u}rst}, P. and {Oksala}, M.~E.},
        title = "{Photometric signatures of corotating magnetospheres of hot stars governed by higher-order magnetic multipoles}",
      journal = {\aap},
         year = 2022,
        month = mar,
       volume = {659},
          eid = {A37},
        pages = {A37},
          doi = {10.1051/0004-6361/202141997},
archivePrefix = {arXiv},
       eprint = {2201.11435},
 primaryClass = {astro-ph.SR},
       adsurl = {https://ui.adsabs.harvard.edu/abs/2022A&A...659A..37K}
}

@ARTICLE{2022ApJ...925..125D,
       author = {{Das}, Barnali and {Chandra}, Poonam and {Shultz}, Matt E. and {Wade}, Gregg A. and {Sikora}, James and {Kochukhov}, Oleg and {Neiner}, Coralie and {Oksala}, Mary E. and {Alecian}, Evelyne},
        title = "{Discovery of Eight ``Main-sequence Radio Pulse Emitters'' Using the GMRT: Clues to the Onset of Coherent Radio Emission in Hot Magnetic Stars}",
      journal = {\apj},
         year = 2022,
        month = feb,
       volume = {925},
       number = {2},
          eid = {125},
        pages = {125},
          doi = {10.3847/1538-4357/ac2576},
archivePrefix = {arXiv},
       eprint = {2109.04043},
 primaryClass = {astro-ph.SR},
       adsurl = {https://ui.adsabs.harvard.edu/abs/2022ApJ...925..125D}
}

@ARTICLE{2002ApJ...568..965O,
       author = {{Owocki}, Stanley P. and {Puls}, Joachim},
        title = "{Ion Runaway Instability in Low-Density, Line-driven Stellar Winds}",
      journal = {\apj},
         year = 2002,
        month = apr,
       volume = {568},
       number = {2},
        pages = {965-978},
          doi = {10.1086/339037},
       adsurl = {https://ui.adsabs.harvard.edu/abs/2002ApJ...568..965O}
}

@ARTICLE{2021ApJ...921....9D,
       author = {{Das}, Barnali and {Chandra}, Poonam},
        title = "{Ultra-wideband, Multiepoch Radio Study of the First Discovered ``Main-sequence Radio Pulse Emitter'' CU Vir}",
      journal = {\apj},
         year = 2021,
        month = nov,
       volume = {921},
       number = {1},
          eid = {9},
        pages = {9},
          doi = {10.3847/1538-4357/ac1075},
archivePrefix = {arXiv},
       eprint = {2107.00849},
 primaryClass = {astro-ph.SR},
       adsurl = {https://ui.adsabs.harvard.edu/abs/2021ApJ...921....9D}
}

@ARTICLE{2022MNRAS.516.2812C,
       author = {{Chojnowski}, S. Drew and {Hubrig}, Swetlana and {Labadie-Bartz}, Jonathan and {Rivinius}, Thomas and {Sch{\"o}ller}, Markus and {Niemczura}, Ewa and {Nidever}, David L. and {Stutz}, Amelia M. and {Hummel}, C.~A.},
        title = "{Trumpler 16-26: a new centrifugal magnetosphere star discovered via SDSS/APOGEE H-band spectroscopy}",
      journal = {\mnras},
         year = 2022,
        month = oct,
       volume = {516},
       number = {2},
        pages = {2812-2823},
          doi = {10.1093/mnras/stac2396},
archivePrefix = {arXiv},
       eprint = {2208.12722},
 primaryClass = {astro-ph.SR},
       adsurl = {https://ui.adsabs.harvard.edu/abs/2022MNRAS.516.2812C}
}

@ARTICLE{2022MNRAS.517.5756D,
       author = {{Das}, Barnali and {Chandra}, Poonam and {Shultz}, Matt E. and {Leto}, Paolo and {Mikul{\'a}{\v{s}}ek}, Zden{\v{e}}k and {Petit}, V{\'e}ronique and {Wade}, Gregg A.},
        title = "{Testing a scaling relation between coherent radio emission and physical parameters of hot magnetic stars}",
      journal = {\mnras},
         year = 2022,
        month = dec,
       volume = {517},
       number = {4},
        pages = {5756-5769},
          doi = {10.1093/mnras/stac3123},
archivePrefix = {arXiv},
       eprint = {2210.14746},
 primaryClass = {astro-ph.SR},
       adsurl = {https://ui.adsabs.harvard.edu/abs/2022MNRAS.517.5756D}
}

@ARTICLE{2021A&A...654A.119P,
       author = {{Przybilla}, N. and {Fossati}, L. and {Jeffery}, C.~S.},
        title = "{HD 144941: the most extreme helium-strong star}",
      journal = {\aap},
         year = 2021,
        month = oct,
       volume = {654},
          eid = {A119},
        pages = {A119},
          doi = {10.1051/0004-6361/202141625},
archivePrefix = {arXiv},
       eprint = {2110.11267},
 primaryClass = {astro-ph.SR},
       adsurl = {https://ui.adsabs.harvard.edu/abs/2021A&A...654A.119P}
}

@ARTICLE{2021MNRAS.507.1283S,
       author = {{Shultz}, M.~E. and {Kochukhov}, O. and {Labadie-Bartz}, J. and {David-Uraz}, A. and {Owocki}, S.~P.},
        title = "{Detection of an extremely strong magnetic field in the double-degenerate binary merger product HD 144941}",
      journal = {\mnras},
         year = 2021,
        month = oct,
       volume = {507},
       number = {1},
        pages = {1283-1295},
          doi = {10.1093/mnras/stab2162},
archivePrefix = {arXiv},
       eprint = {2107.11211},
 primaryClass = {astro-ph.SR},
       adsurl = {https://ui.adsabs.harvard.edu/abs/2021MNRAS.507.1283S}
}

@ARTICLE{2025A&A...700A..14S,
       author = {{Shultz}, M.~E. and {Berry}, I. and {Bohlender}, D. and {Catanzaro}, G. and {Giarrusso}, M. and {Klement}, R. and {Labadie-Bartz}, J. and {Leone}, F. and {Leto}, P. and {Neiner}, C. and {Owocki}, S.~P. and {Rivinius}, Th. and {ud-Doula}, A. and {Wade}, G.~A.},
        title = "{Discovery of the binary nature of the magnetospheric B-type star {\ensuremath{\rho}} Oph A}",
      journal = {\aap},
         year = 2025,
        month = aug,
       volume = {700},
          eid = {A14},
        pages = {A14},
          doi = {10.1051/0004-6361/202554023},
archivePrefix = {arXiv},
       eprint = {2505.08007},
 primaryClass = {astro-ph.SR},
       adsurl = {https://ui.adsabs.harvard.edu/abs/2025A&A...700A..14S}
}

@ARTICLE{2015JATIS...1a4003R,
       author = {{Ricker}, George R. and {Winn}, Joshua N. and {Vanderspek}, Roland and {Latham}, David W. and {Bakos}, G{\'a}sp{\'a}r {\'A}. and {Bean}, Jacob L. and {Berta-Thompson}, Zachory K. and {Brown}, Timothy M. and {Buchhave}, Lars and {Butler}, Nathaniel R. and {Butler}, R. Paul and {Chaplin}, William J. and {Charbonneau}, David and {Christensen-Dalsgaard}, J{\o}rgen and {Clampin}, Mark and {Deming}, Drake and {Doty}, John and {De Lee}, Nathan and {Dressing}, Courtney and {Dunham}, Edward W. and {Endl}, Michael and {Fressin}, Francois and {Ge}, Jian and {Henning}, Thomas and {Holman}, Matthew J. and {Howard}, Andrew W. and {Ida}, Shigeru and {Jenkins}, Jon M. and {Jernigan}, Garrett and {Johnson}, John Asher and {Kaltenegger}, Lisa and {Kawai}, Nobuyuki and {Kjeldsen}, Hans and {Laughlin}, Gregory and {Levine}, Alan M. and {Lin}, Douglas and {Lissauer}, Jack J. and {MacQueen}, Phillip and {Marcy}, Geoffrey and {McCullough}, Peter R. and {Morton}, Timothy D. and {Narita}, Norio and {Paegert}, Martin and {Palle}, Enric and {Pepe}, Francesco and {Pepper}, Joshua and {Quirrenbach}, Andreas and {Rinehart}, Stephen A. and {Sasselov}, Dimitar and {Sato}, Bun'ei and {Seager}, Sara and {Sozzetti}, Alessandro and {Stassun}, Keivan G. and {Sullivan}, Peter and {Szentgyorgyi}, Andrew and {Torres}, Guillermo and {Udry}, Stephane and {Villasenor}, Joel},
        title = "{Transiting Exoplanet Survey Satellite (TESS)}",
      journal = {Journal of Astronomical Telescopes, Instruments, and Systems},
         year = 2015,
        month = jan,
       volume = {1},
          eid = {014003},
        pages = {014003},
          doi = {10.1117/1.JATIS.1.1.014003},
       adsurl = {https://ui.adsabs.harvard.edu/abs/2015JATIS...1a4003R}
}

@ARTICLE{2025A&A...701A.171F,
       author = {{Frost}, A.~J. and {Sana}, H. and {Le Bouquin}, J.-B. and {Perets}, H.~B. and {Bodensteiner}, J. and {Igoshev}, A.~P. and {Banyard}, G. and {Mahy}, L. and {M{\'e}rand}, A. and {Ram{\'\i}rez-Agudelo}, O.~H.},
        title = "{An interferometric study of B star multiplicity}",
      journal = {\aap},
         year = 2025,
        month = sep,
       volume = {701},
          eid = {A171},
        pages = {A171},
          doi = {10.1051/0004-6361/202554344},
archivePrefix = {arXiv},
       eprint = {2505.02300},
 primaryClass = {astro-ph.SR},
       adsurl = {https://ui.adsabs.harvard.edu/abs/2025A&A...701A.171F}
}

@ARTICLE{2022A&A...667A.163M,
       author = {{Miret-Roig}, N. and {Galli}, P.~A.~B. and {Olivares}, J. and {Bouy}, H. and {Alves}, J. and {Barrado}, D.},
        title = "{The star formation history of Upper Scorpius and Ophiuchus. A 7D picture: positions, kinematics, and dynamical traceback ages}",
      journal = {\aap},
         year = 2022,
        month = nov,
       volume = {667},
          eid = {A163},
        pages = {A163},
          doi = {10.1051/0004-6361/202244709},
archivePrefix = {arXiv},
       eprint = {2209.12938},
 primaryClass = {astro-ph.GA},
       adsurl = {https://ui.adsabs.harvard.edu/abs/2022A&A...667A.163M}
}

@ARTICLE{1995A&A...301..823B,
       author = {{Babel}, J.},
        title = "{Multi-component radiatively driven winds from A and B stars. I. The metallic wind of a main sequence A star.}",
      journal = {\aap},
         year = 1995,
        month = sep,
       volume = {301},
        pages = {823},
       adsurl = {https://ui.adsabs.harvard.edu/abs/1995A&A...301..823B}
}

@article{Owocki2026,
    author = {{Owocki}, S. and {Krticka}, J.  and {Shultz}, M.~E. and {ud-Doula}, A.},
    title = "{Low effective-temperature diminution of Hα emission from
metallic centrifugal magnetospheres fed by ion-runaway winds}",
    journal =  {\mnras, submitted},
    year = 2026
}

@article{Gunderson2026,
    author = {{Gunderson}, S. and {Golay}, W.  and {Codd}, J. and {Huenemoerder}, D.},
    title = "{A Multiwavelength View of $\rho$ Oph II: Disentangling the Variability of the Multi-Star Component C}",
    journal =  {\apj, submitted},
    year = 2026
}





\bsp	
\label{lastpage}

\end{document}